\documentclass[acmsmall]{acmart}
\AtBeginDocument{%
  }

\setcopyright{acmlicensed}
\acmJournal{POMACS}
\acmYear{2025} \acmVolume{9} \acmNumber{1}
\acmArticle{7} \acmMonth{3}
\acmDOI{10.1145/3711700}

\received{October 2024}
\received[revised]{January 2025}
\received[accepted]{January 2025}

\author{Weizhao Tang}
\email{wtang2@andrew.cmu.edu}
\affiliation{\institution{Carnegie Mellon University}\country{USA}}
\orcid{0009-0002-3676-4582}

\author{Rachid El-Azouzi}
\email{rachid.elazouzi@univ-avignon.fr}
\affiliation{%
  \institution{Avignon University}
  \country{France}
}
\orcid{0000-0002-4756-0887}

\author{Cheng Han Lee}
\email{leech@allium.so}
\affiliation{\institution{Allium}\country{USA}}
\orcid{0009-0006-3885-699X}

\author{Ethan Chan}
\email{ethan@allium.so}
\affiliation{\institution{Allium}\country{USA}}
\orcid{0009-0004-5252-310X}

\author{Giulia Fanti}
\email{gfanti@andrew.cmu.edu}
\affiliation{\institution{Carnegie Mellon University}\country{USA}}
\orcid{0000-0002-7671-2624}

\usepackage{macro}

\definecolor{brick}{rgb}{.7, .3, .1}

\colorlet{revclr}{violet!70}
\title{Game Theoretic Liquidity Provisioning in Concentrated Liquidity Market Makers}
\date{}

\begin{abstract}
Automated marker makers (AMMs) are decentralized exchanges that enable the automated trading of digital assets. Liquidity providers (LPs) deposit  digital tokens, which  can be used by traders to execute trades, which generate fees for the investing LPs. In AMMs, trade prices are determined algorithmically, unlike classical limit order books. Concentrated liquidity market makers (CLMMs) are a major class of AMMs that offer liquidity providers flexibility to decide not only \emph{how much} liquidity to provide, but \emph{in what ranges of prices} they want the liquidity to be used. This flexibility can complicate strategic planning, since fee rewards are shared among LPs. We formulate and analyze a game theoretic model to study the incentives of LPs in CLMMs. Our main results show that while our original formulation admits multiple Nash equilibria and has complexity quadratic in the number of price ticks in the contract, it can be reduced to a game with a unique Nash equilibrium whose complexity is only linear. We further show that the Nash equilibrium of this simplified game follows a waterfilling strategy, in which low-budget LPs use up their full budget, but rich LPs do not. Finally, by fitting our game model to real-world CLMMs, we observe that in liquidity pools with risky assets, LPs adopt investment strategies far from the Nash equilibrium. Under price uncertainty, they generally invest in fewer and wider price ranges than our analysis suggests, with lower-frequency liquidity updates. In such risky pools, by updating their strategy to more closely match the Nash equilibrium of our game, LPs can improve their median daily returns by \$116, which corresponds to an increase of 0.009\% in median daily return on investment (ROI). At maximum, LPs can improve daily ROI by 0.855\% when they reach Nash equilibrium. In contrast, in stable pools (e.g., of only stablecoins), LPs already adopt strategies that more closely resemble the Nash equilibrium of our game.
\end{abstract}



\begin{document}

\maketitle

\section{Introduction}

Automated market makers (AMMs) are decentralized exchanges (DEXes) that allow users to exchange cryptocurrency via a smart contract that algorithmically manages liquidity and exchange rates \cite{sok2023}.  As a dominant class of DEXes \cite{dex-rank}, AMMs play an outstanding role more generally as decentralized applications \cite{dapp-rank}.   Today, AMMs drive billions of dollars in daily trading volume on several blockchains \cite{ripple-amm,coingecko}. 

The core functionality of AMMs is facilitated by \emph{liquidity pools}, 
i.e.,  blockchain smart contracts that store and manage cryptocurrency tokens for trading. 
Most liquidity pools store two types of tokens, which we denote $X$ and $Y$; we focus on the two-token class of AMMs in this work. 
A typical trade proceeds in three steps: 
\begin{enumerate}[leftmargin=*]
    \item A trader proposes to pay $\Delta x$ amount of the $X$ token and asks for a quote.
    \item The pool tells the trader they will obtain $\Delta y$ amount of the $Y$ token (assuming there is sufficient liquidity in the pool). 
    \item The trader decides whether to execute the token swap, in which case they also must pay a \emph{trading fee} (in units of $X$) that is proportional to the amount of added tokens.  
\end{enumerate}
The trading fee is used by the AMM to incentivize \emph{liquidity providers} (LPs),
who initially deposit liquidity tokens into the pool to support trades. 

An AMM can broadly be characterized by three policies, which are implemented algorithmically: 
(1) the \emph{exchange rate} of $\Delta x$ for $\Delta y$ (and vice versa) based on the state of the pool,  
(2) the \emph{liquidity investment policy} (i.e., what constitutes a valid deposit/withdrawal), and
(3) the \emph{trading fee reward mechanism}, i.e., how trading fees are allocated to LPs. 
In this paper, we consider two canonical types of LPs, which we call {Legacy automated market makers} (Legacy AMMs) and {concentrated liquidity market makers} (CLMMs). 

\textbf{Legacy AMMs.}
Legacy AMMs  determine the exchange rate via a constant product market maker (CPMM) \cite{sok2023}. CPMMs ensure that the product of pool reserves always remains constant. 
LPs that invest in legacy AMMs must deposit both $X$ and $Y$ tokens, and fee rewards are split among investing LPs proportionally to their initial investment (details in \S\ref{sec:amm:legacy}). 
Although legacy AMMs are extremely widely used \cite{uv2,sushi,balancerv2,curve,cake,orca,raydium}, 
they are known to suffer from the price slippage problem: 
the price of a large trade is significantly worse than that of a small trade; this is due to the hyperbolic shape of the price curve \cite{adams2024dontletmevslip}.

\textbf{Concentrated Liquidity Market Makers (CLMMs).}
To mitigate slippage, a number of AMMs \cite{uv3,sushi,balancerv3,orca}  adopted a scheme called \emph{concentrated liquidity}, first introduced in Uniswap v3 in 2021 \cite{uv3}. 
In a CLMM, an LP decides not only the amount of liquidity they want to add, but also the price range in which the liquidity is active. 
As a result, the LP  only earns fees from trades when the external token price lies in the LP's range of investment.
Investing in a narrower price range grants the LP a higher share of the fees from transactions within that range. 
In CLMMs, LPs suffer less price slippage in price ranges with high liquidity. 
However, CLMMs force LPs to 
choose which price range(s) to invest in, as well as the amount of investment. 
The consequences of these choices are complex --- 
even if an LP could accurately predict future price ranges, 
they would still need to compete against other LPs over their share of the trading fee.

To date, it remains unclear how strategic LPs should invest funds in CLMMs. 
Prior literature has studied incentives in AMMs, including optimal liquidity provision \cite{milionis2023myersonian,cartea2024decentralized}, but no prior work has simultaneously modeled and analyzed the following three  properties of CLMMs: (1) LPs can only invest up to a fixed budget, (2) LPs in CLMMs can invest different amounts in different price ranges, and most importantly, (3) LPs compete against each other for fees, and thus must take into account other LPs' strategies and their budgets \cite{Frtisch23}. 
Most existing works focus on the study of a single LP's strategy \cite{Fan2022} or on the case where LPs are identical \cite{concave-pro-rata-game, Fan2022, Heimbach2023, Bayraktar24,he2024liquiditypooldesignautomated, fan2024strategicliquidityprovisionuniswap}, 
 and in  both cases the budget is assumed to be unlimited~\cite{concave-pro-rata-game}. 
For legacy AMMs,  \cite{concave-pro-rata-game} proposes a framework using symmetric games to show the uniqueness of the symmetric  Nash equilibrium.  However, the game in this work does not incorporate budget constraints, which is an important practical consideration.  It also does not capture LP strategies involving complex combinations of liquidity positions in CLMMs when LPs have different capacity investments.

Our goal in this work is to study the incentives of LPs in CLMMs under a game-theoretic model. 
In particular, we want to understand the following questions:
\begin{quote}
    \emph{Do there exist equilibrium investment strategies for CLMMs, and if so, what are their characteristics in relation to LPs's investment capacity?
    How do the strategies of real-world LPs compare to those at Nash equilibrium? }
\end{quote}

To answer these questions, we make the following contributions: 
\begin{enumerate}
    \item \textbf{Game theoretic model:} 
    We model strategic liquidity provisioning as a game played by rational and selfish LPs. 
    Each LP is constrained by their own budget, rewarded by trading fees, and penalized by impermanent loss (the opportunity cost of not choosing to hold the tokens in hand). 
    Analysis of this game is complex due to its large strategy space, as LPs can invest in a set of price ranges that is quadratic in the number of feasible price range endpoints. 
    However, we prove that this complex game is equivalent to a much simpler game in which each LP's investment can be broken into a much smaller (linear) set of \emph{atomic} price ranges (Thm. \ref{thm:complex-simple}). 

    \item \textbf{Nash equilibrium analysis:} 
    We prove the unique existence of a Nash equilibrium in our simplified game (Thm \ref{thm:ne-simple-unique}). 
    We show that the Nash equilibrium exhibits a waterfilling pattern (Thm. \ref{thm:waterfill}), which reveals a division of LPs by their budget --- poor LPs exhaust their budgets, while rich LPs spend equally. 
    We characterize properties of the Nash equilibrium as a function of LP budgets.

    \item \textbf{Real-World Data Analysis:}
    On the Ethereum blockchain, we compare our theoretical results to the actions of real LPs on two types of liquidity pools -- stable pools and risky pools. 
    We compare LPs' real-world liquidity provisions against their (simulated) best-response actions and Nash equilibrium actions under our game formulation. 
    In stable pools, we show that real LPs deploy strategies similar to the Nash equilibrium of our game, and return on investment at equilibrium is much lower than that in risky pools. 
    In risky pools, we show that most real LPs prefer to invest in few ($<5$) price ranges, each with wide price coverage. 
    While this behavior is far from our predicted Nash equilibrium strategies, we can partially predict LP behavior by solving a modified version of our game with stale information, suggesting that many LPs have not yet taken full advantage of the data available on the blockchain. 
    We show that in risky pools, LPs' median daily return on investment (ROI) can be improved by 0.009\%
    by updating their strategy to a heuristic derived from our game. 
    In dollars, this corresponds to an increase in median daily utility of \$116 and average daily utility of \$222. 
    The difference between their real-world strategy and the true Nash equilibrium reveals a more substantial potential increase of 0.855\%  in daily ROI and \$13,352 increase in net daily profit.
    We have released our code for computing the equilibrium strategy on \href{https://github.com/WeizhaoT/CLMM-game}{GitHub}.
\end{enumerate}

\section{Model}
\label{sec:amm:model}

\subsection{Automated Market Maker Basics}
\label{sec:amm:amm}
An AMM 
supports trades between two cryptocurrency tokens $X$ and $Y$ via a liquidity pool, which stores reserves of both tokens.
The \emph{liquidity tokens} stored in the pool are provided by \emph{liquidity providers} (LPs).
Let $x$ and $y$ denote the liquidity token amounts of $X$ and $Y$, respectively, stored in the pool at a given point in time.
AMMs are also associated with a \emph{fee rate} $\gamma \ge 0$, which determines the fees paid by traders. 
To trade $X$ for $Y$ tokens in the AMM, for example,  a trader pays $(1+\gamma)\Delta x$, the amount of $X$ they are willing to pay; the AMM algorithmically determines the amount of $Y$ the trader gets back, denoted by $\Delta y$. 
After the trade, the AMM now keeps $x + (1+\gamma) \Delta x$ $X$ tokens and $y-\Delta y)$ $Y$ tokens. 
These tokens are stored separately -- 
an independent fee reserve keeps $\gamma \Delta x$ $X$ tokens that do \textbf{not} provide liquidity, while all the remaining tokens belong in the liquidity pool. 
The tokens in the fee reserve will be distributed among the LPs as a reward for their investment.
The total pool reserves thus change by amounts $(\Delta x, \Delta y)=((1+\gamma) \Delta x, -\Delta y)$ (or vice versa if trading $Y$ for $X$). 
The precise mechanism for allocating fees to LPs is the central topic of this paper; we discuss different allocation mechanisms in Section \ref{sec:amm:model}.

\paragraph*{Bonding Curves}
Among the AMMs, we focus on constant function market makers (CFMMs), which are characterized a publicly-known function $\phi$, namely a \emph{bonding curve}. 
Most pools \cite{uv2, uv3,sushi,orca,raydium} ensure $\phi$ to be monotonically decreasing, convex and differentiable. 
To determine $\Delta y$ from $\Delta x$, we solve the following system of equations: 
\begin{equation}
    y = \phi(x), \qquad y - \Delta y = \phi(x+\Delta x).
\end{equation}

\begin{wrapfigure}{R}{.3\linewidth}
    \begin{center}
        \includegraphics[width=\linewidth]{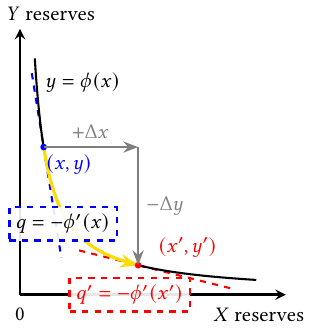}
    \end{center}
    \caption{Example AMM bonding curve and trade.}
    \label{fig:amm-curve}
\end{wrapfigure}

Fig. \ref{fig:amm-curve} demonstrates a bonding curve along with an example trade.
From the initial token amounts $(x, y)$ in the pool, a trader paid $\Delta x = x' - x$ to the AMM, and got $\Delta y = y - y'$ in return.
The token amounts shift to $(x', y')$ along the yellow curve.

\paragraph*{Prices} In addition to the token amounts in the liquidity pool,
another important state of an AMM is the \emph{pool price} $q$, a.k.a. marginal price.
Pool price $q$ is defined as the value of a unit of token $X$ in the amount of $Y$ in an infinitesimal trade under the current liquidity allocation.
Formally, $q = - \mathrm{d}y/\mathrm{d}x = -\phi'(x)$.
Typically, a trade changes not only the token amounts $(x, y)$ in an AMM, but also the pool price.
For example, in Fig. \ref{fig:amm-curve}, the pool price changes from $q$ to $q'$ after a trade.
In contrast, the \emph{average price} of the trade is $\Delta x / \Delta y$, which is typically different than the pool price, and worse for the trader (this can be seen by Jensen's inequality, as the bonding curve is convex).
The difference between the average price and the initial pool price $q$ is called \emph{price slippage}.
Traders prefer lower  slippage.

In this work, we will assume $Y$ is a fiat-pegged token (e.g., USDC in Ethereum), which can be traded for government-issued currency at a fixed rate.
\footnote{We extend our results to tokens with variable prices in App. \ref{amm:app:model-general}.}   
For simplicity, we let ``dollar'' represent any such government-issued currency.
Hence, the dollar price of $Y$ is always $1$ and the dollar price of $X$ is exactly $q$.
We also assume there exists a large external market with marginal price $q_{\mathtt{ext}}$, which can be modeled by a stationary ergodic process with a known distribution $\pi$ \cite{Fan2022}. 
If at some time, the pool price $q = q_0$ deviates such that $q_0 \neq q_{\mathtt{ext}}$,
an arbitrageur can profit from an AMM trade that shifts $q$ towards $q_{\mathtt{ext}}$. 
The arbitrageur's gain is maximized at $q = q_{\mathtt{ext}}$ \cite{amm101}\footnote{We assume the fee rate $\gamma$ is small enough to make this approximation.}. 
Hence, We assume the arbitrageur is fast enough to keep $q = q_{\mathtt{ext}}$ at all times. 
As a result, the pool price $q$ follows the same stationary ergodic process.

\paragraph*{Liquidity}
To support trades, a pool must be supplied with liquidity tokens, which are provided by liquidity providers (LPs). 
During a liquidity addition, an LP adds $\Delta x$ $X$ tokens and $\Delta y$ $Y$ tokens to the liquidity pool, which shifts pool reserves from $(x, y)$ to $(x+\Delta x, y + \Delta y)$. 
$\Delta x$ and $\Delta y$ cannot be determined arbitrarily; instead, they must be added proportionally. 
In other words, there exists ratios $r_{XY}, r_{YX} \ge 0$, such that $\Delta y = r_{YX} \Delta x$ or $\Delta x = r_{XY} \Delta y$. 
These ratios depend on the pool price $q$. 
Generally, the higher $q$ is, the more valuable $X$ tokens are, and the fewer $X$ tokens the LP provides with the same number of $Y$ tokens. 
As a result, the LP's share of liquidity tokens, which are initially $(\Delta x, \Delta y)$, will change as $q$ changes in trades. 
Therefore, when they remove liquidity, they retrieve different amounts of tokens from what they added to the pool, which depend on the pool price at removal. 
This makes it inconvenient to describe the LP's liquidity share with the tuple $(\Delta x, \Delta y)$. 

In response, AMM designers \cite{uv3} introduce a virtual amount \emph{liquidity}, which we denote by $L$. 
$L$ is fully control by the LP who provided it, and must remain constant regardless of the pool price or other LPs' liquidity. 
$L$ is proportional to both $\Delta x$ and $\Delta y$, which implies the existence of ratios $r_X(q)$ and $r_Y(q)$ as functions of pool price $q$,
such that when an LP deposits or withdraws, it must be in quantities $\Delta x = r_X(q) L$ and $\Delta y = r_Y(q) L$.

In summary, 
we consider AMMs that are uniquely 
characterized by their bonding curve $\phi$, as well as the ratios $r_X$ and $r_Y$. 
Within this class of AMMs, we focus on two of the most widely-implemented AMMs: legacy AMMs and concentrated liquidity AMMs (CLMMs).

\subsection{Legacy AMM}
\label{sec:amm:legacy}
\begin{figure}[!htb]
    \centering
    \begin{subfigure}[t]{0.48\linewidth}
        \centering
        \includegraphics[width=.7\linewidth]{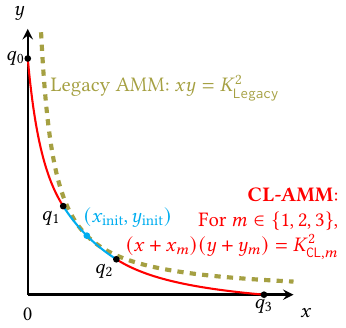}
        \caption{Bonding curves of Legacy AMM and CLMM. Both AMMs were initially funded with $x_{\textrm{init}}$ $X$ tokens and $y_{\textrm{init}}$ $Y$ tokens. CLMM concentrates its liquidity in price ranges $(q_0, q_1), (q_1, q_2)$ and $(q_2, q_3)$, which results in lower curvature than Legacy AMM. }
        \label{fig:CLMM}
    \end{subfigure}
    \hfill
    \begin{subfigure}[t]{0.48\linewidth}
        \centering
        \includegraphics[width=.65\linewidth]{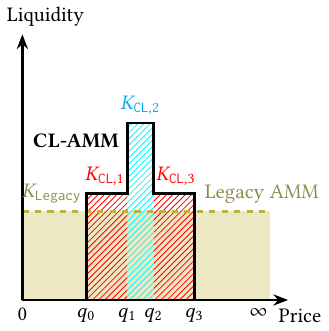}
        \caption{Histogram of liquidity corresponding to the bonding curves. Legacy AMM's liquidity is constant $K_{\texttt{Legacy}}$ over all prices in $\R_{> 0}$, while CLMM's liquidity is flexibly distributed ($K_{\texttt{CL}, m}$ in $(q_{m-1}, q_m)$ for $m \in \{1,2,3\}$) as per the LPs' choices.}
        \label{fig:cl-hist}
    \end{subfigure}
    \caption{Comparison between Legacy AMM and CLMM in their bonding curves and liquidity distributions. }
    \Description[test]{test test}
    \label{fig:amm-model}
\end{figure}

Legacy AMMs (including Uniswap v2 and more \cite{uv2,sushi,balancerv2,curve,cake,orca,raydium}) implement a hyperbolic bonding curve $y = \phi(x) = K_{\texttt{Legacy}}^2/x$, where $K_{\texttt{Legacy}}$ is the total liquidity from all LPs. 
It is shown as the dashed golden curve in Fig. \ref{fig:CLMM}.
When an LP deposits or withdraws liquidity at price $q$,
their token-liquidity ratios are $r_X(q) = 1/\sqrt{q}$ and $r_Y(q) = \sqrt{q}$, such that $L$ units of liquidity are worth $L/\sqrt{q}$  tokens of $X$ plus $L\sqrt{q}$ tokens of $Y$. 

\subsection{CLMM}
Concentrated liquidity market makers (CLMMs) \cite{uv3,sushi,balancerv3,orca} implement more sophisticated mechanisms and were introduced to create more efficient liquidity markets.
In addition to liquidity $L$, an LP also needs to select a price range $(a, b)$ when depositing liquidity, such that the LP's liquidity will be frozen if the price $q$ is outside the range $(a, b)$.
As a result, the tuple $(L, a, b)$ forms a \emph{liquidity position}.
When the LP wants to withdraw liquidity, they must decrease $L$ from an existing liquidity position they own.
When the price is $q$, the actual token amounts $(\Delta x, \Delta y)$ of position $(L, a, b)$ are given by
\begin{equation}
    \Delta x = L \left( \frac{1}{\sqrt{\widehat q}} - \frac{1}{\sqrt{b}} \right), \qquad \Delta y = L \left( \sqrt{\widehat q} - \sqrt{a} \right), \qquad \text{where~} \widehat q \triangleq \max\{a, \min\{b, q\}\}.
    \label{eqn:liqv3}
\end{equation}

By \eqref{eqn:liqv3}, $\Delta y = 0$ when $q \le a$ and $\Delta x = 0$ when $q \ge b$.
In other words, when the $X$ token is cheaper than $a$, the liquidity will be stored as pure $X$ tokens; when $X$ is more valuable than $b$, the liquidity will be stored as pure $Y$ tokens.
Further, let $\mathtt{USD}$ denote the dollar value of liquidity position $(L, a, b)$ at price $q$.
We have
\begin{align}
    \mathtt{USD} = L \left(\sqrt{\widehat q} - \sqrt{a} + \dfrac{q}{\sqrt{\widehat q}}   - \dfrac{q}{\sqrt{b}}\right) \triangleq L \cdot \epsilon_{a, b}(q).
    \label{eqn:epsilon}
\end{align}

Hence, we may regard $\epsilon_{a, b}(q)$ as the dollar price of liquidity over price range $(a, b)$ at pool price $q$.
We note that when $q \ge b$, $\epsilon_{a, b} \equiv \sqrt{b} - \sqrt{a}$ becomes a constant because the liquidity is purely $Y$ tokens with constant dollar value.
However, when $q \le a$, $\epsilon_{a, b}(q) = q\left(a^{-\frac{1}{2}} - b^{-\frac{1}{2}}\right)$ increases in $q$.
This asymmetry results from pegging $Y$ to the dollar.

In practical implementations, both $a$ and $b$ are chosen from a finite set of \emph{ticks} $T = (t_i)_{i=0}^M$, where $t_0 < t_1 < \cdots < t_M $.
Collectively, all the liquidity positions aggregate into a liquidity histogram, as shown in Fig. \ref{fig:cl-hist}.
The histogram corresponds to a piecewise-hyperbolic bonding curve in Fig. \ref{fig:CLMM}.

Generally, the higher the liquidity, the lower the curvature on the bonding curve, and the less price slippage.
For example, in Fig. \ref{fig:cl-hist}, the CLMM concentrates liquidity within the price range $[q_0, q_3]$ which is higher than that of the legacy AMM $L_0$.
Consequently, when traders trade within $[q_0, q_3]$, they suffer from a lower price slippage in the CLMM than in the legacy AMM, while the LPs spend the same amount of tokens $(x_{\textrm{init}}, y_{\textrm{init}})$ on liquidity.
See App. \ref{amm:app:clamm-phi} for the analytical form of the bonding curve.

\begin{remark}
A CLMM collapses into a legacy AMM if $T = \{0, \infty\}$, i.e., every LP only chooses price range $(0, \infty)$ for their liquidity positions.
\end{remark}

\subsection{Incentive Mechanisms for Liquidity Providers}
\label{sec:amm:reward-cost}

Liquidity providers (LPs) 
are incentivized by the fee reward from traders.
Recall that each time a trader pays $\Delta x$ $X$ tokens (or $\Delta y$ $Y$ tokens) 
to the liquidity pool, they additionally pay a fee of $\gamma \Delta x$ $X$ tokens (or $\gamma \Delta y$ $Y$ tokens), which is divided among the LPs.

\paragraph*{Legacy AMMs}
In a legacy AMM, the fee share of each LP is simply proportional to their provided liquidity.
For instance, if an LP provided $\ell$ liquidity out of total liquidity $L$, they get a reward of $(\gamma \Delta x \cdot \ell/L)$  tokens of $X$ after the trade. ($Y$ token fee is analogous.)

\paragraph*{CLMMs}
In a CLMM, the fee sharing policy is more sophisticated.
Let $q$ and $q'$ be the pool prices before and after a trade, and without loss of generality, we assume $q < q'$.\footnote{Since the bonding curve $\phi$ is differentiable and convex, given $q$ and one of $q', \Delta x, \Delta y$, we can derive the other two. }
Let there be $N$ LPs indexed by $[N]$.\footnote{Throughout the paper, we use notation $[k] \triangleq \{1, 2, \cdots, k\}$ for $k \in \mathbb{Z}_{> 0}$. }
Further, we let $L_{n, (a, b)}$ denote the liquidity that LP $n$ provides on price range $(a, b)$.
The space of all the possible price ranges, namely the set of \emph{general ranges}, is denoted by $\gp \triangleq \{ (a, b) \in T^2 | a < b \}$ with cardinality $|\gp| = M(M+1)/2$.
In contrast, we define a set of \emph{atomic ranges} $\ap \triangleq \{ (t_{m-1}, t_m) | m \in [M] \}$, which only includes indivisible price ranges, i.e., ranges covering exactly two neighboring ticks.

\begin{wrapfigure}[14]{R}{.3\linewidth}
    \begin{center}
        \includegraphics[width=\linewidth]{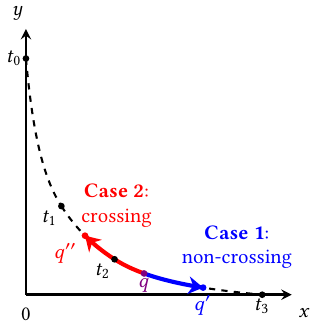}
    \end{center}
    \vskip-2ex
    \caption{Trades in CLMM.}
    \label{fig:crossing}
\end{wrapfigure}

\textbf{Case 1.} If the trade is a \emph{non-crossing} trade (e.g., the \textcolor{blue}{blue} trade in Fig. \ref{fig:crossing}), i.e., there exists an atomic range $(t_{m-1}, t_{m})$ such that $t_{m-1} \le q < q' \le t_{m}$,
then the fee is shared only among LPs with \emph{active} positions, which refers to liquidity positions $(L, a, b)$ satisfying $a \le t_{m-1} < t_{m} \le b$.
Each LP $n$ is weighted by their \emph{active} liquidity $K_{n, m}$, the sum of liquidity of all $n$'s positions covering atomic range $(t_{m-1}, t_{m})$.
Formally, $n$'s weight equals $K_{n, m} / \sum_{i \in [N]} K_{i, m}$, where
\begin{equation}
    K_{n, m} \triangleq \sum_{(a,b) \in \gp: (a, b) \supseteq (t_{m-1}, t_m)} L_{n, (a, b)} = \sum_{i=0}^{m-1} \sum_{j=m}^M L_{n, (t_i, t_j)}.
    \label{eqn:liq-atomic}
\end{equation}

\begin{figure}[!htb]
    \centering
    \includegraphics[width=\linewidth]{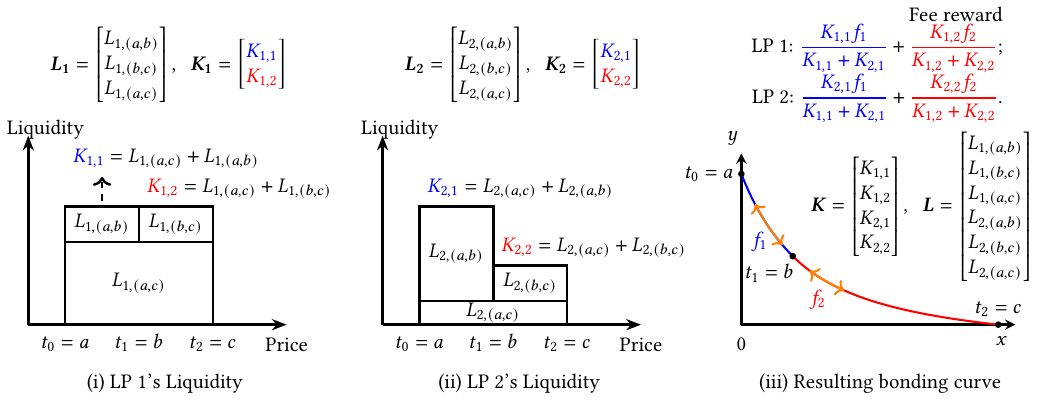}
    \caption{$N=2$ LPs sharing fee across $M=2$ atomic intervals. Interval $m \in \{1,2\}$ has a fee reward of $f_m$. 
    Both LPs invested in 3 liquidity positions, one for each of the general price ranges $(a,b)$, $(b,c)$, and $(a,c)$.
    Recall that $L_{n, (a, b)}$ denotes LP $n$'s liquidity in interval $(a,b)$ (which can span multiple ticks), whereas $K_{n, m}$ denotes all of $n$'s active liquidity in the atomic range  $(t_{m-1},t_m)$.
    }
    \Description[test]{test test}
    \label{fig:fee-share}
\end{figure}

Fig. \ref{fig:fee-share} (iii) shows an example of fee sharing.

\textbf{Case 2.} If the trade spans more than one atomic range (e.g., the \textcolor{red}{red} trade in Fig. \ref{fig:crossing}), i.e., there exists $i, j$ such that $1 \le i \le j \le M$ and $q < t_i < \cdots < t_j < q'$,
the CLMM decomposes the trade into a sequence of $j-i+2$ non-crossing trades that shift the pool price from $q$ to $t_i$, $t_i$ to $t_{i+1}$, ..., and $t_j$ to $q'$, respectively.
The trading fee is separately computed for each non-crossing trade.

\paragraph*{Liquidity Vectors}
For each LP $n \in [N]$, we define vectors $\Lb_n \triangleq ( L_{n, (a, b)})_{(a, b) \in \gp}$ and $\Kb_n \triangleq ( K_{n, m} )_{m \in [M]}$, which concatenate an LP's liquidity and active liquidity, respectively, for all of the price ranges.
Further, we define globally concatenated vectors $\Lb \triangleq ( \Lb_n )_{n \in [N]} \in \R^{N|\gp|}$ and $\Kb \triangleq ( \Kb_n )_{n \in [N]} \in \R^{NM}$, which concatenate across LPs.
Fig. \ref{fig:fee-share} shows an example of this notation.

Eq. \eqref{eqn:liq-atomic} defines a linear relation between $\Lb_n$ and $\Kb_n$.
We condense \eqref{eqn:liq-atomic} into a linear map $\theta: \R^{|\gp|} \to \R^{|\ap|} = \R^M$, such that $\Kb_n = \theta(\Lb_n)$ if and only if \eqref{eqn:liq-atomic} holds for $n$ and \textbf{all} $m \in [M]$. 
Essentially, $\theta$ \emph{summarizes} a complex composition of general-range liquidity positions by LP $n$, represented by $\Lb_n$, into a simpler active liquidity histogram $\Kb_n$. 
In addition, to summarize relations $\Kb_n = \theta(\Lb_n)$ for all $n \in [N]$, 
we define linear map $\Theta: \R^{N|\gp|} \to \R^{NM}$, such that $\Kb = \Theta(\Lb)$ if and only if \eqref{eqn:liq-atomic} holds for all $m \in [M]$ and \textbf{all} $n \in [N]$. 
Like $\theta$, $\Theta$ is also a liquidity histogram summarizer, except that $\Theta$ maps to histograms globally for every LP, while $\theta$ only does this for a single LP. 

\subsubsection{Impermanent Loss}
\label{sec:amm:cost}
LPs  suffer from \emph{impermanent loss}, a phenomenon by which the total dollar value of an LP's  liquidity tokens always decreases when the pool price changes \cite{fan2024strategicliquidityprovisionuniswap,loesch2021impermanentlossuniswapv3}. As a result, absent the fee reward, an LP earns more by withholding liquidity tokens instead of investing them into an AMM.
We define impermanent loss as the LP's opportunity cost of choosing to deposit liquidity instead of holding tokens.
For LPs to be incentivized to deposit liquidity, their fee rewards must exceed their impermanent loss. 

Formally, let $q$ be the initial price when an LP provided liquidity with token amounts $(\Delta x, \Delta y)$.
The dollar value of the LP's initial funds is $\vinit \triangleq q \Delta x + \Delta y$.
At a new price $q'$, the LP's liquidity turns into token amounts $(\Delta x', \Delta y')$ with dollar value $\vlp \triangleq q' \Delta x' + \Delta y'$.
Hypothetically, if they had held the tokens, the dollar value of their tokens would be $\vhold \triangleq q' \Delta x  +\Delta y$.
The total impermanent loss is $\vhold - \vlp$, which is always non-negative (Prop. \ref{thm:positive-il}). 
Intuitively, the loss is proportional to the LP's initial investment $\vinit$. 
Suppose $\vinit$ is invested in liquidity position $(L, a, b)$, which implies $\vinit = L \epsilon_{a, b}(q)$ and further, $\vhold - \vlp \propto L$.
This motivates the definition of \emph{impermanent loss rate}.

\begin{definition}
    Let $\Delta x, \Delta y, q, \Delta x', \Delta y', q'$ be defined as above.
    The \emph{impermanent loss rate} $\hat \tau$ is 
    \begin{equation}
        \hat \tau \triangleq \frac{\vhold - \vlp}{L} = \epsilon_{a, b}(q) \cdot \frac{q'(\Delta x - \Delta x') + (\Delta y - \Delta y')}{q \Delta x + \Delta y},
        \label{eqn:il}
    \end{equation}
    \label{def:il}
\end{definition}

Note that by the proportional relations, $\hat \tau$ is a constant irrelevant to $\Delta x, \Delta y, \Delta x', \Delta y'$. 
Given this constant, we may derive the total impermanent loss in dollars by
\[
    \text{Impermanent Loss} = L \cdot \hat \tau. 
\]

In CLMMs, $\hat \tau$ is a function of $a, b, q$ and $q'$. 
The analytical form of $\hat \tau$ is in Prop. \ref{thm:il-analytic-general} of App. \ref{amm:app:model-general}.

\section{Game Theoretic Analysis}
\label{sec:amm:ne}

As introduced in \S\ref{sec:amm:reward-cost}, a liquidity provider (LP) is both incentivized by the share of trading fee and disincentivized by the impermanent loss. 
Thus, to understand the LP's investment decisions, it is essential to analyze their net income. 
Since the fee reward of LP $n \in [n]$ depends on the liquidity configuration of all LPs, it turns out that the optimal decision of each LP depends on the  strategies made by other LPs and, since LPs are selfish, are faced with a non-cooperative game \cite{Chapter4NoncooperativeGames}. 

The complex interactions and decision making processes of the AMM are analyzed using the analytical tools of non-cooperative game theory. In particular, we formulate a  non-cooperative game in normal form, $\mathtt{G}([N], \as, \util)$ that is characterized by three main
elements: (i) a set of LPs $[N]$ in AMM, (ii) action of each LP $n\in [N]$ $\Zb_n \in \as_n$, where $\as_n$ denotes the \emph{action space}
of LP $n$, and (iii) a utility function $\util_n$ for each LP $n\in [N]$ that evaluates $n$'s net gains, which $n$ aims to maximize. 
Each $\util_n$ is dependent on not only $n$'s action $\Zb_n$, but also other LPs' actions $\Zb_{-n}$. 
We denote this by $\util_n(\Zb_n; \Zb_{-n})$. 
In the end, we are interested in the \emph{Nash equilibria} of the game, where the LP actions are stable such that no LP finds it beneficial to change its liquidity on each range. 
We formally define a Nash equilibrium as follows \cite{ne}: 

\begin{definition}[Nash equilibrium]
   Consider the proposed non-cooperative game in normal form $\mathtt{G}(P, \as, \util)$. A vector of strategies $\Zb^* = (\Zb^*_n)_{n \in P}$ is called a \emph{Nash equilibrium} if and only if\footnote{Throughout this paper, we use $\boldsymbol{v}_{-n}$ to denote the concatenated vector of $(\boldsymbol{v}_i)_{i \in [N], i \neq n}$. }
   \begin{equation}
       \util_n\left(\Zb^*_n; \Zb^*_{-n}\right) \ge \util_n\left(\Zb_n; \Zb^*_{-n}\right), 
       \qquad \forall \Zb_n \in \as_n, n\in[N].
   \end{equation}
    Further, if the inequality is strict, $\Zb^*$ is a \emph{strict Nash equilibrium}.
    \label{def:ne}
\end{definition}

Notice that in general, LPs' strategy spaces in CLMMs are complex, allowing them to take out liquidity positions on price ranges with arbitrary lower and upper bounds, even spanning multiple tick marks. 
Our first main result show that surprisingly, this game is equivalent to a so-called \emph{atomic} game, in which LPs are only permitted to own liquidity positions over atomic price intervals.
This significantly simplifies the analysis of CLMMs. 

\subsection{The Original Game}
\label{sec:amm:orig-game}

We study a one-shot game among the LPs of an AMM. 
There are $N$ LPs as players in the game, which are indexed by set $[N]$. 
The game has two phases. 

In Phase 1, each
LP $n \in [N]$ chooses an \emph{action} $\Lb_n = ( L_{n, (a, b)} \ge 0)_{(a, b) \in \gp}$, where $L_{n, (a, b)}$ is the liquidity deposited by $n$ into price range $(a, b) \in \gp$. 
By \eqref{eqn:epsilon}, the dollar cost of such liquidity is $L_{n, (a, b)} \cdot \epsilon_{a, b}(q)$, where $q$ is the initial pool price. 
With a total budget of $B_n$ dollars, LP $n$ is subject to the following budget constraint:
\begin{equation}
    \sum_{(a, b) \in \gp} L_{n, (a, b)} \epsilon_{a, b}(q) \le B_n. 
    \label{eqn:orig-budget-constraint}
\end{equation}

In Phase 2, traders make trades, pay trading fees to LPs, and shift the pool price $q$ according to the stationary ergodic process we defined in \S\ref{sec:amm:amm}.
This process determines the utility function $\util_n$ of each LP $n$,
Under this budget constraint, each LP $n$ aims to maximize their utility function $\util_n$,
which equals their expected fee reward $\Fc_n$ minus their expected impermanent loss 
$\Cc_n$: 
\begin{equation}
   \util_n = \Fc_n - \mathbb \Cc_n. \notag 
\end{equation}

\begin{remark}
We assume that Phase 2 is a period of time when LPs do not change their liquidity positions.
In the end, at each (atomic) price range $m$, traders collectively pay $V_m$ dollars in trades, which shifts the liquidity token amounts along the bonding curve and does not include the fees. 
We model $V_m$ as a constant.
Then the fee reward $f_m = \gamma V_m$ (recall that $\gamma$ denotes the fee rate) is also a constant. 
To simplify notation, our formulation uses only $f_m$, and leaves $\gamma$ and $V_m$ out of the picture. 
\end{remark}

A real CLMM can be modeled by a dynamic game, which is a sequence of infinitely many instances of this one-shot game. 
We assume that the parameters of any one-shot instance in the sequence do not depend on prior instances. 
Under this setting, the dynamic game can be analyzed by considering each one-shot instance of the game independently. 
This approximation is reasonable, because our evaluation (\S\ref{sec:amm:alternative-games}) shows that most real-world LPs 
change their allocations daily, or even less frequently.

\subsubsection{Fee Reward}
For the expected \textbf{fee reward} $\Fc_n$, under the assumption of ergodicity of the price process, 
we consider $f_m$ as the average fee reward  over each atomic range $(t_{m-1}, t_m)$ to be shared by LPs whose liquidity positions cover that range.  This modeling choice is made to simplify analysis,
as we do not want to model the individual trades that generate transaction fees. Following {\S\ref{sec:amm:reward-cost}}, we define LP $n$'s total share of expected fee reward as
The expected \textbf{fee reward} $\Fc_n$ is the summation of LP $n$'s share of fees over all atomic ranges, where each atomic range has constant fee $f_m$. 
Under the fee allocation policy from {\S\ref{sec:amm:reward-cost}},  we have
\begin{equation}
    \Fc_n = \sum_{m \in [M]}  f_m \cdot \frac{K_{n, m}^\alpha}{\sum_{i \in [N]} K_{i, m}^\alpha} = \sum_{m \in [M]} f_m \cdot \frac{\left(\sum_{i=1}^m \sum_{j=m}^M L_{n, (t_{i-1}, t_j)}\right)^\alpha}{\sum_{k \in [N]} \left(\sum_{i=1}^m \sum_{j=m}^M L_{k, (t_{i-1}, t_j)}\right)^\alpha}. 
    \label{eqn:reward}
\end{equation}
Recall  from \S\ref{sec:amm:reward-cost} that $K_{n, m}$ denotes the total investment of LP $n$ at the $m$th atomic price range, summed over all investments, whereas $L_{n, (a, b)}$ denotes the investment of LP $n$ at price range $(a,b)$, which need not be atomic. 
Note that we introduced the variable $\alpha > 0$ in both the numerator and the denominator of our fee reward mechanism.  
Most CLMMs in practice use $\alpha = 1$ (e.g., Uniswap v3 \cite{uv3}). 
However, allowing for different values of $\alpha$ enables us to analyze a more general class of fee mechanisms.
For example, when $\alpha > 1$, LPs are encouraged to concentrate their liquidity on the few positions where they have the greatest advantage. 
In the limit as $\alpha \to \infty$, only the LP with the largest investment in each price range receives all fee rewards. 
If $\alpha = 0$, all LPs with \emph{any} liquidity in a given price range are rewarded equally; hence, they are incentivized to allocate an infinitesimal amount of liquidity to  every price range. 

\subsubsection{Impermanent Loss}
For the expected \textbf{impermanent loss} $\mathbb \Cc_n$, we first let $q'$ denote the price by the end of the game.
By Def. \ref{def:il}, the impermanent loss of LP $n$'s per-range liquidity $L_{n, (a, b)}$ equals $L_{n, (a, b)} \cdot \hat\tau_{a, b}(q', q)$ dollars.  
The total expected impermanent loss of LP $n$ is given by
\begin{align}
  \Cc_n =
    \sum_{(a, b) \in \gp} \E_{q' \sim \pi}\left[L_{n, (a, b)} \cdot  \hat\tau_{a, b}(q', q)\right] 
    & = \sum_{(a, b) \in \gp} L_{n, (a, b)} \E_{q' \sim \pi}\left[\hat\tau_{a, b}(q', q) \right] \nonumber  \\
    & \triangleq \sum_{(a, b) \in \gp} L_{n, (a, b)} \bar \tau_{a, b}.
    \label{eqn:il-1}
\end{align}
where $\pi$ is the stationary distribution of the price process $q'$\footnote{$\pi$'s support may or may not intersect with the ticks $T$}, and the (expected) impermanent loss rate $\bar \tau_{a, b} = \E_{q' \sim \pi}\left[\hat\tau_{a, b}(q', q)\right]$ is a constant related to only $a, b, q$ and the stationary distribution $\pi$.
For notational simplicity, we suppress the dependence on $q$ and $\pi$ in $\tau_{a, b}$. 

\subsubsection{Overall Definition}
In summary, we define the game above as an \textbf{original game}, which is denoted by
$\og([N], \oas, \outil)$. 
Recall that $[N]$ denotes the set of players (LPs), and for each LP $n \in [N]$, $\oas_n$ and $\outil_n$ are their action space and utility function, respectively. 
With budget $B_n$, 
\begin{align}
    \oas_n &\triangleq \left\{ \Lb_n \in \R_{\ge 0}^{|\gp|} \middle|  \textstyle \sum_{(a, b) \in \gp} L_{n, (a, b)} \epsilon_{a, b}(q) \le B_n.  \right\}. 
        \label{eqn:orig-action-space} \\
    \outil_n & \triangleq \underbrace{\sum_{m \in [M]} f_m \frac{\left(\sum_{i=1}^m \sum_{j=m}^M L_{n, (t_{i-1}, t_j)}\right)^\alpha}{\sum_{k \in [N]} \left(\sum_{i=1}^m \sum_{j=m}^M L_{k, (t_{i-1}, t_j)}\right)^\alpha}}_{\text{Fee Reward } \Fc_n} - \underbrace{\sum_{(a, b) \in \gp}  \bar \tau_{a, b} L_{n, (a, b)}}_{\text{Impermanent Loss } \mathbb \Cc_n}. 
        \label{eqn:orig-util}
\end{align}
We note that the concept of Nash equilibrium in the original game is particularly important from a dynamic standpoint: in a CLMM, an LP changes its liquidity for certain ranges repeatedly in response to other LPs' strategies and price changes. The stability points of such  systems are exactly those at which no LP finds it advantageous to change  its liquidity on each range. An interesting question is whether the Nash equilibrium exists and is unique, and how the system actually converges to this Nash equilibrium. The following proposition establish the existence of the NE.  

\begin{proposition}
    There exists at least one Nash equilibrium in $\og([N], \oas, \outil)$. 
    \label{thm:orig-ne-exist}
\end{proposition}

By Prop. \ref{thm:orig-ne-exist} (proof in App. \ref{amm:app:orig-ne-exist}), a Nash equilibrium exists.
Undesirably, the Nash equilibrium is not necessarily unique, which makes it impossible to predict LP $n$'s action $\Lb_n$ at Nash equilibrium. 
However, through simulation, we observed empirically that each LP $n$'s liquidity histogram $\theta(\Lb_n)$ is the same regardless of which Nash equilibrium $\Lb_n$ we consider. 
This observation motivates us to theoretically establish the uniqueness of $\theta(\Lb_n)$, which is part of a unique Nash equilibrium in a simplified game that we call the \emph{atomic game}.

\subsection{The Atomic Game}
\label{sec:amm:atom-game}

Intuitively, instead of determining $\Lb_n$, we let each LP $n$ directly determine the total active liquidity per atomic range $\Kb_n$. 
This reduces the dimensionality of $n$'s action space from $|\gp| = M(M+1)/2$ to $|\ap| = M$. 
Namely, the \textbf{atomic game} $\ag([N], \aas, \autil)$ is defined as follows. 
$[N]$ denotes the set of players (LPs). 
For each LP $n$, $\aas_n$ and $\autil_n$ denote their action space and utility function, respectively. 
Formally,
\begin{align}
    \aas_n &\triangleq \left\{ \Kb_n \in \R_{+}^{M} \middle|  \textstyle \sum_{m \in [M]} K_{n, m} \epsilon_{t_{m-1}, t_m}(q) \le B_n  \right\}, 
    \label{eqn:atom-budget-constraint} \\
    \autil_n &\triangleq \sum_{m \in [M]} f_m \frac{K_{n, m}^\alpha}{\sum_{i \in [N]} K_{i, m}^\alpha} - \sum_{m \in [M]} \tau_{m} K_{n, m}.
    \label{eqn:atom-utility}
\end{align}

Here $\tau_m \triangleq \bar \tau_{t_{m-1}, t_m} = \E_{q' \sim \pi}\left[\hat\tau_{t_{m-1}, t_m}(q', q)\right]$ is a constant. 
By Thm. \ref{thm:ne-simple-unique} (proof in App. \ref{amm:app:ne-simple-unique}), 
a Nash equilibrium \emph{uniquely} exists in an atomic game. 

\begin{theorem}[Uniqueness]
    When $0 < \alpha \le 1$, there always exists a unique Nash equilibrium in the atomic game. 
    \label{thm:ne-simple-unique}
\end{theorem}

\begin{remark}
Because of the diagonal strict concavity property \cite{rosen1965dsc}, the unique equilibrium can be solved by various iterative algorithms, such as relaxation \cite{relax-1}, mirror descent learning \cite{zhou2017mirror} and no-regret learning \cite{mertikopoulos2019learning}. 
Moreover, if the same game is played repeatedly and LPs learn according to any of the algorithms, then the actions taken by LPs will converge to the Nash equilibrium. 
\end{remark}

\subsection{Connection Between the Original and Atomic Games}

Intuitively, an atomic game $\ag$ is formulated by simplifying the action space of each LP in the original game $\og$, without changing parameters such as the LPs' budgets and the total fee rewards. 
We define the notion of \emph{twin games} 
to describe such a connection in Def. \ref{def:twin}. 
A pair of twin games satisfies: 
1) there is a correspondence between the action spaces of both games --- the atomic action space can be obtained from the original action space;
2) each player's utility is equal for strategies $\Lb$ of $\og$ and $\Kb$ of $\ag$, if $\Kb = \Theta(\Lb)$, i.e., if they have the same liquidity per atomic price range, so 
$K_{n, m} = \sum_{i=0}^{m-1} \sum_{j=m}^M L_{n, (t_i, t_j)}$ for all $n \in [N]$ and $m \in [M]$ (recall the definition of $\theta$ and $\Theta$ in \S\ref{sec:amm:reward-cost}). 
The notion of twin games is different from that of equivalent games \cite{thompson1952equivalence}, which states that the games' normal forms are equal. In our case, twin games do not necessarily even have the same strategy space. 

\begin{definition}[Twin games]
    Let $\og([N], \oas, \outil)$ and $\ag([N], \aas, \autil)$ be an original game and an atomic game, respectively. 
    We call $\og$ and $\ag$ \emph{twin games} if all the following conditions hold for all $n \in [N]$. 
    \begin{enumerate}
        \item If $\Lb_n \in \oas_n$, then $\Kb_n = \theta(\Lb_n) \in \aas_n$. 
        \item If $\Kb_n \in \aas_n$, then there exists $\Lb_n \in \oas_n$ that satisfies $\Kb_n = \theta(\Lb_n)$.  
        \item If for all $i \in [N]$, $\Lb_i \in \oas_i$ and $\Kb = \Theta(\Lb)$, then $\outil_n(\Lb_n; \Lb_{-n}) = \autil_n(\Kb_n; \Kb_{-n})$. 
    \end{enumerate}
    \label{def:twin}
\end{definition}

Our goal is to establish a connection between the Nash equilibria of the original game $\og$ and the atomic game $\ag$. 
Next we give our first main result: 
Thm. \ref{thm:complex-simple} (proof in App. \ref{amm:app:twin}) states that, 
given a Nash equilibrium $\Lbt$ of the original game $\og$, 
we can directly derive an equilibrium $\Kbt = \Theta(\Lbt)$ of the atomic game $\ag$. 
Throughout this paper, we conventionally use tilde vectors such as $\widetilde{\Lb}, \widetilde{\Kb}$ to represent actions or derived quantities at Nash equilibrium.

\begin{theorem}[Original game and atomic game are twin games]
    Consider an original game $\og$ and an atomic game $\ag$ with the same parameters $\left(N, M, T, \pi, \alpha, (B_n)_{n \in [N]}, (f_m)_{m \in [M]}\right)$.
    Then $\og$ and $\ag$ are twin games.
    In addition, 
    \begin{enumerate}
        \item For every Nash equilibrium $\Lbt$ of $\og$, $\Theta(\Lbt)$ is a Nash equilibrium of $\ag$. 
        \label{item:simplify}
        \item For every Nash equilibrium $\Kbt$ of $\ag$ and strategy $\Lbt$ of $\og$ with $\Kbt = \Theta(\Lbt)$, $\Lbt$ is a Nash equilibrium.
        \label{item:complexify}
    \end{enumerate}
    \label{thm:complex-simple}
\end{theorem}

\begin{figure}[!htb]
    \centering
    \includegraphics[width=\linewidth]{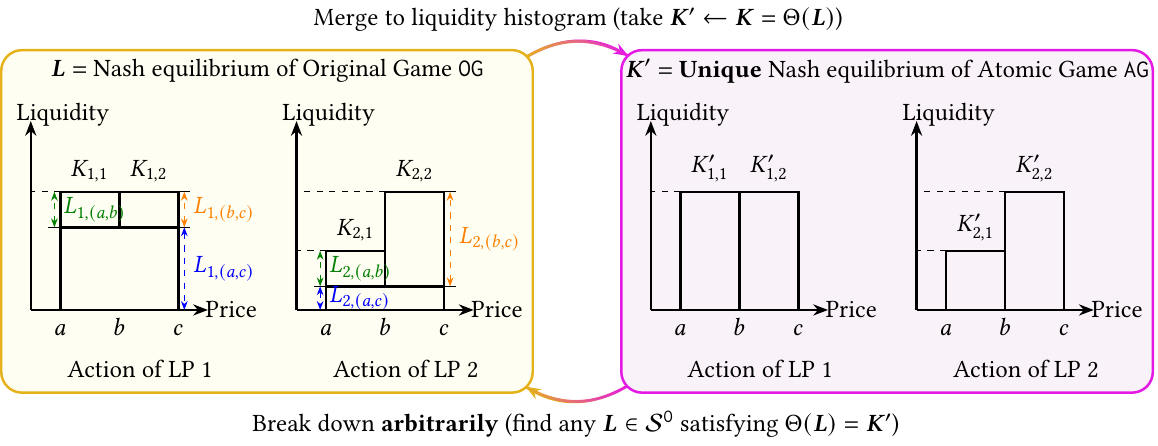}
    \caption{Illustration of Theorem \ref{thm:complex-simple} for an example game with two LPs and two atomic price ranges. In the original game ($\og$) (left), each LP has three possible ranges in which it can invest: $(a,b)$, $(b,c)$, and $(a,c)$. A Nash equilibrium can have allocations in all three ranges. Thm. \ref{thm:complex-simple} implies that from any such Nash equilibrium, we can derive (through the mapping $ \Theta$) a corresponding strategy for the atomic game (right), in which each LP invests only in atomic price ranges $(a,b)$ and $(b,c)$. This strategy turns out to be the \emph{unique} Nash equilibrium for the atomic game, and has the same liquidity in each atomic price range as the original equilibrium of the $\og$. Similarly, for a given Nash equilibrium of the atomic game, any corresponding liquidity allocation in the $\og$ with the same total liquidity per atomic range per LP is guaranteed to be a Nash equilibrium.} 
    \label{fig:original-to-atomic}
    \Description[]{}
\end{figure}

Hence, regardless of \emph{which} Nash equilibrium $\Lbt$ of $\og$ we consider, the following three quantities are unique and equal to that of the atomic game equilibrium $\Kbt$:
1) the atomic liquidity $\Theta(\Lbt)$ (i.e. the amount of liquidity the LPs have invested in each atomic range); 
2) the amount of total budget used by each LP; and
3) the utility of each LP. 
Fig. \ref{fig:original-to-atomic} illustrates Thm. \ref{thm:complex-simple} via an example.

Since the above three quantities are the only properties of interest in this work, there is no need to characterize the multiple Nash equilibria of $\og$. 
Instead, if we find the unique Nash equilibrium of $\ag$, we  obtain all the quantities of interest for \emph{all} Nash equilibria of $\og$. 
Also, note that given the Nash equilibrium $\Lbt$ of the simple game, \emph{any} strategy for the original game that is consistent with $\Lbt$ (i.e., has the same atomic liquidity allocation and budget usage per LP) is also a Nash equilibrium of the original game.

\paragraph*{Special case of Legacy AMMs} 
To apply our game theoretic model to a Legacy AMM, we only need to set the price ticks as $T = \{0, \infty\}$. 
As a consequence, when an LP selects a price range, their choice is unique, which makes $\gp = \ap = \{ (0, \infty) \}$. 
The original game and the atomic game collapse into the same game.

\subsection{Properties of the Unique Equilibrium}
\label{sec:amm:ne-properties}

While it is difficult to express the unique Nash equilibrium of an atomic game in closed form, 
we study properties of the equilibrium theoretically. 
First of all, we introduce notation $A_n$ which represents the total dollar budget used by each LP. 
We require $A_n \le B_n$ by the budget constraint defined in action space \eqref{eqn:atom-budget-constraint}. 

\begin{figure}[!tb]
    \centering
    \includegraphics[width=\linewidth]{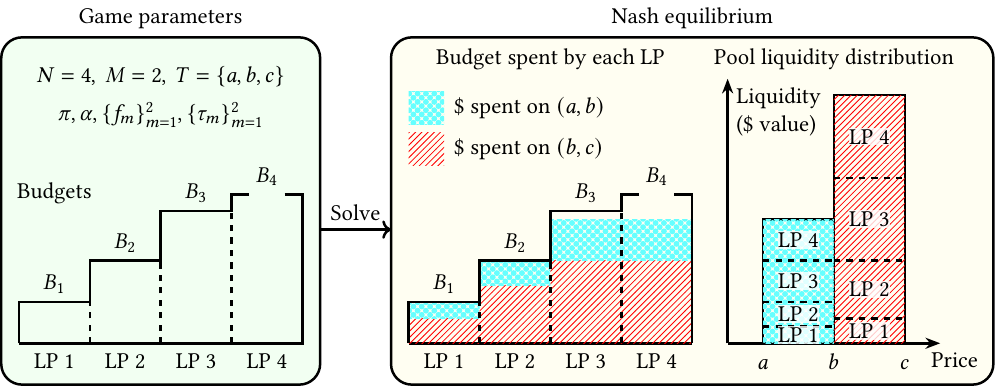}
    \caption{The waterfilling pattern of LPs' strategies at Nash equilibrium among $N=4$ LPs across $M=2$ atomic ranges. Each LP's budget is represented by the height of their corresponding bar.}
    \label{fig:waterfill}
    \Description[test]{test}
\end{figure}

\paragraph*{The waterfilling pattern} We start by showing that the Nash equilibrium of the simple game follows a waterfilling pattern. 
For example, Fig. \ref{fig:waterfill} shows the strategies of four LPs at Nash equilibrium. 
To understand the pattern,  imagine a stair-shaped water tank, which only allows water to be filled from the top.
Each step corresponds to the allocation of one LP, and the height of the step corresponds to that LP's budget. 
After filling a volume of water of $A \triangleq \sum_{n \in [N]} A_n$ into the tank, 
the resulting water level at each LP $n$'s step of the tank is exactly $A_n$. 
In this example, LPs 1 and 2 used up their budgets (i.e., their steps are fully shaded), while LPs 3 and 4 did not. 

More generally, Prop. \ref{thm:waterfill} (proof in App. \ref{amm:app:waterfill}) shows that \emph{any} Nash equilibrium of the atomic game follows a waterfilling pattern. 

\begin{proposition}[Waterfilling]
    Let $\alpha \in (0, 1]$ and $\Kbt$ be the Nash equilibrium. Define $\Nc_+ \triangleq \{n \in [N]|A_n < B_n\}$.
    \begin{enumerate}
        \item For each Range $m \in [M]$, there exists $h_m > 0$, where for all $n \in \Nc_+$, $\Kt_{n, m} = h_m$ and for all $n \notin \Nc_+$, $\Kt_{n, m} \le h_m$. 

        \item There exists $h > 0$, where for all $n \in [N]$, $A_{n} = \min\{h, B_n\}$. 
    \end{enumerate}
    \label{thm:waterfill}
\end{proposition}
The waterfilling pattern  described in Proposition~\ref{thm:waterfill} also implies that all LPs who   do \emph{not} invest their entire budget at Nash equilibrium use the same strategy, i.e., $\Kt_{n, m}=\Kt_{n', m}$, for all $m\in [M]$ and $n, n'\in\Nc_+$. 
Given the total investment amount $A$, the waterfilling  principle determines the investment of each individual LP, $A_n$. However, to determine $A$ (and consequently, $h$), it is necessary to solve for the Nash equilibrium (to the best of our knowledge). This can be done using a number of algorithms, such as those highlighted in \S\ref{sec:amm:atom-game}.

\paragraph{Budget Dominance}
At Nash equilibrium of the atomic game, we can rank the LPs by their total budgets. 
If LP $i$ is ``richer'' (has more budget) than LP $j$, 
then $i$ will put no less budget than $j$ in \textbf{every} atomic range. 
For example, in Fig. \ref{fig:waterfill}, the waterfilling pattern implies that both LPs 1 and 2 must have less budget than LPs 3 and 4, and the latter must have spent equal amount of their budgets on liquidity. 
In addition, if budgets $B_1 < B_2 < B_3 < B_4$, their used budgets follow the same hierarchy: $A_1 \le A_2 \le A_3 \le A_4$. 
Prop. \ref{thm:ne-compare} formalizes this notion, showing that there is a monotonicity among LPs in their use of price ranges: 
an LP with a higher budget can use more price ranges, and  deposit  more liquidity in utilized  price ranges. 
In particular, if two LPs have the same budget, their strategies are identical.

\begin{proposition}
    Let $\alpha \in (0, 1]$ and consider LPs $i, j \in [N]$. Let $\Kbt$ denote the Nash equilibrium of the atomic game. 
    If budgets $B_i < B_j$, then for all price ranges $m \in [M]$, the liquidity allocations of $i$ and $j$ satisfy $\Kt_{i, m} \le \Kt_{j, m}$, with equality only when $\Kt_{i, m} = \Kt_{j, m}= 0$ for  $i, j\in\Nc_+$.  Moreover, when $B_i=B_j$, it holds that for all atomic ranges $m \in [M]$, 
    their provide equal utility: $\Kt_{i, m} = \Kt_{j, m}$.
    \label{thm:ne-compare}
\end{proposition}


\paragraph*{Other observations}
We collect some other relevant observations. 
First (Prop. \ref{thm:positive-liq}, proof given by Cor. \ref{thm:positive-liq-alpha}), if the index $\alpha < 1$, every LP provides \emph{non-zero} liquidity on every range; this may not hold for $\alpha = 1$. 

\begin{proposition}[Positive liquidity at equilibrium]
    Let $0 < \alpha < 1$. At equilibrium, every LP provides \emph{non-zero} liquidity on every range. That is, $\Kbt$, $\Kt_{n, m} > 0$ for all $ m, n \in [M] \times [N]$. 
    \label{thm:positive-liq}
\end{proposition}

Second Prop. \ref{thm:const-util} (proof in App. \ref{amm:app:const-util}) shows that if all LPs spend the same amount of budget while at least one LP does not fully use their budget,
all LPs' utilities are the same constant regardless of impermanent loss.  
Therefore, the Nash equilibrium is symmetric.
An interesting point in this particular scenario is that the LPs' utility increases as $\alpha$ decreases. 
This is because LPs respond by reducing their investment, thereby reducing the impermanent loss while keeping the same fee reward from the trades.

\begin{proposition}
    Let $\alpha \in (0, 1]$ and $\Kbt$ be the Nash equilibrium. 
    If for all $ i, j \in [N]$, $A_{i} = A_{j}$ and there exists $n_0 \in [N]$ such that $A_{n_0} < B_{n_0}$, 
    then for every LP $n \in [N]$, its utility $\util_n = \displaystyle\frac{(1-\alpha) N + \alpha}{N^2} \sum_{m \in [M]} f_m$.
    \label{thm:const-util}
\end{proposition}

\section{Experiments}

In this section, we compare our game theoretic model with data from real-world CLMMs. 
\footnote{See our analytic tool in \url{https://github.com/WeizhaoT/CLMM-game}.}
We focus on four Uniswap v3 liquidity pools on the Ethereum blockchain, listed in Table \ref{tab:pools}.
There exist two classes of pools by the tokens they reserve: 1) \textbf{Stable Pool}, which reserves two stablecoins (USDT and USDC, in T5); and 2) \textbf{Risky Pool}, which reserves a stablecoin (USDC) and a risky asset (WETH or WBTC, in all other 4 pools).  
USDT and USDC are pegged to the US dollar, while WETH and WBTC stand for wrapped Ether and wrapped Bitcoin, respectively. 
For each liquidity pool, we crawl three types of events: 1) \texttt{Swap}, a trade of one type of token for another; 2) \texttt{Mint}, an addition of liquidity; and 3) \texttt{Burn}, a removal of liquidity. 
We collect \texttt{Mint} and \texttt{Burn} events from the entire lifetime of each pool to ensure that our estimate of liquidity in the pool is complete, but we consider \texttt{Swap} events only from Jan 1, 2024 to June 30, 2024 to focus on the most recent period of time. 
Our data was collected from the Allium platform \cite{allium}.

\begin{table}[!htb]
    \centering\small
    \begin{tabular}{*{6}{c}}
    \toprule
    Nickname & Token X & Token Y & Fee Rate & \# Events & Address \\
    \midrule
    B30 & WBTC & USDC & 0.3\% & 43172 & 0x99ac8ca7087fa4a2a1fb6357269965a2014abc35 \\
    E100 & USDC & WETH & 1\% & 9126 & 0x7bea39867e4169dbe237T55c8242a8f2fcdcc387 \\
    E30 & USDC & WETH & 0.3\% & 247477 & 0x8aT599c3a0ff1de082011efddc58f1908eb6e6d8 \\
    E5 & USDC & WETH & 0.05\% & 1428340 & 0x88e6a0c2ddd26feeb64f039a2c41296fcb3f5640 \\
    T5 & USDC & USDT & 0.05\% & 20697 & 0x7858e59e0c01ea06df3af3d20ac7b0003275d4bf \\
    \bottomrule
    \end{tabular}
    \caption{Metadata of Uniswap v3 liquidity pools.}
    \label{tab:pools}
    \vspace{-0.2in}
\end{table}

\subsection{Game Setup}
\label{sec:amm:game-setup}
We assume that each LP can change their action daily. 
Hence, we split the blockchain data events into 182 daily chunks and construct a game for each chunk. 
In each game, we specify the game parameters including the budgets of player LPs $(B_1, \cdots, B_N)$, the price ticks $(t_0, \cdots, t_M)$, the average fee reward of each range $(f_1, \cdots, f_M)$ , the share of non-player LPs in each range $(\chi_1, \cdots, \chi_M)$ (more details below and in App. \ref{amm:app:fee-assume-3}), and the impermanent loss $(\tau_1, \cdots, \tau_M)$. 
Next, we explain how we determine these parameters. 

\subsubsection{Players and Their Budgets}
\begin{figure}[tb]
    \centering
    \includegraphics[width=.9\linewidth]{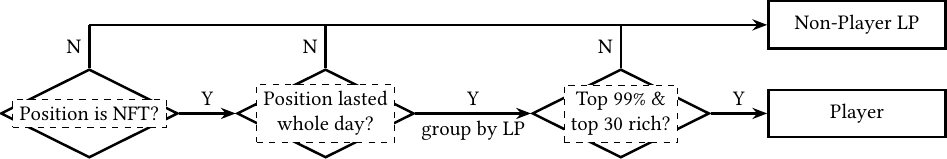}
    \caption{Logical categorization of player and non-player LPs. 
    We include LPs who hold NFT positions that lasted a full day, and are both among the top 30 investors that hold at least 99\% of daily investments.
    }
    \label{fig:player-flow}
    \Description{}
    \vspace{-0.1in}
\end{figure}

To make a meaningful comparison between the data and our model, we need to filter out certain LPs and determine their effective budgets. 
We determine the identity of the players and their liquidity of each daily game following the flow chart in Fig. \ref{fig:player-flow}. 
First, we collect all liquidity positions from \texttt{Mint} events at the beginning of the day. 
When a position is minted, the creator typically receives a non-fungible token (NFT) which is unique and tractable, or a fungible token if they create it through a vault service such as Arrakis Finance \cite{arrakis}.
It is difficult to track ownership and transfers of fungible tokens, so we regard the position as being owned by a non-player LP. 
Next, we filter out all liquidity positions that are minted or burned during the day; this is our formulation assumes players hold their action throughout the game. 
All remaining liquidity positions that meet our criteria are treated as being created by player LPs. 

For each remaining position, we designate the transaction sender who minted the position as the owner LP.\footnote{We make this simplification since the ownership of liquidity position NFTs may be rarely transferred. }
By grouping the positions by owner, we obtain a list of player LPs along with their actions and their total expenditures on these positions. 
However, the number of such LPs is large, and can prevent the Nash equilibrium calculation from converging. Hence, we set up a third filter for investment.
In detail, we sort the LPs by their investment and find the top 30 LPs. 
Within the 30 LPs, if the top $N < 30$ LPs already hold $\ge 99\%$ of total investment, we exclude the last $30-N$ LPs. 
Otherwise, we keep all of them.
Finally, we set each remaining LP's budget to their total investment over the time period in question.

\paragraph*{Weights of Player LPs}
Despite our seemingly strict conditions for player LP selection, the remaining players still exert a significant impact on their respective liquidity pools. 
Tab. \ref{tab:player-percentage} lists the fraction of total budget and earned fees that are covered by our player LPs in each liquidity pool.  
It also shows the number of uncovered price ranges, i.e., ranges that were covered in the dataset, but not by our player LPs.
We observe a clear division: 

\begin{table}[!htb]
    \centering\small
    \begin{tabular}{*{5}{c}}
    \toprule
    Nickname & Total Value Locked (TVL, \$) & Budget Percentage & Fee Percentage & Uncovered/Ranges \\
    \midrule
    B30 & $15.0 \pm 2.6$ Million & $93.54 \pm 3.73 \%$ & $88.25 \pm 16.17 \%$ & 0/1271 \\
    E100 & $1.72\pm 0.31$ Million & \textcolor{orange}{$89.35 \pm 5.79 \%$} & \textcolor{orange}{$34.44 \pm 9.06 \%$} & \textcolor{orange}{390}/833 \\
    E30 & $74.9\pm 18.7$ Million & $85.57 \pm 7.73 \%$ & $93.68 \pm 12.38 \%$ & 0/1493 \\
    E5 & $135 \pm 20$ Million & $78.08 \pm 4.49 \%$ & $65.22 \pm 17.73 \%$ & 0/9452 \\
    T5 & $6.35\pm 1.79$ Million & $97.61 \pm 5.72 \%$ & $96.94 \pm 9.89 \%$ & 0/609 \\
    \bottomrule
    \end{tabular}
    \caption{Preprocessed pool information. Statistics aggregated over days are shown in $\text{mean} \pm \text{std}$. \vspace{-4ex}}
    \label{tab:player-percentage}
\end{table}

\begin{itemize}[leftmargin=*]
    \item In pools B30, E30, E5 and T5, trading activity is more frequent and long-term LPs are more incentivized to join the pool. 
    Every price range is covered by player LPs, which implies our estimate of budget percentage is accurate. 
    In these pools, player LPs are highly influential with at least 79\% of funds and 88\% collected fees. 
    In particular, the stable pool T5 has comparatively high budget and fee percentages by player LPs. 
    This suggests that when the price is stable, the actions of LPs are also stable and long-lasting, while short-term LPs are less incentivized than in pools with risky tokens. 

    \item In the E100 pool, the fee rate is comparatively high, and relatively little liquidity is held by long-term LPs. 
    Instead, there appears to be a higher rate of so-called \emph{just-in-time} (JIT) LPs, which opportunistically move their allocations in response to ongoing trades. 
    Right before a swap transaction is validated by the blockchain, a JIT LP can quickly deploy a high liquidity position over the narrowest price range that covers the swap.
    This grants the LP a high share of the fees without paying as much as long-term LPs. 
    After taking a large share of the trading fee, the LP immediately withdraws its liquidity from the pool. 
    As a result, we observe a significant number of uncovered ranges, and an abnormally 
    low portion of fees collected by player LPs. 
\end{itemize}

\subsubsection{Price Ticks}
We do not use the full set of possible ticks in the system because there are 1.7M ticks, which are both unnecessary and computationally burdensome to describe a liquidity distribution in practice. 
Instead, we obtain a histogram of liquidity for both player LPs and non-player LPs. 
Then, we adopt the \textbf{union} of price tick sets of these histograms. 
If adding a tick $c$ requires partitioning the liquidity $L$ of a price range $(a, b)$, 
we may safely partition it into same liquidity $L$ over two separate ranges $(a, c)$ and $(c, b)$. 

\subsubsection{Fee Reward}
We compute the fee reward from all trades within the daily time window. 
Each trade is characterized by the starting price $q$ and the ending price $q'$. 
If $q < q'$, the fee is paid in $Y$ tokens; otherwise, the fee is paid in $X$ tokens. 
We collect the fees in both tokens before converting them together to US dollars at the end of the day. 
As described in \S\ref{sec:amm:reward-cost}, fees are collected separately for each relevant atomic interval. 
The vector of fee rewards result from summing up the fees for each atomic range across all the trades. 
Further, we categorize the fee as collected by players and non-players (refer to App. \ref{amm:app:non-player} for details). 

The fee reward distribution incurred by a trade may change as the liquidity distribution changes. 
For simplicity in the game theoretic simulations, we explicitly assume that the aggregated fee reward in each price range remains constant even as the pool liquidity varies. 

\subsubsection{Impermanent Loss Rate} 
\label{sec:il-empirical}

Eq. \eqref{eqn:il} defines the impermanent loss rate $\hat \tau$ under two strong assumptions, which we relax in our experiments. 
First, it assumes the price $p_Y$ of the fiat token $Y$  to be constant, so that when both pool tokens are fiat tokens (such as USDC and USDT), impermanent loss is zero everywhere. 
In practice, $p_Y$ may slightly fluctuate between 0.99 and 1.01. 
We relax the assumption and consider a model with generalized $p_Y$ in App. \ref{amm:app:model-general}. 
Second, we previously assumed the fiat price of $X$ satisfies $p_X = q \cdot p_Y$. 
In practice, all three prices $q, p_X$ and $p_Y$ can be read from real-world data (and do not necessarily exactly follow that relation, though they are close). 
In our experiments, $q$ is parsed from the blockchain, while $p_X$ and $p_Y$ are crawled from Coingecko \cite{coingecko}.

After relaxing the assumptions, our impermanent loss can be computed as follows. 
At the end of Day $d$, we let $q_d, p_{X, d}$ and $p_{Y, d}$ denote the prices above. 
From liquidity and $q_d$, we may derive the corresponding token amounts $x_d$ and $y_d$ from \eqref{eqn:single-liq}. 
Hence, the impermanent loss rate can be written as 
\begin{align}
    \hat \tau = \epsilon_{a, b}(q_{d-1}) \frac{\vhold - \vlp}{\vinit} = \epsilon_{a, b}(q_{d-1}) \frac{p_{X, d} (x_{d-1} - x_d) + p_{Y, d} (y_{d-1} - y_d)}{p_{X, d-1} \cdot x_{d-1} + p_{Y, d-1} \cdot y_{d-1}}. 
    \label{eqn:tau-strawman}
\end{align}

{
However, we cannot directly apply this  definition, because $p_{X, d}$ does not always equal $q p_{Y,d}$, which can cause impermanent loss $\hat\tau < 0$. 
In other words, the price range induces impermanent \emph{gain} proportional to the liquidity,  attracting a huge proportion of LPs' budgets, even if the fee reward is low. 
The root cause of this phenomenon (which does not occur in practice) is slippage: in any trade, a quantity $x$ of $X$ tokens does not always have fiat value $x p_X$, because if $x$ is large enough, any market will exhibit significant price slippage such that the overall price of the trade is less than $p_X$. 
In other words, an LP can never fully take advantage of impermanent gain because the market will quickly counter any attempts to arbitrage the favorable position.
On the other hand, accurately valuing  a given quantity of $X$ tokens is extremely complex, since it requires the information of the market depth of centralized exchanges (CEXes), and an accurate estimate of the delay when these tokens are sold at price $p_X$, if it happens in the future. 

As these complexities are difficult to model, we model the value of cryptocurrency tokens by a constant price,  while 
ensuring that $\hat \tau \ge 0$.
This condition is enforced with  ad hoc calibrations that map the observed prices to a ``damped" price that more realistically reflects LPs' ability to react to market prices.
One such calibration is to zero-cap $\hat\tau$, i.e., to set $\hat \tau \leftarrow \max\{0, \hat\tau\}$. 
The shortcoming of this calibration is that price ranges are not treated equally, since only negative impermanent loss rates are changed. 
Hence, we choose the second option, which is to generate ``shifted'' dollar prices $\bar p_X$ and $\bar p_Y$ such that $\bar p_X = q \bar p_Y$. 
We would like this price shift to be reflectionally symmetric for $X$ and $Y$, i.e., $p_X p_Y = \bar p_X \bar p_Y$. 
This gives: 

\begin{align}
    \left(\bar p_{X}, \bar p_{Y} \right) = \left( \sqrt{p_{X} p_{Y} q}, \sqrt{p_{X} p_{Y} q^{-1}} \right). 
\end{align}

In experiments, we use impermanent loss rate $\hat \tau$  replacing $p_{X, d}, p_{Y, d}$ with $\bar p_{X, d}, \bar p_{Y, d}$ in \eqref{eqn:tau-strawman}. 
We show in Fig. \ref{fig:price-bend} (App. \ref{amm:app:price-bend}) that the error induced by this price shift is generally less than 1\%.

\subsection{Comparison Between Nash Equilibrium and Ground Truth}
\label{sec:amm:gap-ne-gt}

Under this setup, we study how closely real LPs match the Nash equilibria of our proposed game. 
To start, we consider the following three types of LP actions:

\begin{enumerate}[leftmargin=*,label=\Alph*)]
    \item \textbf{Ground truth (\gt):} The LP actions that were actually taken in the real world. 

    \item \textbf{Nash equilibrium (\ne):} The LP actions under Nash equilibrium. We compute Nash equilibria with the relaxation algorithm \cite{relax-1} throughout this work. 

    \item \textbf{Best response (\br):} 
    The optimal action of each LP (that maximizes utility) when every other LP maintains their action in the real world (i.e., adopts \gt).  
    To compute utility under \br, we assume other LPs use \gt actions instead of their own \br actions.
    This typically results in higher utility than every other action the LP takes. 
\end{enumerate}

To evaluate how close two actions are from each other, we introduce a metric named \textbf{overlap}. 
Let $\Kb_n^{\mathtt{A}}$ denote LP's liquidities under action \texttt{A}. 
Between two actions \texttt{A1} and \texttt{A2}, 
\begin{align}
    \mathtt{overlap}\left(\Kb_n^{\mathtt{A1}}, \Kb_n^{\mathtt{A2}}\right) = 
     \frac{\displaystyle \sum_{m \in [M]} \epsilon_{t_{m-1}, t_m}(q) \left|K_{n,m}^{\mathtt{A1}} - K_{n, m}^{\mathtt{A2}}\right| + \left|\sum_{m \in [M]} \epsilon_{t_{m-1}, t_m}(q)\left( K_{n,m}^{\mathtt{A1}} - K_{n, m}^{\mathtt{A2}}\right)\right|}{2 B_n}. 
\end{align}

Intuitively, we consider each price range as a bucket containing some liquidity \textbf{in dollars}. 
Besides the price range buckets, we set up one additional bucket containing \emph{unused} budget. 
As a result, the per-range liquidity allocation can be mapped to a categorical (per-bucket) dollar value distribution. 
Our overlap metric is equivalent to the total variation distance of these dollar value distributions. 
By this definition, \texttt{overlap} must be in $[0, 1]$, and higher \texttt{overlap} implies better similarity of the actions.

\begin{minipage}{.96\linewidth}
\begin{minipage}[b]{0.48\linewidth}
    \centering
    \includegraphics[width=\linewidth]{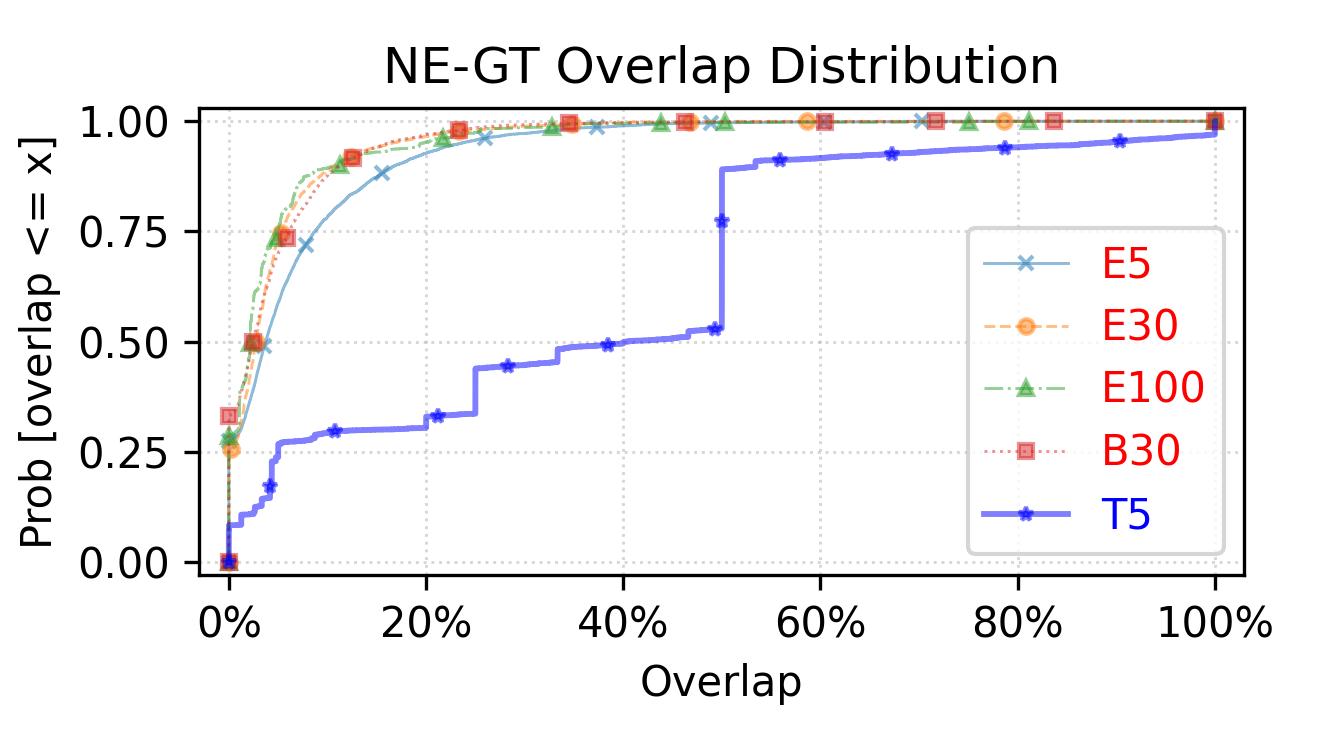}
    \vskip-3ex
    \captionof{figure}{Distribution of \ne-\gt overlaps across the liquidity pools.}
    \label{fig:ne-olap-all-pools}
\end{minipage}
\hfill
\begin{minipage}[b]{0.48\linewidth}
    \centering
    \includegraphics[width=\linewidth]{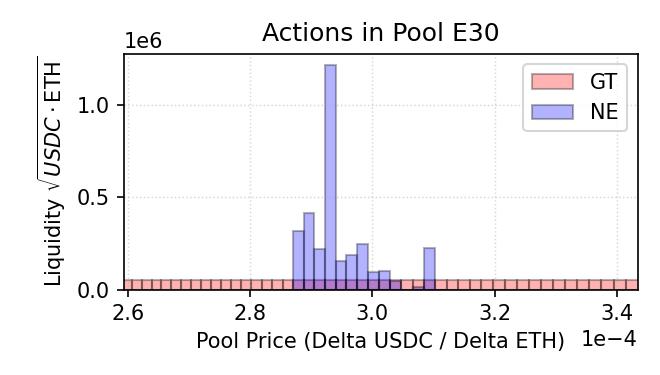}
    \vskip-3ex
    \captionof{figure}{A qualitative example of \gt and \ne actions of a typical LP in a day. }
    \label{fig:qual-gt-ne}
\end{minipage}
\vspace{3ex}
\end{minipage}

After aggregating all (\ne, \gt) action pairs over days and player LPs, we obtain a distribution of the \ne-\gt overlap in each liquidity pool, as shown in Fig. \ref{fig:ne-olap-all-pools}. 

\begin{finding}
    In the stable pool (T5), most LPs exhibit high overlap with our Nash equilibrium strategy. In risky pools, most LPs exhibit low overlap, preferring to invest in fewer price ranges with a larger price span. 
\end{finding}

In the stable pool T5, we observe a high overlap (with 40.4\% median and 34.5\% mean), 
while across all risky pools, we observe a low overlap: the 75\% quantiles are all below 9\%.

We look more deeply into the E30 pool to better understand the phenomenon. 
Fig. \ref{fig:qual-gt-ne} shows the \gt and \ne actions by a typical player LP as liquidity histograms. 
While at Nash equilibrium, LPs typically invest different amounts in multiple atomic price ranges, 
real-world LPs often choose to open a single liquidity position with wide price coverage and relatively low liquidity with the same amount of budget. 
This motivates us to look at 1) the number of liquidity positions required by each action, and 2) the span (max-min ratio) of prices of each action (Fig. \ref{fig:pos-count-span}).  
By the CDFs in these two figures, Fig. \ref{fig:qual-gt-ne} reveals two trends:
\begin{enumerate}[label=\arabic*., leftmargin=*]
    \item \textbf{LPs tend to mint (i.e., invest in) very few liquidity positions} in \gt, with over 65\% minting only \emph{one} position. The amount is much smaller than at \ne. 
    Additionally, \br also requires much fewer liquidity positions than \ne, 
    because in many atomic ranges, liquidity by other player LPs (following \gt) has already saturated, which makes it unprofitable to further add any amount of liquidity to them. 

    \item \textbf{\gt LPs tend to invest in a much wider range of prices} than the price ranges that actually generate fee income, which are accurately captured by \br and \ne. 
\end{enumerate}

\begin{figure}[!htb]
    \centering
    \includegraphics[width=\linewidth]{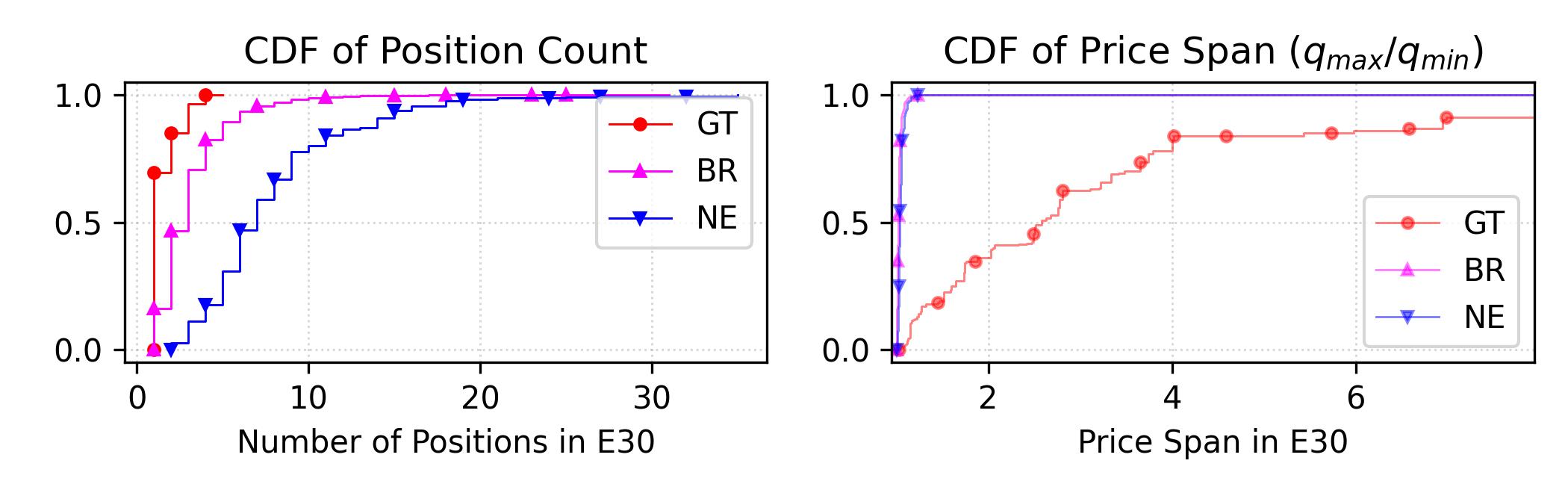}
    \caption{Distribution of number of positions and price spans (max / min) of \gt and \ne. }
    \label{fig:pos-count-span}
\end{figure}

These trends, however, do not apply to the stable pool T5. The CDFs of \gt, \br and \ne are similar. We refer to App. \ref{amm:app:stable-pool} for relevant plots. 

\subsection{Strategies Based on Games Instantiated with Historical Data}
\label{sec:amm:alternative-games}

By instantiating our game with the present day's data (Sec. \ref{sec:amm:game-setup}), we essentially assume LPs can accurately predict 
1) the ending price of the day $q_d$, 
2) the fee reward distribution, and 
3) the non-player LP investments. 
In practice, this assumption is clearly unachievable in risky pools. 
To better understand LPs real strategies, we turn to game models that rely only on historical information. 

We consider three strategies that use historical information. 
In our utility evaluation, we allow one strategic agent to adopt each of these strategies, while keeping the other players' strategies fixed according to the data.

\begin{itemize}[leftmargin=*]
\item \textbf{\yday:} (which stands for yesterday) simply repeats the previous day action, if it exists for the LP. 
If an LP has a different budget $B'$ at the present day from previous day's budget $B$, we scale the liquidity of each price range by $B'/B$. 
\item \textbf{Reactive game (\rne):} LPs proactively monitor the pool information from \emph{only the previous day} and instantiate our game accordingly (details below). They adopt the Nash equilibrium strategy of that game, and we abbreviate it reactive-Nash equilibrium (\rne). 
\item \textbf{Inert game (\ine):} LPs roughly estimate the same game parameters from aggregated information \emph{over a window of 7 days} (details below). They again use the Nash equilibrium strategy, and we abbreviate it inert-Nash equilibrium (\ine). 
\end{itemize}

In both games above, we consider the same set of player LPs (with their budgets) as in \S\ref{sec:amm:gap-ne-gt}. 
Tab. \ref{tab:exp-game-setups} lists our constructions of the five basic game parameters for the inert game and the reactive game. 
Among them, the fluctuation parameter $r$ is a proxy parameter of impermanent loss $\tau$. 
In detail, we assume the starting price $q$ (a known constant) and the ending price $q'$ (to be estimated) satisfy $q' = Rq$, 
where $R$ is a random variable with $\log R \sim \mathtt{Unif} (-\log r, \log r)$. 
Following \S\ref{sec:amm:reward-cost}, $\tau$ can be computed from a resulting distribution of $q'$. 

\begin{table}[!htb]
    \centering\small
    \begin{tabular}{m{5em}<{\centering}m{8.4em}m{26.6em}}
    \toprule
        & Reactive Game & Inert Game \\
    \midrule
        Price Ticks & Same as previous day & Two ticks $(a, b) = (\underline{t} / E, E\overline{t})$, where $\underline{t}, \overline{t}$ are the minimum and maximum ticks over 7 days, respectively. $E \in [1, \infty)$ is the \emph{expansion factor} chosen for each pool (further discussed in App. \ref{amm:app:expansion}). \\ \hline
        Fees & Same as previous day & Average of daily total fee (sum of daily per-range fee)  \\ \hline
        Non-Player Investment & Same as previous day & Average of total daily investment (sum of daily per-range investment) \\ \hline
        Fluctuation Parameter $r$ & $=1.1$ & Average of $\max\{q'/q, q/q'\}$ for starting and ending price pairs $(q, q')$  \\
    \bottomrule
    \end{tabular}
    \caption{Game construction for reactive and inert actions.}
    \label{tab:exp-game-setups}
\end{table}

\begin{finding}
    In all liquidity pools, the YDay strategy (i.e., repeating the previous day's action) has over 99.5\% overlap with the ground truth (\gt) strategy in mean and 25\% quantile. 
\end{finding}

Fig. \ref{fig:box-all} shows the distributions of overlaps with \gt and utilities. 
We refer to App. \ref{amm:app:raw-data} for the raw data from experiments and App. \ref{amm:app:roi-gap} for plots of return over investments and optimality gaps. 
The top row of Fig. \ref{fig:box-all} shows the overlap  of different strategies with the \gt of the these ``reactive" LPs. 
This yields our second main observation:

\begin{figure}
    \centering
    \includegraphics[width=\linewidth]{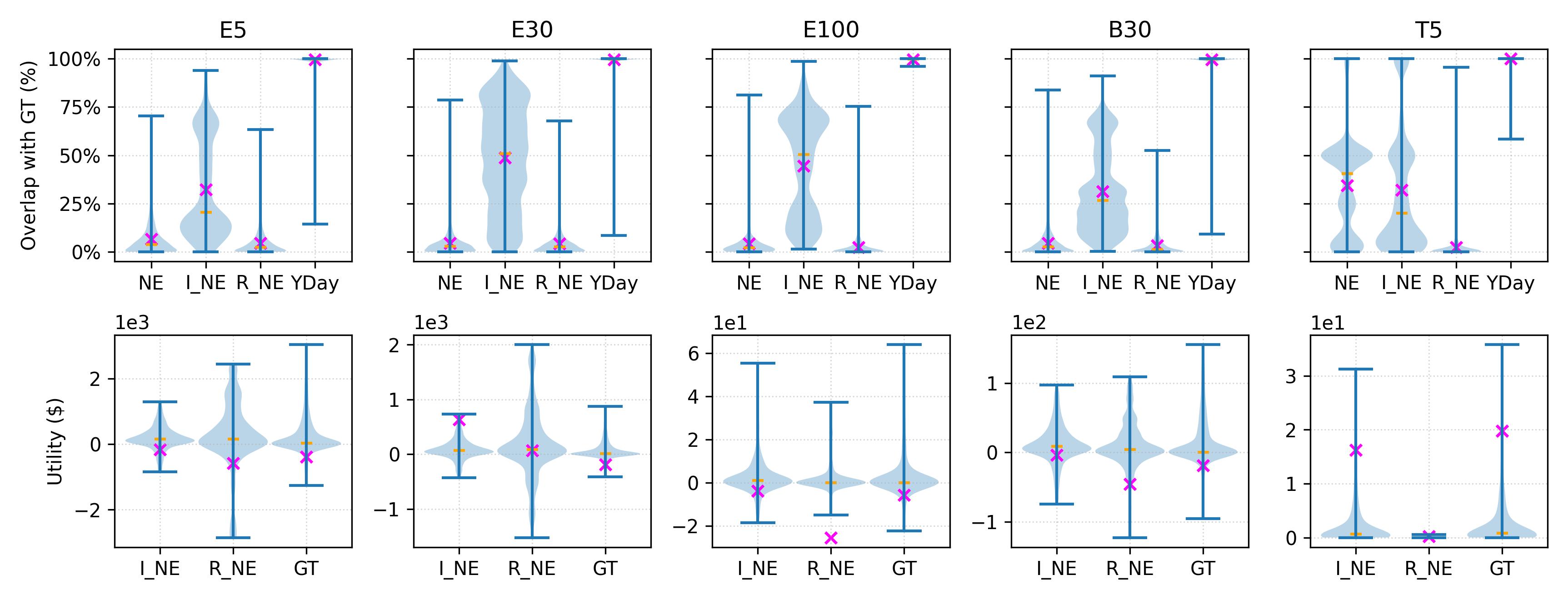}
    \caption{Violin plot of overlap and utility distributions. 
    Each marker (\textcolor{magenta}{$\times$}) represents the mean and each horizontal line (\textcolor{orange}{$-$}) represents the median. 
    We list the means, standard deviations and quartiles in Tab. \ref{tab:olap-gt} and \ref{tab:util}. 
    }
    \label{fig:box-all}
    \Description{}
\end{figure}

\begin{finding}
    In all risky pools, \ine (game computed from 7-day history) has an overlap with \gt at least 28\% higher than both \ne and \rne (game computed from 1-day history). 
    In the stable pool T5, \ne instead has the highest overlap with \gt, with mean and median close to 40\%. 
\end{finding}

Fig. \ref{fig:box-all} (top row) suggests that in response to the uncertainty within risky pools, LPs tend to use stale information and take simpler actions, as recommended in the inert game model (\ine). 
In contrast, in the stable pool T5 where price hardly fluctuates, LPs take actions with about 40\% overlap to the Nash equilibrium.
Meanwhile, LPs in T5 are only worse than the best utility by only 0.005\% in median of their budgets (Fig. \ref{fig:roi-gap} and Tab. \ref{tab:gap}), which is at least $40 \times$ smaller than the risky pools.
This suggests that real-world LPs are learning to play the game and their actions are converging towards rationality.

Finally, we compare the ground truth \gt, the strategy computed from the inert game (\ine) and the strategy computed by the reactive game (\rne) by their utilities.
The utility of each action is calculated assuming all other LPs are using \gt, except for \nea. 
We present our third finding:

\begin{finding}
    In all risky pools, the Nash equilibrium strategy of the inert game (i.e., \ine) has up to \$116 higher daily median utility than \gt, which corresponds to improving the average daily utility by \$222 and median daily ROI by 0.009\%.  
    Even greater gains may be possible with improved strategies: the median daily ROI for \ne is 0.855\% higher than for \gt. 
    In contrast, in the stable pool T5, \ine and \rne have lower utility than \gt, and the advantage of \ne over \gt is smaller than in risky pools.
\end{finding}

This finding is particularly informative for both risky pools and stable pools.
In stable pools, the actions of LPs are highly similar to the Nash equilibrium, which results in a narrow optimality gap in utility maximization when other LPs follow the ground truth.  
This implies that LPs have learned well to maximize utility under competitions, such that each LP has little space for further improvement. 
In risky pools, the finding reveals that many real-world LPs can increase their median daily ROI by about 0.01\% after adopting very simple strategies such as \ine.
In other words, if other LPs maintain the same strategy, an LP can statistically improve their ROI by the following 3 steps:
1) Collect pool data from past 7 days to obtain the parameters for the inert game;
2) Identify player LPs and set their budgets from existing liquidity positions in the pool by estimation based on the their durations; and 
3) Solve the Nash equilibrium of the inert game and adopt the resulting strategy. 
We note that the \ine strategy is a heuristic and could be improved further; the median daily ROI for \ne is 0.855\% higher than for \gt, which suggests much room for improvement. Designing such improved heuristics is an important  question for future work.

\section{Related Work }

The incentive structures of AMMs, as well as the behavior of agents, have been widely studied. Prior work has considered how to  measure  agent utility, design AMMs, and invest strategically. 

\paragraph*{Measuring LP utility}
Prior work has considered different ways of quantifying utility and behavior of LPs in AMMs \cite{heimbach2021behavior,berg2022}. 
The release of Uniswap v3 popularized the concept of leveraged liquidity provision, and a large part of the literature focuses on the analysis of liquidity positions and the profitability challenge faced by liquidity providers \cite{Fan2022}. 
Indeed, empirical studies have shown that a high number of LP positions are unprofitable \cite{berg2022,loesch2021impermanentlossuniswapv3}, and many works have proposed solutions based on novel AMM designs \cite{nadkarni2024adaptive,lvr,aoyagi2020liquidity,angeris2022does,Fan2022,jeong22,baron2023uniswapliquidityprovisiononline,bastankhah2024thinking}.

To help explain these phenomena, several papers have studied models that explain LP gains and losses \cite{engel2022presentation,deng2023static,clark2020replicating,lvr,lipton2024unified}.  
One of the most classic measures of LP losses is impermanent loss, which we have used in our models \cite{harvey2021defi,capponi2021adoption}. 
An alternative measure of loss was proposed by Milionis \emph{et al.} called loss-versus-rebalancing (LVR) \cite{lvr}, defined as the difference between two terms:  the first is profits of a so-called \emph{rebalancing strategy}, in which the LP continually adjusts their holdings on a centralized exchange to match the exogenous price. 
The second is the final pool price. Together, these two terms capture the net utility lost due to investing in the AMM instead of actively rebalancing a position. 
The main difference between impermanent loss and LVR is that the latter measures gains with respect to a rebalancing strategy, whereas impermanent loss measures gains with respect to simply holding tokens. 
In expectation, the risk-neutral expectation of these two are the same \cite{lvr}. 
Note that neither LVR nor impermanent loss alone explain how to determine LP investment strategies.

\paragraph*{Analyzing profitable LP investment strategies}

Prior work has considered optimal liquidity provision for a single LP \cite{capponi2021adoption,cartea2024decentralized,cartea2024decentralized}. 
For example, Capponi \emph{et al.} \cite{capponi2021adoption} argue that under a simple model, when price is too volatile or trading is not profitable enough, LPs are incentivized to not invest \emph{any} liquidity.
However, their predictions (e.g., LPs should not invest when prices are volatile \cite{capponi2021adoption}) do not hold true in practice \cite{heimbach2021behavior}.
Milionis \emph{et al.} \cite{milionis2023myersonian} analyze a Myersonian model of traders and a monopolistic liquidity provider to determine the optimal AMM structure for a profit-maximizing market maker.

Another body of literature discusses liquidity provision of LPs specifically in CLMMs by focusing on a single  LP's choice of concentrated liquidity interval.  In \cite{Goyal23}, a framework is proposed to analyse how liquidity provision changes with changing asset price mean and volatility. 
In \cite{Fan2022}, the authors develop a model for LP profits and losses that incorporates profits from fees  accrued by traders as well as impermanent losses due to price deviation.  Their formulation, like ours, assumes that 
the fees  received by LPs do not depend on the total liquidity of all LPs.  They formulate the liquidity allocation problem as a convex stochastic optimization, and use it to study contract design in Uniswap v3. Heimbach et al. \cite{Heimbach2023} also formalize the same problem and strategy analysis in a Black-Scholes stochastic market model.

To our knowledge, no existing work on AMM liquidity provision models all of: (1) agents with a limited investment capacity (i.e., budget), (2) the complex strategy space of concentrated liquidity, in which liquidity provisions for different price intervals can differ \cite{Frtisch23}, and (3) competition between LPs.  

\paragraph*{Relation to network resource management}
Our game model  is  similar to resource allocation problems dealing with user heterogeneity and congestion management in networks \cite{kelly98,johari2004}. In \cite{kelly98}, Kelly proposed a market mechanism in which each user submits a bid to the network operator, then the network operator determines the price of resources after collecting all bids and allocates resources to each user in proportion to their bid.  
The aim of these mechanisms is to establish a resource-sharing mechanism  with non-cooperative users in order to maximize social welfare while keeping users' utilities private. 
Starting with the seminal work of \cite{johari2004}, a series of papers studied the existence and uniqueness of pure Nash equilibria of the induced games \cite{832485,8766145} and calculated  their inefficiency using the price of anarchy \cite{johari2004}.

The link between our game model and congestion management in networks   using Kelly's mechanism is that the bid represents the provision of liquidity at each interval, and the system's incentive mechanism (i.e., fees distributed to LPs)  is the same rule as that proposed by Kelly's mechanism. 
However, our work differs from prior literature in several respects: the budget constraint is absent in prior works, and its presence invalidates  strategies calling for arbitrarily high payments.
Prior work has studied the price of anarchy under a single resource (or independent resources) in the presence of a budget constraint for each user \cite{10.1145/3219166.3219186}. 
However, our problem is different in several ways: (1) we consider multiple resources (i.e., each price range) under a single budget constraint, (2) the utility of each user is known in our case, and (3) our main focus is on characterizing Nash equilibria and their properties, whereas \cite{10.1145/3219166.3219186} focuses on the liquid price of anarchy -- the ratio between optimal liquid welfare and liquid welfare obtained in the worst equilibrium, and does not characterize Nash equilibria as a function of budget.

\section{Conclusion}
\label{sec:amm:conclusion}

In this work, we theoretically respond to the challenge of competitive liquidity providing with high-dimensional action space in CLMMs. 
We first introduce our core game theoretic model --- the atomic game --- and then show the uniqueness of its Nash equilibrium. 
We further theoretically establish the waterfilling pattern of LPs' actions at equilibrium. 
By fitting our game model to real-world CLMMs, we show that in stable pools, our model is a good predictor of how LPs invest. However, in risky pools, we identify a gap between equilibrium actions and actions actually adopted: LPs tend to provide liquidity in a single wide price range, and infrequently update their liquidity allocations. 
Finally, we propose the inert game model, where a single LP can statistically improve their return over interest by adopting actions at Nash equilibrium. 
Our work leaves several open questions. One is incorporating LPs that invest at different time scales, such as just-in-time LPs, into the game model.
Another is to understand how to design more profitable strategies from historical data. 
Given such strategies, we envision automated services that can redeploy liquidity in a more strategic fashion.

\section*{Acknowledgments}
The authors gratefully acknowledge the support of IC3 and their industry sponsors, as well as the CyLab Secure Blockchain Initiative and their industry sponsors, including Ripple Labs for valuable comments. We thank Allium for providing access to their blockchain data platform. 

\bibliographystyle{plain}
\bibliography{bibfile}

\newpage
\appendix


\section{Notations}
\label{amm:app:notations}

\begin{table}[!htb]
    \centering
    \begin{tabular}{cp{.8\linewidth}}
        \hline
        Notation             & Description                                                                                                       \\
        \hline
        $[\cdot]$            & Set of integers from $1$ to $\cdot$.~~ $[k] = \{1, 2, \cdots, k\}$                                                \\
        $x, y$               & Amounts of $X$ and $Y$ tokens in a liquidity pool                                                                 \\
        $q$                  & (Initial) Pool price, or marginal price of $X$ (value of a unit of $X$ in $Y$ tokens)                             \\
        $q', \pi$            & Future pool price (random variable), and its distribution                                                         \\
        $p_X, p_Y$           & Dollar price of $X$ and $Y$ tokens                                                                                \\
        $\gamma$             & Fee rate of AMM                                                                                                   \\
        $\phi$               & The AMM bonding curve                                                                                             \\
        $M$                  & Number of price ticks ($|T|$)                                                                                     \\
        $N$                  & Number of LPs                                                                                                     \\
        $T$                  & Set of price ticks $\{t_m| m \in \{0, 1, \cdots, M\}\}$                                                           \\
        $a, b$               & Endpoints of a price range                                                                                        \\
        $\proj{q}{a}{b}$     & Projection of $q$ onto interval $[a, b]$; $\max\{a, \min\{b, q\}\}$                                               \\
        $\gp$                & Set of general price ranges $\{(a, b) | a \in T, b \in T, a<b \}$                                                 \\
        $\ap$                & Set of atomic price ranges $\{(t_{m-1}, t_m)| m \in [M]\}$                                                        \\
        $\epsilon_{a, b}(q)$ & Dollar price of a unit liquidity in price range $(a, b)$ at price $q$                                             \\
        $\hat \tau_{a, b}(q', q)$ & Impermanent loss rate of price range $(a, b)$ in CLMM when price changes from $q$ to $q'$                       \\
        $\bar \tau_{a, b}$   & Expectation of $\igb_{a, b}(q', q) \epsilon_{a, b}(q)$ where $q' \sim \pi$                                        \\
        $\tau_{m}$           & $\bar \tau_{t_{m-1}, t_m}$                                                                                        \\
        $\chi_{m}$           & Average liquidity of non-player LPs in range $m$                                                                                        \\
        $L_{n, (a, b)}$      & Liquidity of LP $n$ on price range $(a, b)$                                                                       \\
        $\Lb_n$              & Liquidity of LP $n$ on each general range $(L_{n, (a, b)})_{(a, b) \in \gp}$                                      \\
        $\Lb$                & Liquidity of each LP on each general range $(L_{n, (a, b)})_{(a, b) \in \gp, n \in [N]}$ or $(\Lb_n)_{n \in [N]}$ \\
        $K_{n, m}$           & Total liquidity of LP $n$'s positions that cover the $m$-th atomic range $(t_{m-1}, t_m)$                         \\
        $\Kb_n$              & Vector $(K_{n, m})_{m \in \ap}$                                                                                   \\
        $\Kb$                & Vector $(K_{n, m})_{m \in \ap, n \in [N]}$ or $(\Kb_{n})_{n \in [N]}$                                             \\
        $B_n$                & Budget of LP $n$ in dollars                                                                                       \\
        $f_m$                & Fee reward of the $m$-th atomic range $(t_{m-1}, t_m)$                                                            \\
        $A_n$                & Budget used by LP $n$ in dollars                                                                                  \\
        \hline
    \end{tabular}
    \caption{Notations}
    \label{tab:notations}
\end{table}

\section{CLMM Mathematics}
\label{amm:app:clamm-phi}

\subsection{Analytical Form of CLMM Bonding Curve}
We derive the bonding curve $\phi$ from the liquidity-amount relation \eqref{eqn:liqv3}.
In detail, for each liquidity position $\left(L_{n, (a, b)}, a, b\right)$ by LP $n$ at price range $(a, b)$,
the token amounts $(x_{n, (a, b)}, y_{n, (a, b)})$ are
\begin{equation}
    x_{n, (a, b)} = L_{n, (a, b)} \left( \frac{1}{\sqrt{\proj{q}{a}{b}}} - \frac{1}{\sqrt{b}} \right), \quad y_{n, (a, b)} = L_{n, (a, b)}  \left( \sqrt{\proj{q}{a}{b}} - \sqrt{a}\right).
    \label{eqn:single-liq}
\end{equation}

We define $J_{(a, b)}$ as the total liquidity on price range $(a, b)$,
and $J_m$ as the total liquidity of positions from \textbf{all} LPs that cover atomic range $(t_{m-1}, t_m)$:
\begin{align}
    J_{(a, b)} & \triangleq \sum_{n \in [N]}L_{n, (a, b)};                                                                                                                   \\
    J_m        & \triangleq \sum_{n \in [N]} K_{n, m} = \sum_{n \in [N]} \sum_{i=1}^{m} \sum_{j=m}^M L_{n, (t_{i-1}, t_j)} = \sum_{i=1}^{m} \sum_{j=m}^M J_{(t_{i-1}, t_j)}.
\end{align}

Thm. \ref{thm:clamm-phi} presents the resulting bonding curve of a CLMM.
Notably, the bonding curve is only dependent on the \textbf{aggregated} liquidity covering each atomic range $\left(J_m\right)_{m \in [M]}$.
The ``degree of freedom'' is just $M$, instead of $NM(M+1)/2$ which is the dimensionality of vector $\Lb$.

\begin{theorem}
    The CLMM with liquidity described above has a decreasing, convex and continuously differentiable bonding curve $\phi$ with domain $[x_m, x_0]$, where for each $m \in [M] \cup \{0\}$,
    \begin{equation}
        x_m \triangleq \sum_{k = m+1}^M \left( t_{k-1}^{-\frac{1}{2}} - t_k^{-\frac{1}{2}} \right) J_k; \qquad y_m \triangleq \sum_{k=1}^{m} \left( t_k^{\frac{1}{2}} - t_{k-1}^{\frac{1}{2}} \right) J_k.
    \end{equation}

    For each $m \in [M] \cup \{0\}$, $\phi(x_m) = y_m$ and $\phi'(x_m) = -t_m$.
    In addition, for each $m \in [M]$ and all $x \in [x_m, x_{m-1}]$,
    \begin{equation}
        \phi(x) = \frac{J_m^2}{x - x_m + J_m t_{m}^{-\frac{1}{2}}} + y_{m-1} - J_m t_{m-1}^{\frac{1}{2}}.
        \label{eqn:clamm-phi}
    \end{equation}
    \label{thm:clamm-phi}
\end{theorem}

\textbf{Proof.}
\textbf{First}, we compute the analytical form of $\phi$ when the price $q$ is in an interval $(t_{m-1}, t_m)$ for some $m \in [M]$.
By summing up \eqref{eqn:single-liq}, the total amount of $X$ liquidity tokens is
\begin{align}
    x & = \sum_{(t_{i-1}, t_j) \in \gp} \sum_{n \in [N]} L_{n, (t_{i-1}, t_j)} \left( \proj{q}{t_{i-1}}{t_j}^{-\frac{1}{2}} - t_j^{-\frac{1}{2}} \right) \notag                                                                    \\
      & = \sum_{(t_{i-1}, t_j) \in \gp, j \ge m} J_{(t_{i-1}, t_j)} \left( \proj{q}{t_{i-1}}{t_j}^{-\frac{1}{2}} - t_j^{-\frac{1}{2}} \right) \notag                                                                               \\
      & = \sum_{j=m}^M \left[  \left( q^{-\frac{1}{2}} - t_j^{-\frac{1}{2}} \right) \sum_{i=1}^m J_{(t_{i-1}, t_j)} +  \sum_{i=m+1}^j \left( t_{i-1}^{-\frac{1}{2}} - t_j^{-\frac{1}{2}} \right) J_{(t_{i-1}, t_j)} \right] \notag \\
      & = \left( q^{-\frac{1}{2}} - t_m^{-\frac{1}{2}} \right) \sum_{j=m}^M \sum_{i=1}^m J_{(t_{i-1}, t_j)} +
    \sum_{j=m}^M \left[ \left( t_m^{-\frac{1}{2}} - t_j^{-\frac{1}{2}} \right) \sum_{i=1}^m J_{(t_{i-1}, t_j)} +  \sum_{i=m+1}^j \left( t_{i-1}^{-\frac{1}{2}} - t_j^{-\frac{1}{2}} \right) J_{(t_{i-1}, t_j)} \right] \notag      \\
      & = \left(q^{-\frac{1}{2}} - t_m^{-\frac{1}{2}}\right) J_m +
    \sum_{j=m+1}^M \left[ \sum_{k=m+1}^j\left( t_{k-1}^{-\frac{1}{2}} - t_{k}^{-\frac{1}{2}} \right) \sum_{i=1}^m J_{(t_{i-1}, t_j)} +  \sum_{i=m+1}^j \sum_{k=i}^j\left( t_{k-1}^{-\frac{1}{2}} - t_{k}^{-\frac{1}{2}} \right) J_{(t_{i-1}, t_j)} \right]
    \notag                                                                                                                                                                                                                         \\
      & = \left(q^{-\frac{1}{2}} - t_m^{-\frac{1}{2}}\right) J_m +
    \sum_{k=m+1}^M \left( t_{k-1}^{-\frac{1}{2}} - t_{k}^{-\frac{1}{2}} \right) \left[ \sum_{j=k}^M \sum_{i=1}^m J_{(t_{i-1}, t_j)} + \sum_{j=k}^M \sum_{i=m+1}^k J_{(t_{i-1}, t_j)} \right]
    \notag                                                                                                                                                                                                                         \\
      & = \left(q^{-\frac{1}{2}} - t_m^{-\frac{1}{2}}\right) J_m +
    \sum_{k=m+1}^M \left( t_{k-1}^{-\frac{1}{2}} - t_{k}^{-\frac{1}{2}} \right)  \sum_{j=k}^M \sum_{i=1}^k J_{(t_{i-1}, t_j)}
    \notag                                                                                                                                                                                                                         \\
      & = \left(q^{-\frac{1}{2}} - t_m^{-\frac{1}{2}}\right) J_m + \sum_{k = m+1}^M \left( t_{k-1}^{-\frac{1}{2}} - t_k^{-\frac{1}{2}} \right) J_k \triangleq
    \phi_X(q) .
    \label{eqn:liq-x}
\end{align}

$\phi_X$ is clearly decreasing in $q$.
Hence, $q\in [t_{m-1}, t_m]$ is equivalent to
\[
    x = \phi_X(q) \in [\phi_X(t_m), \phi_X(t_{m-1})] = [x_{m}, x_{m-1}].
\]

Analogously, the total amount of $Y$ liquidity tokens is
\begin{align}
    y = \left(q^{\frac{1}{2}} - t_{m-1}^{\frac{1}{2}}\right) J_m + \sum_{k=1}^{m-1} \left( t_k^{\frac{1}{2}} - t_{k-1}^{\frac{1}{2}} \right) J_k \triangleq
    \phi_Y(q).
    \label{eqn:liq-y}
\end{align}

$\phi_Y$, on the other hand, is increasing in $q$, which implies that for $x \in (x_{m}, x_{m-1})$, $y = \phi(x) = \phi_Y\left(\phi_X^{-1}(x)\right)$ is decreasing in $x$.
By \eqref{eqn:liq-x} and \eqref{eqn:liq-y}, when $q \in [t_{m-1}, t_m]$, or $x \in [x_{m}, x_{m-1}]$,
\[
    \left(x - x_m + J_m t_{m}^{-\frac{1}{2}}\right)\left( y - y_{m-1} + J_m t_{m-1}^{\frac{1}{2}} \right) = J_m^2.
\]

This proves \eqref{eqn:clamm-phi}.

\textbf{Second}, we show that $\phi$ is globally continuous. Because for all $m \in [M-1]$,
\[
    \lim_{q \to t_{m}^-} \phi_X(q) = x_m = \lim_{q \to t_{m}^+} \phi_X(q); \qquad
    \lim_{q \to t_{m}^-} \phi_Y(q) = y_m = \lim_{q \to t_{m}^+} \phi_Y(q),
\]

we have
\[
    \lim_{x \to x_{m}^-} \phi(x) = \lim_{q \to t_{m}^+} \phi_Y(q) = \lim_{q \to t_{m}^-} \phi_Y(q) = \lim_{x \to x_{m}^+} \phi(x).
\]

This implies the continuity of $\phi$ at $\{x_m\}_{m \in [M-1]}$, and that $\phi(x_m) = y_m$.
Combined with $\phi$'s continuity in intervals $(x_{m}, x_{m-1})$, we know $\phi$ is continuously decreasing in the full domain $[x_M, x_0]$.

\textbf{Third} and finally, we show that $\phi$ is continuously differentiable.
Again, by \eqref{eqn:liq-x} and \eqref{eqn:liq-y},
\[
    \phi'(x) = \deriv{y}{x} = \deriv{y}{q} \cdot \deriv{q}{x} =  \frac{1}{2} J_m q^{-\frac{1}{2}} \cdot \left(- \frac{2}{J_m} q^{\frac{3}{2}}\right) = -q, \qquad \forall x \in (t_m, t_{m-1}).
\]

This implies for all $m \in [M-1]$,
\begin{align*}
    \lim_{x \to x_m^+} \phi'(x) = \lim_{x \to x_m^-} \phi'(x) = -t_m.
\end{align*}

Hence, $\phi'(x) = -q$ holds globally and $\phi$ is continuously differentiable in the full domain $[x_M, x_0]$.
In addition, since $q$ decreases in $x$, $\phi'$ increases in $x$, and $\phi$ is convex. \qeda

\subsection{CLMM Model with General Token Prices}
\label{amm:app:model-general}

\paragraph*{Liquidity Price and Impermanent Loss Rate}

In this section, we generally do \textbf{not} assume that $Y$ is fiat-pegged.
Instead, we let $p_X$ and $p_Y$ denote the dollar prices of $X$ and $Y$, respectively, where $p_Y$ is no longer necessarily a constant.
Still, we assume the existence of fast arbitrageurs who keep $q = p_X/p_Y$.
Our setting in the main text is merely a special case where $p_Y = 1$ and $p_X = q$.
Now a liquidity position in price range $(a, b)$ at price $q$ has price
\begin{equation}
    \epsilon_{a, b}(q) \triangleq \textcolor{red}{p_Y}\left(\sqrt{\proj{q}{a}{b}} - \sqrt{a} + \frac{q}{\sqrt{\proj{q}{a}{b}}} - \frac{q}{\sqrt{b}}\right).
    \label{eqn:eps-general}
\end{equation}

The multiplier $p_Y$ is the only difference between our current setting and our main text setting in \eqref{eqn:epsilon}.

We assume that the price of $X$ and $Y$ change to $p_X'$ and $p_Y'$, respectively, where the fast arbitrageurs ensure the new pool price $q' = p_X'/p_Y'$.
Recall that $(\Delta x, \Delta y)$ is the initial holdings of an LP and $(\Delta x', \Delta y')$ is the LP's new holdings in the pool.
Based on \eqref{eqn:il}, the generalized impermanent loss is defined as
\begin{equation}
    \hat \tau = \frac{\vhold - \vlp}{\vinit} = \frac{p_X' (\Delta x - \Delta x') + p_Y'(\Delta y - \Delta y') }{p_X \Delta x + p_Y \Delta y} = \textcolor{red}{\frac{p_Y'}{p_Y}} \cdot \frac{q'(\Delta x - \Delta x') + (\Delta y - \Delta y')}{q \Delta x + \Delta y}.
    \label{eqn:il-general}
\end{equation}

Based on the generalized definition, we present the impermanent loss rate of range $(a, b)$ when price changes from $(p_X, p_Y)$ to $(p_X', p_Y')$.

\begin{proposition}
    In price range range $(a, b)$ of a CLMMs, when price changes from $(p_X, p_Y)$ to $(p_X', p_Y')$ with $q = p_X/p_Y$ and $q' = p_X'/p_Y'$, the impermanent loss rate is given by
    \begin{equation}
        \hat\tau_{a, b}(q', q) = \textcolor{red}{\frac{p_Y'}{p_Y}} \cdot 
        \left( \sqrt{\proj{q}{a}{b}} + \frac{q'}{\sqrt{\proj{q}{a}{b}}} - \sqrt{\proj{q'}{a}{b}} - \frac{q'}{\sqrt{\proj{q'}{a}{b}}} \right).
        \label{eqn:ilv3-general}
    \end{equation}

    In a legacy AMM with the same price changes, the impermanent loss is
    \begin{equation}
        \hat \tau_{0, \infty}(q', q) = \textcolor{red}{\frac{p_Y'}{p_Y}} \cdot
        \left( \sqrt{q} + \frac{q'}{\sqrt{q}} - 2\sqrt{q'} \right). 
        \label{eqn:ilv2-general}
    \end{equation}
    \label{thm:il-analytic-general}
\end{proposition}

\textbf{Proof.}
Let $L$ denote the liquidity in the price range.
By \eqref{eqn:single-liq}, the initial token amounts are
\begin{equation}
    \Delta x = L \left( \frac{1}{\sqrt{\proj{q}{a}{b}}} - \frac{1}{\sqrt{b}} \right), \qquad \Delta y = L \left(\sqrt{\proj{q}{a}{b}} - \sqrt{a}\right).
\end{equation}

In a CLMM, by \eqref{eqn:eps-general}:
\begin{align*}
    \vinit & =  L p_Y \left( \sqrt{\proj{q}{a}{b}} + \dfrac{q}{\sqrt{\proj{q}{a}{b}}}  - \sqrt{a} - \dfrac{q}{\sqrt{b}} \right);                                                \\
    \vlp   & = L p_Y' \left( \sqrt{\proj{q'}{a}{b}} + \dfrac{q'}{\sqrt{\proj{q'}{a}{b}}}  - \sqrt{a} - \dfrac{q'}{\sqrt{b}} \right);                                            \\
    \vhold & = p_Y' (q' \Delta x + \Delta y) = L p_Y' \left[ q' \left( \frac{1}{\sqrt{\proj{q}{a}{b}}} - \frac{1}{\sqrt{b}} \right) + \sqrt{\proj{q}{a}{b}} - \sqrt{a} \right].
\end{align*}

Plug into \eqref{eqn:il-general}, and
\begin{align}
    \hat\tau_{a, b}(q', q) & =  \epsilon_{a, b}(q) \cdot \frac{\vhold - \vlp}{\vinit} =
    \textcolor{black}{\frac{p_Y'}{p_Y}} \cdot 
    \left( \sqrt{\proj{q}{a}{b}} + \frac{q'}{\sqrt{\proj{q}{a}{b}}} - \sqrt{\proj{q'}{a}{b}} - \frac{q'}{\sqrt{\proj{q'}{a}{b}}} \right).
    \notag
\end{align}

In a legacy AMM, we may simply set $a = 0$ and $b = \infty$. Consequently,
\begin{align*}
    \iga(q', q) = \igb_{0, \infty}(q', q) & = \textcolor{black}{\frac{p_Y'}{p_Y}} \cdot
    \left( \sqrt{q} + \frac{q'}{\sqrt{q}} - \sqrt{q'} - \frac{q'}{\sqrt{q'}} \right)
                                          = \frac{p_Y'}{p_Y} \cdot \left( \sqrt{q} + \frac{q'}{\sqrt{q}} - 2\sqrt{q'} \right).                       \qeda
\end{align*}


\paragraph*{Non-negativity of Impermanent Loss}

\begin{proposition}
    For all $q, q' > 0$ and $a, b$ with $0 \le a < b \le \infty$, $\igb_{a, b}(q', q) \ge 0$.
    \label{thm:positive-il}
\end{proposition}

\textbf{Proof. }
By \eqref{eqn:ilv3-general}, it suffices to prove
\begin{equation*}
    \sqrt{\proj{q}{a}{b}} + \frac{q'}{\sqrt{\proj{q}{a}{b}}} - \sqrt{\proj{q'}{a}{b}} - \frac{q'}{\sqrt{\proj{q'}{a}{b}}} \ge 0.
\end{equation*}

Equivalently,
\begin{equation}
    \left(\sqrt{\proj{q}{a}{b}} - \sqrt{\proj{q'}{a}{b}}\right)\left(1 - \frac{q'}{\sqrt{\proj{q}{a}{b} \proj{q'}{a}{b}} }\right) \ge 0.
    \label{eqn:positive-product}
\end{equation}

It is trivial to prove the statement if $\sqrt{\proj{q}{a}{b}} = \sqrt{\proj{q'}{a}{b}}$.
Now if $\proj{q}{a}{b} < \proj{q'}{a}{b}$, we must have
\[
    q' > a \thus q' \ge \proj{q'}{a}{b} \thus
    \frac{q'}{\sqrt{\proj{q}{a}{b} \proj{q'}{a}{b}} } \ge \sqrt{\frac{\proj{q'}{a}{b}}{\proj{q}{a}{b}}} > 1.
\]

On the other hand, if $\proj{q}{a}{b} > \proj{q'}{a}{b}$, we must have
\[
    q' < b \thus q' \le \proj{q'}{a}{b} \thus
    \frac{q'}{\sqrt{\proj{q}{a}{b} \proj{q'}{a}{b}} } \le \sqrt{\frac{\proj{q'}{a}{b}}{\proj{q}{a}{b}}} < 1.
\]

In summary, \eqref{eqn:positive-product} always holds. \qeda

\paragraph*{Decomposability of Coefficients $\epsilon$ and $\epsilon \cdot \igb$}

\begin{lemma}[Decomposability]
    For all $a, b$ with $0 \le a < b \le \infty$ and $c \in (a, b)$, we have
    \begin{align}
        \epsilon_{a, b}(q)                    & = \epsilon_{a, c}(q) + \epsilon_{c, b}(q),                                       &  & \forall q > 0; \label{eqn:decomp-eps}    \\
        \igb_{a, b}(q', q) \epsilon_{a, b}(q) & = \igb_{a, c}(q', q) \epsilon_{a, c}(q) + \igb_{c, b}(q', q) \epsilon_{c, b}(q), &  & \forall q, q' > 0. \label{eqn:decomp-il}
    \end{align}
    \label{thm:decomp}
\end{lemma}

\textbf{Proof.}
By definition of interval projections, we have the set equality
\begin{align*}
    \left\{\proj{q}{a}{c}, \proj{q}{c}{b}\right\}
     & = \begin{rcases}
             \begin{dcases}
            \{ \max\{a, q\}, c \}, & q \le c \\
            \{ c, \min\{q, b\} \}, & q > c
        \end{dcases}
         \end{rcases}
    = \left\{ c, \proj{q}{a}{b} \right\}.
\end{align*}

Hence, \eqref{eqn:decomp-eps} holds because
\begin{align*}
     & ~ \quad \epsilon_{a, c}(q) + \epsilon_{c, b}(q)                                                                             \\
     & = \sqrt{\proj{q}{a}{c}} + \frac{q}{\sqrt{\proj{q}{a}{c}}} - \sqrt{a} - \frac{q}{\sqrt{c}} +
    \sqrt{\proj{q}{c}{b}} + \frac{q}{\sqrt{\proj{q}{c}{b}} } - \sqrt{c} - \frac{q}{\sqrt{b}}                                       \\
     & = \left( \sqrt{\proj{q}{a}{c}} + \sqrt{\proj{q}{c}{b}} - \sqrt{a} - \sqrt{c} \right)
    + q \left( \frac{1}{\sqrt{\proj{q}{a}{c}}} + \frac{1}{\sqrt{\proj{q}{c}{b}}} - \frac{1}{\sqrt{c}} - \frac{1}{\sqrt{b}} \right) \\
     & = \left( \sqrt{\proj{q}{a}{b}} + \sqrt{c} - \sqrt{a} - \sqrt{c} \right)
    + q \left( \frac{1}{\sqrt{\proj{q}{a}{b}}} + \frac{1}{\sqrt{c}} - \frac{1}{\sqrt{c}} - \frac{1}{\sqrt{b}} \right)              \\
     & = \sqrt{\proj{q}{a}{b}} + \frac{q}{\sqrt{\proj{q}{a}{b}}} - \sqrt{a} - \frac{q}{\sqrt{b}}                                   \\
     & = \epsilon_{a, b}(q).
\end{align*}

Similarly for \eqref{eqn:decomp-il}, by \eqref{eqn:ilv3-general},
\begin{align*}
     & \qquad \igb_{a, c}(q', q) \epsilon_{a, c}(q) + \igb_{c, b}(q', q) \epsilon_{c, b}(q)                                                                                                                                                                                                                                                                                       \\
     & = \begin{aligned}[t]
              & p_Y' \left(\sqrt{\proj{q'}{a}{c}} + \frac{p}{\sqrt{\proj{q'}{a}{c}} } - \sqrt{\proj{q}{a}{c}} - \frac{p}{\sqrt{\proj{q}{a}{c}}} \right) + \\
              & p_Y' \left( \sqrt{\proj{q'}{c}{b}} + \frac{p}{\sqrt{\proj{q'}{c}{b}} } - \sqrt{\proj{q}{c}{b}} - \frac{p}{\sqrt{\proj{q}{c}{b}}} \right)
         \end{aligned} \\
     & = p_Y' \left( \sqrt{\proj{q'}{a}{b}} + \frac{p}{\sqrt{\proj{q'}{a}{b}} } - \sqrt{\proj{q}{a}{b}} - \frac{p}{\sqrt{\proj{q}{a}{b}}}
    + \sqrt{c} + \frac{p}{\sqrt{c}} - \sqrt{c} - \frac{p}{\sqrt{c}} \right)                                                                                                                                                                                                                                                                                                       \\
     & = p_Y' \left( \sqrt{\proj{q'}{a}{b}} + \frac{p}{\sqrt{\proj{q'}{a}{b}} } - \sqrt{\proj{q}{a}{b}} - \frac{p}{\sqrt{\proj{q}{a}{b}}} \right)                                                                                                                                                                                                                                 \\
     & = \igb_{a, b}(q', q) \epsilon_{a, b}(q).
    \qeda
\end{align*}

\section{Generalized Game Assumptions, Twinship, and Nash Equilibrium Existence}


\subsection{Assumption of Fee Reward and LP Shares}
\label{amm:app:fee-assume-3}

From this section on, we specify assumptions of fee reward and LP share which strike a balance between the main text and  \S\ref{amm:app:twin}.
In detail, we assume that
\begin{enumerate}
    \item $f_m$ is a constant for all price ranges, as we assumed in the main text.
    \item The weight of each LP is given by
        \begin{align}
          \bar w_{n, m}(\Lb_n; \Lb_{-n})
           & = \frac{\left(\sum_{i=1}^m \sum_{j=m}^M L_{n, (t_{i-1}, t_j)}\right)^\alpha}{\chi_m + \sum_{k \in [N]} \left(\sum_{i=1}^m \sum_{j=m}^M L_{k, (t_{i-1}, t_j)}\right)^\alpha} \notag \\
           & = \frac{K_{n, m}^\alpha}{\chi_m + \sum_{i \in [N]} K_{i, m}^\alpha} = w_{n, m}(\Kb_n; \Kb_{-n}). \label{eqn:weight-chi}
        \end{align}
        
        When $K_{n, m} = 0$, we define $w_{n, m}(\Kb_n; \Kb_{-n}) = 0$. 
        
        Note that $\chi_m$ is a non-negative constant that represents the total weights of non-player LPs that are better characterized as a part of the environment.
        One typical example of non-player LPs is just-in-time (JIT) LPs,
        who add liquidity just before a transaction and withdraws immediately after the transaction.
\end{enumerate}

Before diving further, we summarize the different assumptions on the fee reward and LP shares in Tab. \ref{tab:fee-assumption}.

\begin{table}[!htb]
    \centering
    \begin{tabular}{cccc}
        \hline
                                  & Main text                                 & \S\ref{amm:app:twin}                                            & Since \S\ref{amm:app:fee-assume-3} \\
        \hline
        $f_m$                     & Constant                                  & Function of $\kappa_m$                                      & Constant                      \\
        $w_{n, m}, \bar w_{n, m}$ & \eqref{eqn:weight-chi} where $\chi_m = 0$ & General function satisfying \eqref{eqn:weight-decomposable} & \eqref{eqn:weight-chi}        \\
        \hline
    \end{tabular}
    \caption{Fee reward and share assumptions. }
    \label{tab:fee-assumption}
\end{table}

\subsection{Concavity and Continuity of Atomized Utility}

In this section, we discuss the basic properties -- concavity and continuity of utility function $\autil_n$ in an atomized game.
For simplicity, we define the total weight of range $m$ as 
\begin{equation}
    \nu_m \triangleq \chi_m + \sum_{i\in[N]} K_{i, m}^\alpha, \qquad \nu_{-n, m} \triangleq \chi_m + \sum_{i \textcolor{red}{\neq n}} K_{i, m}^\alpha.
\end{equation}

\begin{lemma}
    If $\alpha \in (0, 1]$, $\autil_n(\Kb_n; \Kb_{-n})$ is concave when $\Kb_{-n}$ is constant and $\Kb_n \in \aas_n$.
    \label{thm:concave-util}
\end{lemma}

\textbf{Proof.}
By definition, we can decompose the utility into $M$ univariate functions: $\autil_n(\Kb_n; \Kb_{-n}) = \sum_{m \in [M]} V_{n, m}(K_{n, m})$, where
\[
    V_{n, m}(K_{n, m}) = \begin{cases}
        \displaystyle \frac{f_m K_{n, m}^\alpha}{\nu_{-n, m} + K_{n, m}^\alpha} - \tau_m K_{n, m}, & K_{n, m} \in (0, \infty); \\
        0, & K_{n, m} = 0. 
    \end{cases}
\]

To prove $\autil_n(\Kb_n; \Kb_{-n})$ is concave, it suffices to show the concavity of each component $V_{n, m}$. 
In the subdomain $(0, \infty)$, $V_{n, m}$ is concave when $\alpha \in (0, 1]$ by the second-order condition:
\[
    \frac{\mathrm{d}^2 V_{n, m}}{\mathrm{d} K_{n, m}^2} = 
    \frac{\alpha f_m \nu_{-n, m} K_{n, m}^{\alpha - 2} \left[ (\alpha - 1) \nu_{-n, m} - (\alpha + 1) K_{n, m}^\alpha \right]}{\nu_m^3} \le 0.
\]

To show $V_{n, m}$ is globally concave over $[0, \infty)$, it suffices to show 
\[
    V_{n, m}(0) \le \lim_{t \to 0^+} V_{n, m}(t). 
\]

When $\nu_{-n, m} > 0$, the equality holds. When $\nu_{-n, m} = 0$, 
\[
    V_{n, m}(0) = 0 < f_m = \lim_{t \to 0^+} V_{n, m}(t). 
\]

In conclusion, $V_{n, m}$ is concave for all $m$, which implies $\autil_n$ is concave for all $n \in [N]$. \qeda 

\begin{remark}
$\autil_n$ may be discontinuous near the boundaries. 
However, the following lemma shows that $\autil_n$ must be continuous in the neighborhood of a Nash equilibrium. 
\end{remark}

\begin{lemma}
    When $\alpha \in (0, 1]$, at any Nash equilibrium $\Kbt$ of an atomized game $\ag([N], \aas, \autil)$, $\nut_{-n, m} > 0$ for all $n \in [N]$ and $m \in [M]$. 
    As a result, $\outil_n$ is continuous in $n$'s action under other LPs' actions $\Kbt_{-n}$.
    \label{thm:continuous-utility}
\end{lemma}

\textbf{Proof.}
For each $m\in[M]$, the claim is trivial when $\chi_m > 0$, so we focus on when $\chi_m = 0$. 

Assume otherwise, i.e., $\nut_{-n, m} = 0$ for some $n$ and $m$. 
We concentrate on $n$'s utility in range $m$, since actions on other ranges does not interfere with it. 
In this case, $n$ cannot find a strategy that achieves max utility, because by setting $K_{n, m}$ closer (but not equal) to $0$, $n$ still gets full share of fee $f_m$ while decreasing their impermanent loss. 
As $K_{n, m} \to 0$, $n$'s utility in range $m$ approaches $f_m$. 
On the other hand, $n$'s utility in range $m$ when $K_{n, m} = 0$ is $0 < f_m$. 
This implies the utility supremum $f_m$ cannot be achieved at any action. 
Hence, when $\nut_{-n, m} = 0$, the actions cannot be at Nash equilibrium. 

In conclusion, $\nut_{-n, m} > 0$ for all $n$ and $m$. \qeda

\begin{corollary}
    When $\alpha \in (0, 1)$ and $\Kbt$ is a Nash equilibrium of an atomized game, $\Kt_{n, m} > 0$ for all $n \in [N]$ and $m \in [M]$. 
    \label{thm:positive-liq-alpha}
\end{corollary}

\textbf{Proof.} 
We prove by contradiction. Assume $\Kt_{n, m} = 0$ for some $n \in [N]$ and $m \in [M]$. 

\textbf{Case 1.} $\Kt_{n, k} = 0$ for all $k \in [M]$. We let $n$ alternatively use action $\widehat K_{j, m} = \epsilon < B_i$, while $\widehat K_{j, k} = \Kt_{j, k} = 0$ for all $k \neq m$. 
Other LPs do not change their actions.
The resulting utility difference is
\begin{align}
    \Delta \autil_n \triangleq \autil_n\left(\widehat \Kb_n\right) - \autil_n\left(\Kbt_i\right) 
    = \frac{f_m}{\nut_{-n, m}} \epsilon^\alpha + o(\epsilon^\alpha). 
    \label{eqn:positive-grad-1}
\end{align}

\textbf{Case 2.} There exists $k \in [M]$ such that $\Kt_{n, k} > 0$. 
We let $n$ alternatively use action $\widehat K_{n, m} = \epsilon \in (0, K_{n, k})$, 
$\widehat K_{n, k} = \Kt_{n, k} - \epsilon$, 
and $\widehat K_{n, m'} = \Kt_{n, m'}$ for all $m' \notin \{m, m'\}$. 
Other LPs do not change their actions.
The resulting utility difference is
\begin{align}
    \Delta \autil_n \triangleq \autil_n\left(\widehat \Kb_i\right) - \autil_n\left(\Kbt_i\right) 
    = \frac{f_m}{\nut_{-n, m}} \epsilon^\alpha - \frac{f_k \nut_{-n, k} \Kt_{n, k}^{\alpha-1}}{\left(\nut_{-n, k} + \Kt_{n, k}^\alpha\right)^2} \epsilon + o(\epsilon^\alpha) = \frac{f_m}{\nut_{-n, m}} \epsilon^\alpha + o(\epsilon^\alpha). 
    \label{eqn:positive-grad-2}
\end{align}

Both equations \eqref{eqn:positive-grad-1} and \eqref{eqn:positive-grad-2} indicate the existence of $\epsilon > 0$, such that $\Delta \autil_n > 0$, which denies the possibility of $\Kt_{n, m} = 0$ being a part of a Nash equilibrium. 
\qeda  

\begin{remark}
Prop. \ref{thm:positive-liq} is a special case of Cor. \ref{thm:positive-liq-alpha} where $\chi_m = 0$ for all $m \in [M]$. 
\end{remark}

\subsection{Proof of Prop. \ref{thm:orig-ne-exist}: Existence of Nash Equilibrium in Original Games}
\label{amm:app:orig-ne-exist}

By Thm. 1 in \cite{rosen1965dsc}, it suffices to show that this game is a concave game, i.e., each LP in the game solves a convex program with a concave objective to maximize. 
In other words, for each $n \in [N]$, the action space $\oas_n$ is convex and the utility function $\outil_n$ is concave in $n$'s own actions.

\emph{Convexity of $\oas_n$.} $\oas_n$ is convex because it is the intersection of a finite number of half-planes.

\emph{Concavity of $\outil_n$.} 
Let $\ag([N], \aas, \autil)$ be the twin atomized game of $\og$. 
By Lemma \ref{thm:concave-util}, $\autil_n$ is concave. 
Let matrix $Q$ represent the linear mapping $\theta$; i.e., $\theta(\Lb_n) \equiv Q \Lb_n$, where $Q \in \R^{M \times M(M+1)/2}$. 
We have for all $n \in [N]$ and $\Lb_n \in \oas_n$, $Q \Lb_n \in \aas_n$ and
\[
    \outil_n(\Lb_n) = \autil(Q \Lb_n). 
\]

The concavity of $\outil_n$ can be thus shown by definition. 
Hence, $\og$ is a concave game with at least one Nash equilibrium. \qeda

\subsection{Proof of Thm. \ref{thm:ne-simple-unique}: Unique Nash Equilibrium in Atomized Games}
\label{amm:app:ne-simple-unique}


\begin{theorem}
    In an atomized game $\ag([N], \aas, \autil)$ where the fee rewards are defined in \S\ref{amm:app:fee-assume-3}, 
    there exists a unique Nash equilibrium. 
\end{theorem}

\textbf{Proof.} 
Analogous to Prop. \ref{thm:orig-ne-exist}, $\ag$ is also a concave game with at least one Nash equilibrium. 
By Lemma \ref{thm:continuous-utility}, $\autil_n$ must be continuous (and twice differentiable by its analytical form) in $n$'s own actions $\left(K_{n, 1}, \cdots, K_{n, M} \right)$ at a neighborhood of any Nash equilibrium. 
This enables the following second-order analysis. 

By the diagonal strict concavity (DSC) theorem \cite{rosen1965dsc}, 
it suffices to show $H_m \prec 0$, where 
\begin{align}
    [J_m]_{i, j} &= \dfrac{\partial^2 \autil_n}{\partial K_{j, m} \partial K_{n, m}}, \qquad \forall i, j \in [N]^2; \\
    H_m &\triangleq J_m + J_m\T.
\end{align}

Take derivatives of $\autil_n$, and we have 
\begin{align*}
    \frac{\partial \autil_n}{\partial K_{n, m}} &= \frac{\alpha f_m \nu_{-n, m} K_{n, m}^{\alpha - 1}}{\nu_m^2}, \\
    \frac{\partial^2 \autil_n}{\partial K_{n, m}^2} &= \frac{\alpha f_m \nu_{-n, m} K_{n, m}^{\alpha - 2} \left[ (\alpha - 1) \nu_{-n, m} - (\alpha + 1) K_{n, m}^\alpha \right]}{\nu_m^3},\\
    \frac{\partial^2 \autil_n}{\partial K_{j, m} \partial K_{n, m}} &= \frac{\alpha^2 f_m K_{n, m}^{\alpha-1} K_{j, m}^{\alpha-1} \left(2K_{n, m}^\alpha - \nu_m \right)}{\nu_m^3} \quad (i \neq j), \\
    \frac{\partial^2 \autil_n}{\partial K_{n', j} \partial K_{n'', k}} &= 0 \quad (\forall n' \neq n''). 
\end{align*}

Hence, 
\[
    H_m = \frac{2\alpha f_m }{\nu_m^3} 
    \begin{bmatrix} 
        \nu_{-1, m} K_{1, m}^{\alpha - 2} \left[ \alpha \left( \nu_{-1, m} - K_{1, m}^\alpha \right) - \nu_m \right]  
        & \cdots 
        & \alpha K_{1, m}^{\alpha-1}K_{n, m}^{\alpha - 1} \left(K_{1, m}^\alpha + K_{n, m}^{\alpha} - \nu_m \right) \\ 
        \vdots & \ddots & \vdots \\ 
        \alpha K_{1, m}^{\alpha-1}K_{n, m}^{\alpha - 1} \left(K_{1, m}^\alpha + K_{n, m}^{\alpha} - \nu_m \right) 
        & \cdots &  
        \nu_{-n, m} K_{n, m}^{\alpha - 2} \left[ \alpha \left( \nu_{-n, m} - K_{n, m}^\alpha \right) - \nu_m  \right] 
    \end{bmatrix}
\]

For any $x \in \R^{N} - \{ \zero \}$, 
\begin{align*}
    \frac{\nu_m^3  x\T H_m x}{2\alpha f_m} 
    & = 
    \begin{aligned}[t]
        & \nu_m \sum_{n\in [N]} \left[ (\alpha  - 1) \nu_m K_{n, m}^{\alpha - 2} - (2\alpha - 1) K_{n, m}^{2\alpha - 2} \right] x_n^2  \\ 
        & - \alpha \sum_{n, j \in [N]^2} K_{n, m}^{\alpha - 1} K_{j, m}^{\alpha - 1} \left(\nu_m - K_{n, m}^\alpha - K_{j, m}^{\alpha}\right) x_n x_j        
    \end{aligned}
     \\     
    & =
    \begin{aligned}[t]
        & \nu_m \sum_{n\in [N]} \left[ (\alpha  - 1) \nu_m K_{n, m}^{\alpha - 2} - (2\alpha - 1) K_{n, m}^{2\alpha - 2} \right] x_n^2 \\  
        & - \alpha \left( \sum_{n \in [N]} K_{n, m}^{\alpha-1} \left( \nu_m - K_{n, m}^\alpha \right) x_n \right) \left( \sum_{j \in [N]} K_{j, m}^{\alpha-1} x_j \right). 
    \end{aligned}
\end{align*}

Let $z_n = x_n K_{n, m}^{\alpha-1}$ for $n \in [N]$. 
Coefficient $K_{n, m}^{\alpha-1}$ is always positive because 
1) when $\alpha = 1$, $K_{n, m}^{\alpha-1} > 0$; and 
2) when $\alpha < 1$, $K_{n, m}^{\alpha-1} > 0$ by Cor. \ref{thm:positive-liq-alpha}. 
Hence, we have $\zb \neq 0 \Longleftrightarrow \xb \neq 0$. 

Plug back, and 
\[
    \frac{\nu_m^3  x\T H_m x}{2\alpha f_m} = \nu_m \sum_{n\in [N]} \left[ (\alpha  - 1) \frac{\nu_m}{K_{n, m}^\alpha} - (2\alpha - 1)\right] z_n^2  
    - \alpha \left( \sum_{n \in [N]}  \nu_{-n, m} z_n \right) \left( \sum_{j \in [N]} z_j \right). 
\]

Since $\alpha \le 1$, we have 
\begin{align*}
    \frac{\nu_m^3  x\T H_m x}{2\alpha f_m} 
    &=
    \begin{aligned}[t]
        & (\alpha - 1) \nu_m \sum_{n\in [N]} \frac{\nu_{-n, m}}{K_{n, m}^{\alpha}} z_n^2 \\
        & - \alpha \left[ \nu_m \sum_{n\in[N]} z_n^2 + \nu_m \left( \sum_{n \in [N]} z_n \right)^2 - \left( \sum_{n \in [N]} K_{n, m}^\alpha z_n \right) \left( \sum_{j \in [N]} z_j \right) \right] 
    \end{aligned} \\
    & \le - \alpha \left[ \nu_m \sum_{n\in[N]} z_n^2 + \nu_m \left( \sum_{n \in [N]} z_n \right)^2 - \left( \sum_{n \in [N]} K_{n, m}^\alpha z_n \right) \left( \sum_{j \in [N]} z_j \right) \right] \\ 
    &= - \alpha \sum_{k \in [N]} K_{m, k}^\alpha \cdot \left[ \sum_{n \in [N]} z_n^2 + \left(\sum_{n \in [N]} z_n\right)^2 - z_k \sum_{n \in [N]} z_n \right] \\ 
    & = -\alpha  \sum_{k \in [N]} K_{m, k}^\alpha \cdot \left[ \sum_{n \neq k} z_n^2 + \left(\sum_{n \neq k} z_n\right)^2 \right]  \\
    &< 0, \quad \forall \xb \neq 0. 
\end{align*}

Consequently, the DSC property holds, which implies that for all $\alpha \in (0, 1]$, the game has a unique Nash equilibrium.  \qeda

\subsection{The Original Game and the Atomized Game are Twin Games}
\label{amm:app:twin}

We aim to prove Thm. \ref{thm:complex-simple} in this section.
We begin with relaxations of assumptions we made in the main text, and prove the theorem under a generalized setting.
In particular, we assumed in \S\ref{sec:amm:orig-game} that for each atomic range $m$,
the fee reward $f_m$ is constant and the weight of each LP $n$ is proportional to $K_{n, m}^\alpha$.
In this section, we apply two relaxations:
\begin{enumerate}
    \item The fee reward of each range $f_m$ is a function of the total liquidity
          \[
              \kappa_m \triangleq \sum_{n \in [N]} K_{n, m} = \sum_{n \in [N]} \sum_{i=1}^m \sum_{j=m}^M L_{n, (t_{i-1}, t_j)}.
          \]
    \item The weight of LP $n$ in range $m$, denoted by $\bar w_{n, m}$, is a function of $n$'s liquidity $\Lb_{n}$ and other LPs' liquidities $\Lb_{-n}$, which satisfies
          \begin{equation}
              \Theta(\Lb) = \Theta(\Lbt) \thus \bar w_{n, m}(\Lb_n; \Lb_{-n}) = \bar w_{n, m}(\Lbt_n; \Lbt_{-n}).
              \label{eqn:weight-decomposable}
          \end{equation}

          Therefore, there exists a function $w_{n, m}$, such that with $\Kb = \Theta(\Lb)$,
          \begin{equation}
              w_{n, m}(\Kb_n; \Kb_{-n}) = \bar w_{n, m}(\Lb_n; \Lb_{-n}), \quad \forall n\in [N], \Lb.
              \label{eqn:weight-twin}
          \end{equation}
\end{enumerate}

Under these relaxations, we redefine the original and atomic games as follows.

\paragraph*{Original Game}
An original game is denoted by $\og([N], \oas, \outil)$, where $[N]$ denotes the set of players (LPs), and for each LP $n \in [N]$, $\oas_n$ and $\outil_n$ are their action space and utility function, respectively.
$\oas_n$ is the same as \eqref{eqn:orig-action-space}, while utility $\outil_n$ is redefined as
\begin{align}
    \outil_n & \triangleq \sum_{m \in [M]} f_m(\kappa_m) \bar w_{n, m}(\Lb_n; \Lb_{-n}) - \sum_{(a, b) \in \gp}  \bar \tau_{a, b} L_{n, (a, b)}.
    \label{eqn:orig-util-2}
\end{align}

\paragraph*{Atomized Game}
An atomized game is denoted by $\ag([N], \aas, \autil)$, where $[N]$ denotes the set of LPs, and for each LP $n$, $\aas_n$ and $\autil_n$ denote their action space and utility function, respectively.
$\aas_n$ is the same as \eqref{eqn:atom-budget-constraint}, while utility $\autil_n$ is defined as
\begin{align}
    \autil_n & \triangleq \sum_{m \in [M]} f_m(\kappa_m) w_{n, m}(\Kb_n; \Kb_{-n}) - \sum_{m \in [M]} \tau_{m} K_{n, m}.
    \label{eqn:atom-utility-2}
\end{align}

\begin{theorem}
    Let games $\og$ and $\ag$ share parameters $N, M, T, q, \pi, (f_m)_{m \in [M]}, (B_n)_{n \in [N]}$.
    Let weight functions $\bar w$ and $w$ satisfy \eqref{eqn:weight-twin}.
    Then, $\og$ and $\ag$ are twin games and
    \begin{enumerate}
        \item For every Nash equilibrium $\Lbt$ of $\og$, $\Theta(\Lbt)$ is a Nash equilibrium of $\ag$.
              \label{item:simplify-2}

        \item For every Nash equilibrium $\Kbt$ of $\ag$ and strategy $\Lbt$ of $\og$ with $\Kbt = \Theta(\Lbt)$, $\Lbt$ is a Nash equilibrium.
              \label{item:complexify-2}
    \end{enumerate}
\end{theorem}

\textbf{Proof.}
Define linear mapping $\xi: \R^{|\ap|} \to \R^{|\gp|} = \R^{M} \to \R^{M(M+1)/2}$.
Recall that for a vector $v \in \R^{M(M+1)/2}$, we use elements in $\gp = \{(t_{i-1}, t_j) | i, j \in [M], i \le j \}$ to index its elements.
$\xi$ satisfies that for $v = \xi(u)$ and all $i, j \in [M]$ with $i \le j$,
\[
    v_{(t_{i-1}, t_j)} = \begin{cases}
        u_{i}, & i = j;            \\
        0,     & \text{otherwise}.
    \end{cases}
\]

By definition, $\theta(\xi(u)) = u$ for all $u \in \R^M$.

Now we prove the twinship between $\og$ and $\ag$.
Let $\Lb_n \in \R^{M(M+1)/2} \in \oas_n$, which implies
\begin{align}
    \Lb_n \ge 0; \qquad \sum_{(a, b) \in \gp} L_{n, (a, b)} \epsilon_{a, b}(q) \le B_n.
\end{align}

By definition of $\theta$, we have $\Kb_n = \theta(\Lb_n) \ge 0$.
In addition,
\begin{align}
    \sum_{m \in [M]} K_{n, m} \epsilon_{t_{m-1}, t_m}(q)
     & = \sum_{m \in [M]} \epsilon_{t_{m-1}, t_m}(q) \sum_{i=1}^m \sum_{j=m}^M L_{n, (t_{i-1}, t_j)} \notag          \\
     & = \sum_{i=1}^M \sum_{j=i}^M L_{n, (t_{i-1}, t_j)} \sum_{m=i}^j \epsilon_{t_{m-1}, t_m}(q) \notag              \\
     & = \sum_{i=1}^M \sum_{j=i}^M L_{n, (t_{i-1}, t_j)} \epsilon_{t_{i-1}, t_j}(q) \tag{By Lemma. \ref{thm:decomp}} \\
     & = \sum_{(a, b) \in \gp} L_{n, (a, b)} \epsilon_{a, b}(q). \label{eqn:budget-equal}
\end{align}

Hence, $\Kb_n \in \aas_n$.
Conversely, let $\Kb_n$ be an arbitrary element in $\aas_n$.
We may construct $\Lb_n = \xi(\Kb_n)$, which clearly satisfies $\Lb_n \ge 0$, and $\Lb_n \in \oas_n$ by \eqref{eqn:budget-equal}.
Finally, by definition of $f_m$, $\bar w$ and $w$, utilities defined in
\eqref{eqn:orig-util-2} and \eqref{eqn:atom-utility-2} are equal for all ${\Kb} = \Theta(\Lb)$.
We conclude that $\og$ and $\ag$ are twin games.

Next, we prove the relation between their Nash equilibria by contradiction.

\ref{item:simplify}
If for some Nash equilibrium $\Lbt$ of $\og$ and LP $n \in [N]$ (whose utility is $V$),
$\Kbt_n = \theta (\Lbt_n)$ does not maximize $n$'s utility $\autil_n$,
we take the optimal solution $\Kbt_n'$ with utility $V' > V$.
Since $\Kbt_n' \in \aas_n$, $\Lbt_n' = \xi(\Kbt_n') \in \oas_n$.
Hence, keeping the actions of all other LPs unchanged, $\Lbt_n'$ results in utility $V'$ greater than $V$ of $\Lbt_n$ at Nash equilibrium, a contradiction.

\ref{item:complexify}
Let $\Kbt$ be a Nash equilibrium of $\ag$ and $\Lb \in \oas$ satisfying $\Theta(\Lb) = \Kbt$. 
Suppose $\Lb$ is not a Nash equilibrium of $\og$. 
This implies that $\Lb_n$ is not $n$'s optimal action in $\outil_n$ for some LP $n \in [N]$.
By twinship of $\ag$ and $\og$, let $V$ denote the utility of $n$ with strategy $\Kbt$ in $\ag$ and $\Lb$ in $\og$. 
Since $\Lb_n$ is sub-optimal, there must exist another action $\Lb_n' \in \oas_n$ with utility $V' > V$. 
We define $\Lb''$ with $\Lb_n'' = \Lb_n'$ and $\Lb_k'' = \Lb_k$ for all $k \neq n$. 
We have shown that $\Theta(\Lb'') \triangleq \Kb'' \in \aas$ where $n$'s utility equals $V' > V$. This contradicts with $\Kbt$ being a Nash equilibrium. \qeda

\section{Properties of the Unique Nash Equilibrium in Atomized Game}

Throughout the section, we assume the same parameters $N, M, T, q, \pi, (f_m)_{m \in [M]}, (\chi_m)_{m \in [M]}$, $(B_n)_{n \in [N]}$ that are implicitly relevant to $\aas$ and $\autil$ in an atomized game $\ag([N], \aas, \autil)$. 
We start with the KKT conditions at Nash equilibrium of atomized game. 

\begin{proposition}
    When $\alpha \in (0, 1]$, $\Kbt$ is the unique Nash equilibrium of atomized game $\ag([N], \aas, \autil)$ if and only if there exists $\{\lambda_n\}_{n \in [N]}, \{\mu_{n, m}\}_{n, m \in [N] \times [M]}$, such that the following KKT conditions hold. 
    In addition, such $\{\lambda_n\}, \{\mu_{n, m}\}$ is unique. 
    \begin{alignat*}{3}
        \frac{\alpha f_m \nut_{-n, m} \Kt_{n, m}^{\alpha-1}}{\left(\nut_{-n, m} + \Kt_{n, m}^\alpha\right)^2}  - \tau_m - \lambda_n + \mu_{n, m} & = 0, &&\qquad \forall n, m \in [N] \times [M] \tag{KKT-S} \label{eqn:kkt-stat} \\
        \lambda_n \left( \At_n - B_n \right) = \lambda_n \left( \sum_{m \in [M]} \Kt_{n, m} \epsilon_{t_{m-1}, t_m}(q) - B_n \right) & = 0, &&\qquad \forall n \in [N] \tag{KKT-CS} \label{eqn:kkt-slack} \\
        \mu_{n, m}\Kt_{n, m} & = 0, &&\qquad \forall n, m \in [N] \times [M] \tag{KKT-CS2} \label{eqn:kkt-slack-2} \\
        \At_n = \sum_{m \in [M]} \Kt_{n, m} \epsilon_{t_{m-1}, t_m}(q) & \le B_n, &&\qquad \forall n \in [N] \tag{KKT-PF} \label{eqn:kkt-primal} \\
        \Kt_{n, m} & \ge 0, &&\qquad \forall n, m \in [N] \times [M] \tag{KKT-PF2} \label{eqn:kkt-primal-2} \\
        \lambda_n & \ge 0, &&\qquad \forall n \in [N] \tag{KKT-DF} \\\label{eqn:kkt-dual}
        \mu_{n, m} & \ge 0, &&\qquad \forall n, m \in [N] \times [M] \tag{KKT-DF2} \\\label{eqn:kkt-dual-2}
    \end{alignat*}
    

    \label{thm:kkt}
\end{proposition}

\textbf{Proof.} 
At Nash equilibrium, each LP $n$ attains maximum in a convex program
\begin{equation}
    \begin{aligned}
        \maximize \quad & \sum_{m=1}^M \left( \frac{f_m K_{n, m}^\alpha}{\nu_{-n, m} + K_{n, m}^\alpha} - \tau_m K_{n, m} \right) \\
        \sjt \quad & \Kb_n \ge 0; \\
        & \sum_{m \in [M]} K_{n, m} \epsilon_{t_{m-1}, t_m}(q) \le B_n. 
    \end{aligned}
    \label{eqn:atom-cvx-prog}
\end{equation}

By Prop. \ref{thm:continuous-utility}, $\nu_{-n, m} > 0$ at Nash equilibrium, so we do not need to consider the discontinuity of the objective (utility) at $K_{n, m} = 0$ when $\nu_{-n, m} = 0$.  

Slater's condition holds because the constraint is essentially a $n$-dimensional simplex in $\R^n$, which must contain an interior point.
Hence, we have strong duality in \eqref{eqn:atom-cvx-prog},  
which further implies the derived KKT conditions are both sufficient and necessary. 
The dual variables are unique because they can be given by \eqref{eqn:kkt-stat} and \eqref{eqn:kkt-slack}. 
\qeda

\subsection{Budget Dominance}

The \emph{budget dominance} property requires that if an LP $n$ is richer (has more budget) than another $n'$, then $n$ puts no less utility than $n'$ in \textbf{every} price range. 

\begin{proposition}[Budget Dominance]
    Let $\alpha \in (0, 1]$ and $i, j \in [N]$. 
    At equilibrium $\Kbt$, if $B_i < B_j$, then $\lambda_i \ge \lambda_j$ and for all $m \in [M]$, $\Kt_{i, m} \le \Kt_{j, m}$. 
    If $B_i = B_j$, then $\lambda_i = \lambda_j$ and for all $m \in [M]$, $\Kt_{i, m} = \Kt_{j, m}$. 
    \label{thm:budget-dom}
\end{proposition}

\textbf{Proof.} First, we prove $\lambda_i \ge \lambda_j \Longrightarrow \At_i \le \At_j$ and $\lambda_i = \lambda_j \Longrightarrow \At_i = \At_j$ by discussing two cases.

\begin{enumerate}
    \item $\alpha \in (0, 1)$. 
    By Prop. \ref{thm:positive-liq-alpha}, we always have $\Kt_{n, m} > 0$ and thus $\mu_{n, m} = 0$. 
    By \eqref{eqn:kkt-stat}, 
    \begin{equation}
        \lambda_i - \lambda_j = \frac{f_m}{\nut_m^2} \left[ \left(\nut_m - \Kt_{i, m}^\alpha \right) \Kt_{i, m}^{\alpha-1} - \left(\nut_m - \Kt_{j, m}^\alpha \right) \Kt_{j, m}^{\alpha-1} \right], \qquad \forall m \in [M]. \notag
        \label{eqn:lambda-c}
    \end{equation}
    
    Since function $\psi_m(x) = \left(\nut_m - x^\alpha \right) x^{\alpha-1}$ \emph{strictly} monotonically decreases, we obtain 
    \begin{alignat*}{4}
        \lambda_i \ge \lambda_j & \Longrightarrow \forall m \in [M]~ \left(\psi_m(\Kt_{i, m}) \ge \psi_m(\Kt_{j, m}) \right) & \Longrightarrow \forall m \in [M]~ \left(\Kt_{i, m} \le \Kt_{j, m} \right) & \Longrightarrow \At_i \le \At_j;        \\
        \lambda_i = \lambda_j & \Longrightarrow \forall m \in [M]~ \left(\psi_m(\Kt_{i, m}) = \psi_m(\Kt_{j, m}) \right) & \Longrightarrow \forall m \in [M]~ \left(\Kt_{i, m} = \Kt_{j, m} \right) & \Longrightarrow \At_i = \At_j. 
    \end{alignat*}

    \item $\alpha = 1$. By \eqref{eqn:kkt-stat}, 
    \[
        \lambda_i - \lambda_j = \frac{f_m}{\nut_m^2} \left[ \left(\mu_{i, m} - \Kt_{i, m} \right) - \left(\mu_{j, m} - \Kt_{j, m} \right) \right].
    \]
    
    If $\lambda_i \ge \lambda_j$, we must have $\Kt_{i, m} \le \Kt_{j, m}$. 
    Otherwise, by $\Kt_{i, m} > \Kt_{j, m}$ we get $\mu_{i, m} > \mu_{j, m} \ge 0$. 
    By \eqref{eqn:kkt-slack-2}, we find a contradiction that $\Kt_{i, m} = 0 \le \Kt_{j, m}$. 

    If $\lambda_i = \lambda_j$, $\mu_{i, m} - \Kt_{i, m}  = \mu_{j, m} - \Kt_{j, m}$.
    If $\Kt_{i, m} \neq \Kt_{j, m}$, then one of them must be positive and the  other is zero (otherwise, $\mu_{i, m} = \mu_{j, m} = 0$ contradicts with the equality). 
    Without loss of generality, let $\Kt_{i, m} > 0 = \Kt_{j, m}$. 
    As a result, $\mu_{i, m} = 0 \le \mu_{j, m}$, which further leads to a contradiction 
    $\mu_{i, m} - \Kt_{i, m} < 0 \le \mu_{j, m} - \Kt_{j, m}$. 
    
    Therefore, 
    \begin{alignat*}{3}
        \lambda_i \ge \lambda_j &\thus \forall m \in [M] ~ \left(\Kt_{i, m} \le \Kt_{j, m} \right) & \thus \At_i \le \At_j;       \\
        \lambda_i = \lambda_j &\thus \forall m \in [M] ~ \left(\Kt_{i, m} =\Kt_{j, m} \right) & \thus \At_i = \At_j. 
    \end{alignat*}
\end{enumerate}

By swapping $i$ and $j$, we have the other direction $\lambda_i \ge \lambda_j \Longleftarrow \At_i \le \At_j$, which altogether implies $\lambda_i \ge \lambda_j \eqv \At_i \le \At_j$. 

Next, we show by contradiction that if $B_i < B_j$, then $\At_i \le \At_j$.
Assume otherwise ($\At_i > \At_j$), which implies $0 \le \lambda_i < \lambda_j$. 
By \eqref{eqn:kkt-slack}, $\At_j = B_j > B_i \ge \At_n$, which contradicts with $\At_n > \At_j$.
Therefore,
\[
    B_i < B_j \thus \At_i \le \At_j \thus \lambda_i \ge \lambda_j \thus \forall m \in [M] ~ \left(\Kt_{i, m} \le \Kt_{j, m} \right). 
\]

Finally, we show by contradiction that if $B_i = B_j$, then $\At_i = \At_j$.
Without loss of generality, assume $\At_i > \At_j$, which implies $0 \le \lambda_i < \lambda_j$. 
Again, by \eqref{eqn:kkt-slack}, $\At_j = B_j = B_i \ge \At_i$, which contradicts with $\At_i > \At_j$.
Therefore,
\[
    B_i = B_j \thus \At_i = \At_j \thus \lambda_i = \lambda_j \thus \forall m \in [M] ~ \left(\Kt_{i, m} = \Kt_{j, m} \right). \qeda
\]

\subsection{Waterfill}
\label{amm:app:waterfill}

We prove the waterfill pattern in Thm. \ref{thm:waterfill}, which is a special case of Prop. \ref{thm:waterfill-single-liq} and Prop. \ref{thm:waterfill-budget} by letting $\chi_m = 0$ for all $m$.
The pattern summarizes that at Nash equilibrium, we can split LPs into poor LPs ($\Nc_-$) and rich LPs ($\Nc_+$) by a budget threshold, 
such that all poor LPs exhaust their budgets, while all rich LPs spend the same amount of budget that no poor LP affords. 

\begin{proposition}[Liquidity Waterfill]
    Let $\alpha \in (0, 1]$.
    At equilibrium $\Kbt$ of an atomized game, we define set $\Nc_+ \triangleq \{n|\At_n < B_n\}$.
    For all $m \in [M]$, there exists $h_m > 0$, such that
    \begin{align}
        \Kt_{n, m} \begin{cases}
            = h_m, & n \in \Nc_+; \\
            \le h_m, & n \notin \Nc_+
        \end{cases}. \notag
    \end{align}
    \label{thm:waterfill-single-liq}
\end{proposition}

\textbf{Proof.} 
Consider $\Nc_- = [N] - \Nc_+$. 
By \eqref{eqn:kkt-stat}, 
\[
    0 = \lambda_n = f_m \frac{ \nut_{-n, m} \Kt_{n, m}^{\alpha - 1}}{\left( \Kt_{n, m}^\alpha + \nut_{-n, m}\right)^2} - \tau_m + \mu_{n, m}, \qquad \forall n \in \Nc_+.
\]

Now if there exist $i, j \in \Nc_+$ such that $\Kt_{i, m} < \Kt_{j, m}$, 
we have $\mu_{j, m} = 0$. 
Consequently, 
\begin{align*}
    0 & = \left[ f_m \frac{ \left( \nut_m - \Kt_{i, m}^\alpha \right) \Kt_{i, m}^{\alpha - 1}}{\nut_m^2} - \tau_m + \mu_{i, m} \right] - \left[ f_m \frac{ \left( \nut_m - \Kt_{j, m}^\alpha \right) \Kt_{j, m}^{\alpha - 1}}{\nut_m^2} - \tau_m + \mu_{j, m} \right] \\ 
    & = \frac{f_m}{\nut_m^2}\left[ \left( \nut_m - \Kt_{i, m}^\alpha \right) \Kt_{i, m}^{\alpha - 1} - \left( \nut_m - \Kt_{j, m}^\alpha \right) \Kt_{j, m}^{\alpha - 1} \right] + \mu_{i, m} \\
    & \ge \frac{f_m}{\nut_m^2}\left[ \left( \nut_m - \Kt_{i, m}^\alpha \right) \Kt_{i, m}^{\alpha - 1} - \left( \nut_m - \Kt_{j, m}^\alpha \right) \Kt_{j, m}^{\alpha - 1} \right] > 0. 
\end{align*}

The last inequality results from the strict decreasing monotonicity of function $\psi(x) = \left( \nut_m - x^\alpha \right) x^{\alpha - 1}$.
By this contradiction, we assert that for all $i, j \in \Nc_+$, $\Kt_{i, m} = \Kt_{j, m}$.
In other words, there must exist a constant $h_m$ such that $\Kt_{n, m} = h_m$ for all $n, m \in \Nc_+ \times [M]$.


For all $i \in \Nc_-$, 
\[
    \lambda_i 
    \ge 0 = \lambda_n, \quad \forall m, n \in [M] \times \Nc_+. 
\]

By Prop. \ref{thm:budget-dom}, we have $\Kt_{i, m} \le \Kt_{n, m} = h_m$ for all $(n, m) \in \Nc_+ \times [M]$. \qeda

\begin{corollary}[Budget Waterfill]
    Let $\alpha \in (0, 1]$ and $\Kbt$ be the equilibrium in Prop. \ref{thm:waterfill}. 
    There exists $h > 0$, such that 
    $\At_n = \min\{ h, B_n \} $. 
    \label{thm:waterfill-budget}  
\end{corollary}

\textbf{Proof.}
We prove by summing up $\Kt_{i, m} \le \Kt_{n, m} = h_m$ for all $(i, n, m) \in \Nc_- \times \Nc_+ \times [M]$ over $m$ in Prop. \ref{thm:waterfill-single-liq}. \qeda

\subsection{Proof of Prop. \ref{thm:const-util}: No ``Poor'' LPs}
\label{amm:app:const-util}

By Prop. \ref{thm:waterfill-budget} and \ref{thm:continuous-utility}, for all $n, m \in [N] \times [M]$, $\Kt_{n, m} = h_m > 0$. 
Since $\At_n < B_n$ for all $n$, we have $\lambda_n = 0$ for all $n$ by \eqref{eqn:kkt-slack}. 
Then, by \eqref{eqn:kkt-stat}, for the LP with the highest budget which is not exhausted,  
\[
    \frac{\alpha f_m (N-1)h_m^\alpha \cdot h_m^{\alpha-1}}{(N h_m^\alpha)^2}  - \tau_m = 0. \qquad \forall m \in [M].
\]

This yields 
\begin{equation}
    h_m = \frac{\alpha (N-1)f_m}{N^2 \tau_m}, \qquad \forall m \in [M]. 
\end{equation}

For each $n \in [N]$, 
\[
    U_{n} = \sum_{m \in [M]} \left(\frac{f_m}{N} - \tau_m h_m \right) = \frac{(1-\alpha) N + \alpha}{N^2} \sum_{m \in [M]} f_m. \qeda
\]

\section{Evaluation Supplementaries}

\label{amm:app:eval-supp}

\subsection{Action Preference of Real LP in Stable Pool T5}

\label{amm:app:stable-pool}

\begin{figure}[!htb]
    \centering
    \includegraphics[width=\linewidth]{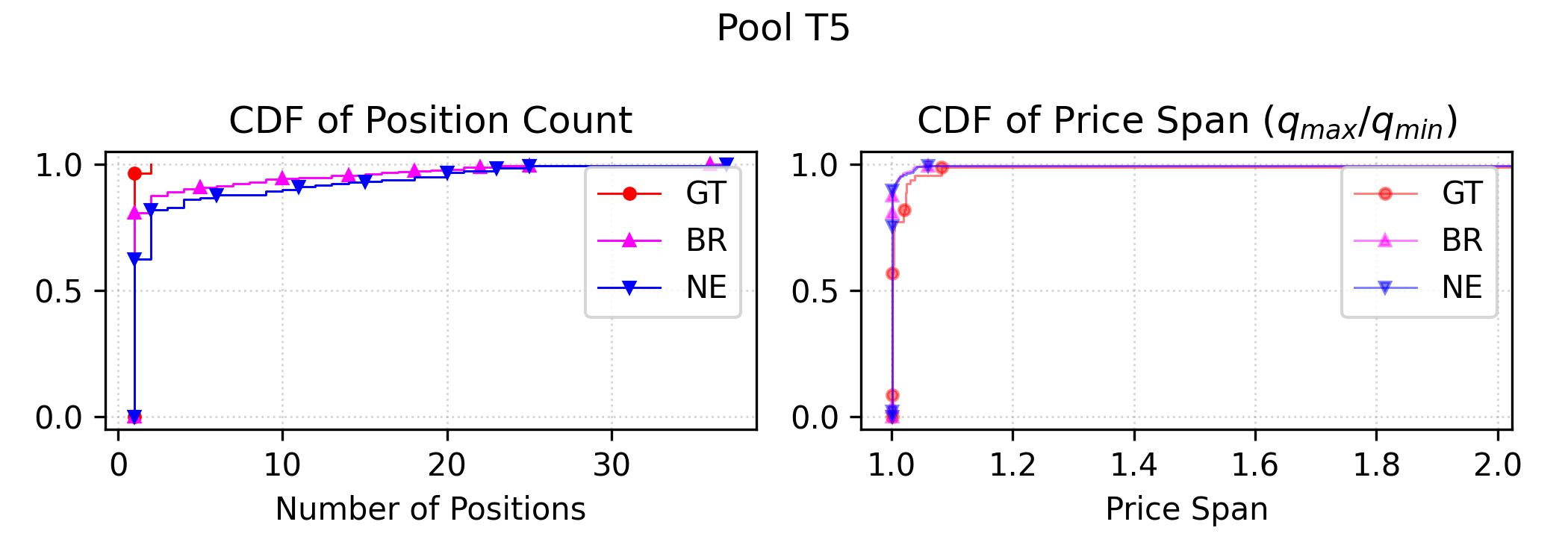}
    \caption{Distribution of number of positions and price spans (max / min) of \gt and \ne. }
    \label{fig:pos-count-span-stable}
\end{figure}

\subsection{Non-Player LP Shares}

\label{amm:app:non-player}

In \S\ref{sec:amm:game-setup}, when we record the fee reward over the price ranges of each trade, we additionally break down the total fee and keep track of the fee share of player LPs and non-player LPs. 
After daily aggregation, we obtain for each price range $m$
1) total liquidity provided by player LPs $\kappa_m$, 
2) total fee reward $f_m$, and 
3) fee reward earned by player LPs $f_m^{\texttt{P}} (\le f_m)$. 
This yields an estimate of non-player LPs' liquidity $\chi_m$: 
\[
    \chi_m = \kappa_m \cdot \frac{f_m - f_m^{\texttt{P}}}{f_m^{\texttt{P}}}. 
\]

Note that it is possible when we observe $f_m > 0$ but $f_m^{\texttt{P}}$ for some price range $m$ in a game. 
We call this price range \emph{uncovered}, where $\chi_m$ is impossible to estimate. 
As a compromise, we do not count $\chi_m$ these ranges when we aggregate the total non-player investments. 
Tab. \ref{tab:player-percentage} shows the number of uncovered ranges for the pools in our scope.

\subsection{Expansion Factor in Inert Game}
\label{amm:app:expansion}

Recall that in the inert game (Tab. \ref{tab:exp-game-setups}), we obtained $\underline{t}$ and $\overline{t}$ as the two ticks that covers all profitable ranges in the past week. 
Then, we expand this price range $(\underline{t}, \overline{t})$ by a factor of $E \ge 1$, resulting in $(\underline{t} / E, E\overline{t})$. 
This is the price range we assume LPs to provide liquidity in the inert game. 
To determine the hyper-parameter $E$, we evaluate the effect of $E$ on utility advantage over ground truth (\gt) and overlap advantage with \gt over Nash equilibrium (\ne). 

Because it is time consuming to calculate Nash equilibria, we instead execute our evaluation based on a proxy strategy, \ibr. 
It assumes the LP joins the game after seeing other LPs actions in \gt, but the LP can only provide liquidity in price range $(\underline{t} / E, E\overline{t})$. 
Inside the range, the LP selects a liquidity value that maximizes the LP's utility. 
This one-dimensional optimization is fast to solve, and it shows an extremely high proximity to \ine, where their differences in liquidity are mostly within the errors in convergence (Fig. \ref{fig:ibr-ine}). 

We summarize the utility and overlap curves in Fig. \ref{fig:expansion}. 
We notice that in risky pools, utility and overlap are sometimes under tradeoff, while in the stable pool T5, both metrics maximize when we do not expand the range at all (by setting $E=1$). 
These curves motivate our final choice of $E$ for each pool in Tab. \ref{tab:expansion}. 

\begin{table}[!htb]
    \centering
    \begin{tabular}{*{6}{c}}
    \toprule
         & B30 & E100 & E30 & E5 & T5 \\
    \midrule
        $E$ Chosen & 3 & 2 & 1.65 & 2 & 1 \\
    \bottomrule
    \end{tabular}
    \caption{Expansion factor of each pool. }
    \label{tab:expansion}
\end{table}

\begin{figure}[!htb]
    \centering
    \includegraphics[width=\linewidth]{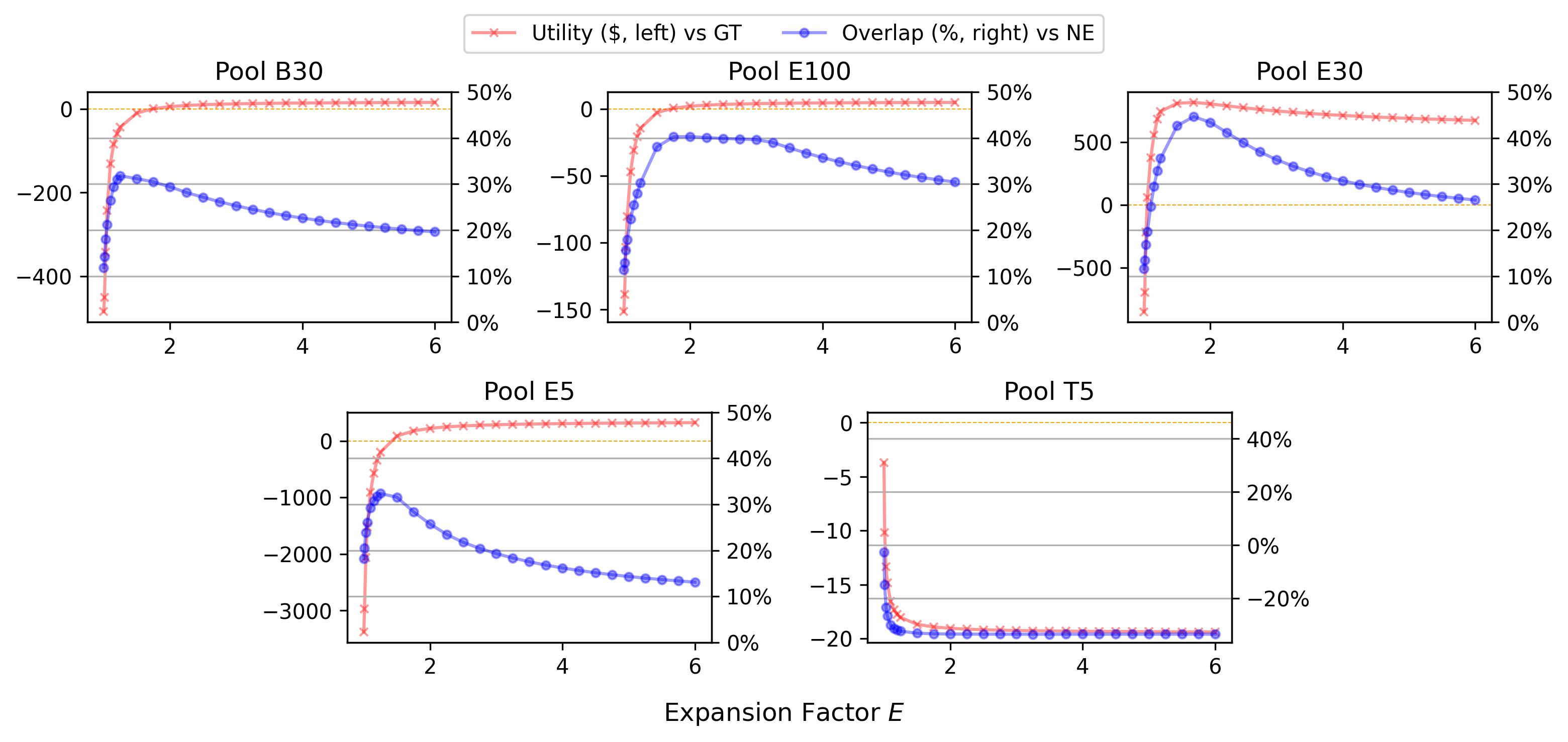}
    \caption{Utility difference between \ibr and \gt and overlap difference between \ibr and \ne under influence of expansion factor $E$.}
    \label{fig:expansion}
\end{figure}

\begin{figure}
    \centering
    \includegraphics[width=\linewidth]{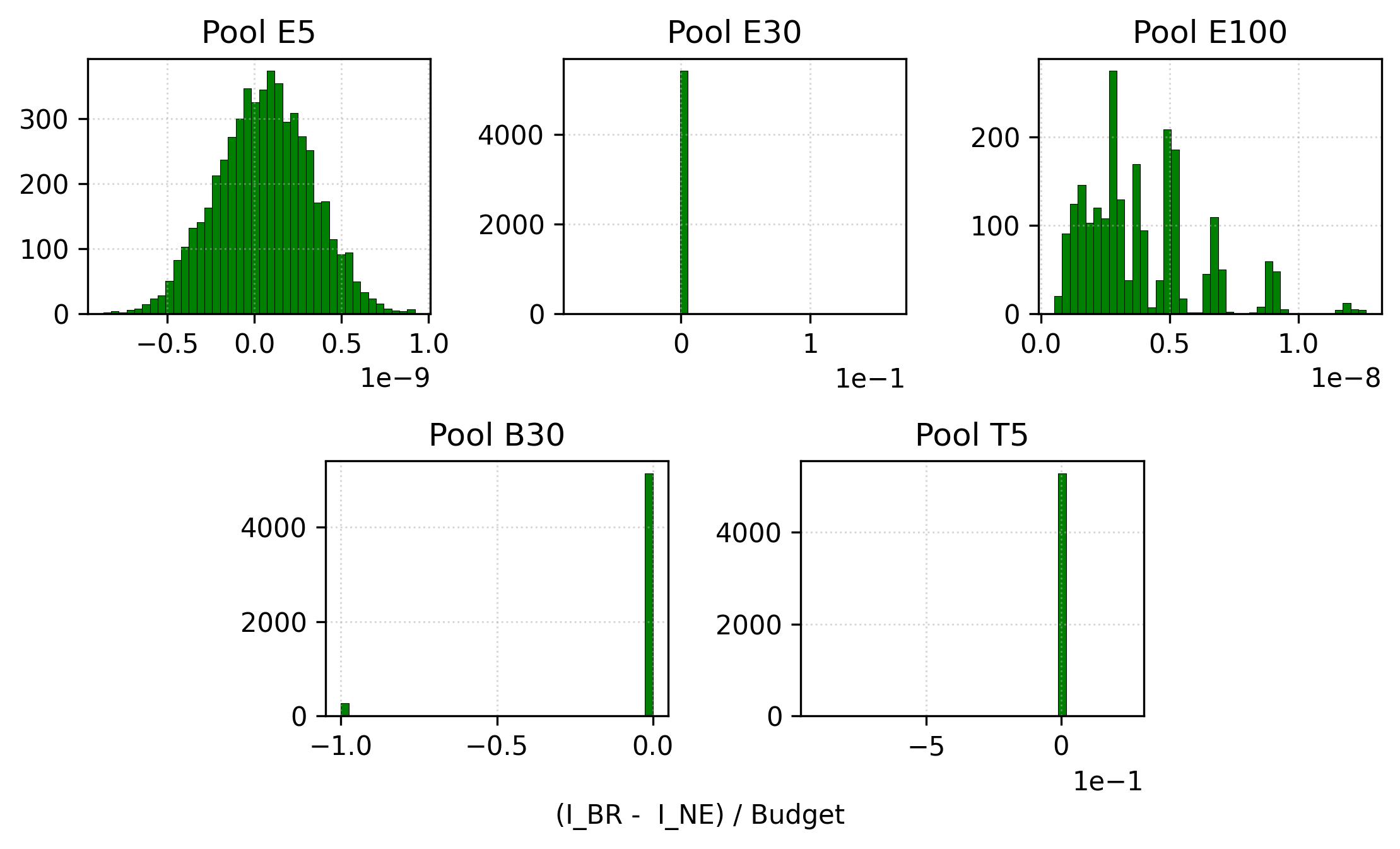}
    \caption{Histogram of normalized difference between \ibr and \ine. Since there is only a single price range, the strategies can be compared by how much budget they used. }
    \label{fig:ibr-ine}
\end{figure}

\subsection{USD Price Estimation Errors}
\label{amm:app:price-bend}

Recall from \S\ref{sec:il-empirical} that we slightly shifted the USD prices of tokens in our evaluation  to more realistically model the profits and utilities seen by LPs. Fig. \ref{fig:price-bend} illustrates the error in prices induced by this calibration. We find that in all of the pools we studied, the error in price is less than 1\%.

\begin{figure}
    \centering
    \includegraphics[width=\linewidth]{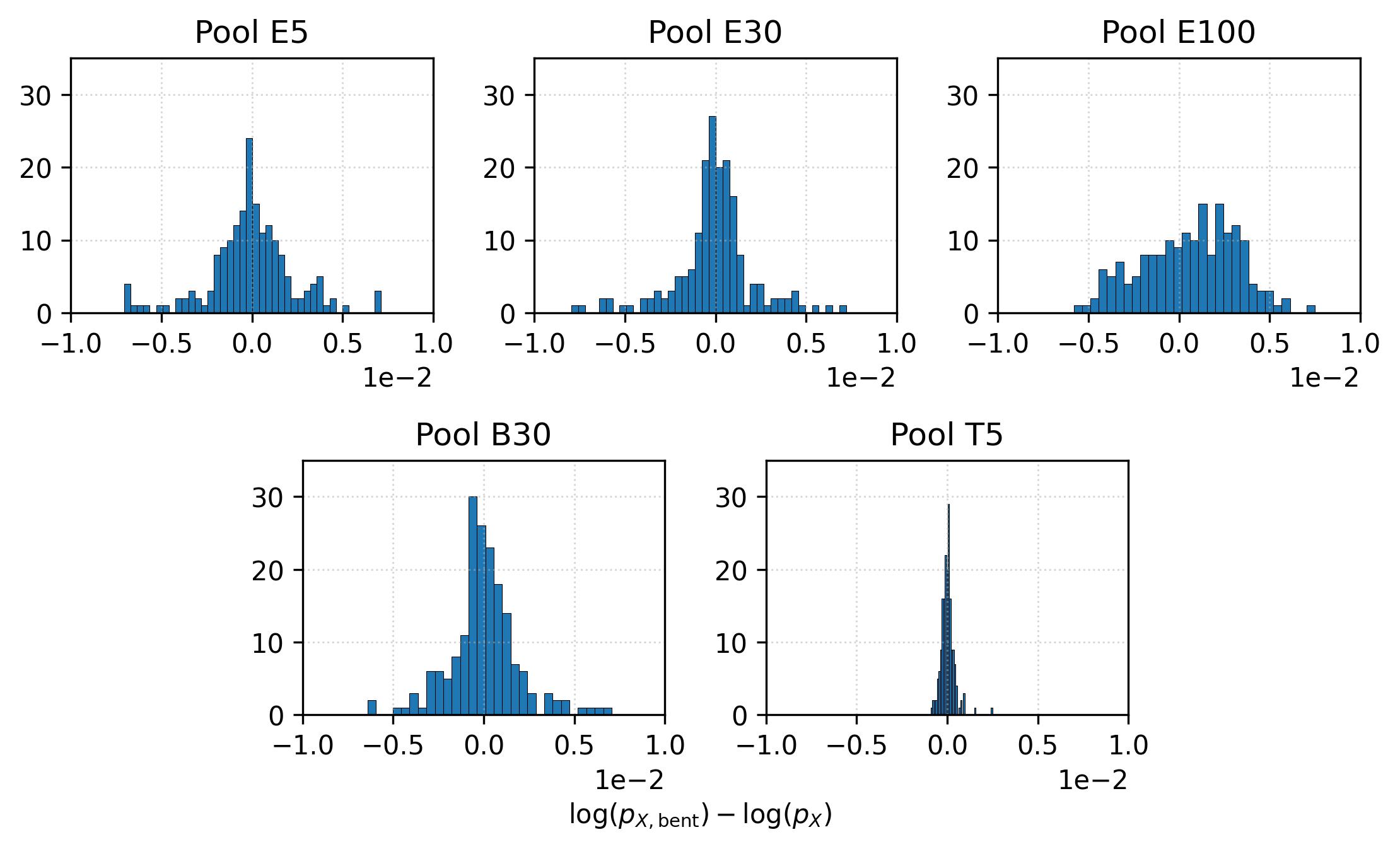}
    \caption{Distribution of $e_X \triangleq \log \bar p_X - \log p_X$. Note that $\log \bar p_Y - \log p_Y = - (\log \bar p_X - \log p_X)$ by definition, and that $\bar p_X \approx (1+e_X) p_X$ when $e_X$ is small. }
    \label{fig:price-bend}
\end{figure}

\subsection{ROI and Optimality Gaps}
\label{amm:app:roi-gap}

We plot the daily return over investment (ROI) and normalized optimality gap (NOG) in Fig. \ref{fig:roi-gap}. 
In detail, they are defined as 
\begin{align}
    \text{ROI}_n &\triangleq \frac{\util_n(\mathtt{action}_n; \gt_{-n})}{B_n}; \\
    \text{NOG}_n &\triangleq \frac{\util_n(\br_n; \gt_{-n}) - \util_n(\mathtt{action}_n; \gt_{-n})}{B_n} = 
    \max_{a \in \as_n}\frac{\util_n(a; \gt_{-n}) - \util_n(\mathtt{action}_n; \gt_{-n})}{B_n}. 
\end{align}

Note that $\gt_{-n}$ represents that all other players take ground truth actions. 
$\mathtt{action}$ represents an action such as \ine and \rne. 
The closer NOG is to 0, the more optimal actions LPs take, and the more rational they are. 

\begin{figure}
    \centering
    \includegraphics[width=\linewidth]{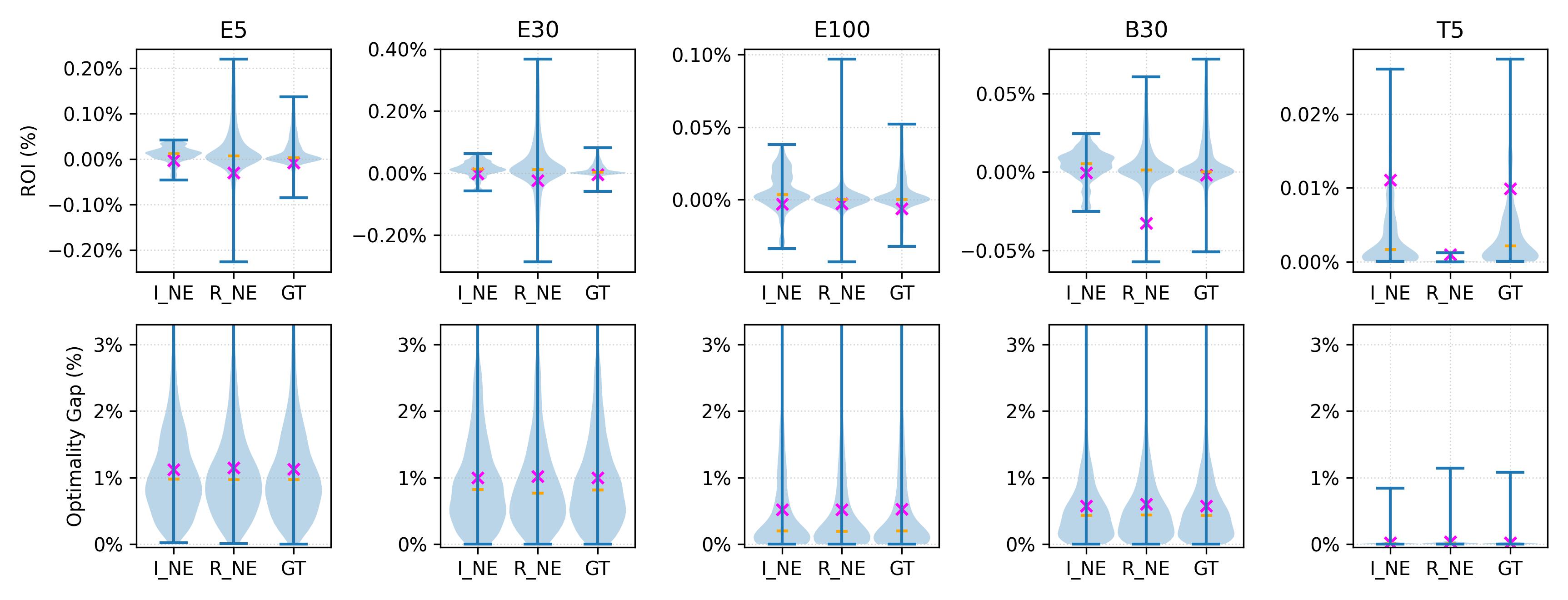}
    \caption{Violin plot of daily return over investment (ROI) and normalized optimality gap. 
    Each marker (\textcolor{magenta}{$\times$}) represents the mean and each horizontal line (\textcolor{orange}{$-$}) represents the median. 
    We list the means, standard deviations and quartiles in Tab. \ref{tab:roi} and \ref{tab:gap}. }
    \label{fig:roi-gap}
\end{figure}

\subsection{Raw Data for Plotting Fig. \ref{fig:box-all} and \ref{fig:roi-gap}}
\label{amm:app:raw-data}

We list Tab. \ref{tab:olap-gt}, \ref{tab:util}, \ref{tab:roi} and \ref{tab:gap}. 
In each table, we list the statistics for each relevant action (\gt, \ne, \ine, etc.) in each liquidity pool. 
The statistics are aggregated from data points collected from each player LP on each day. 
They include mean, standard deviation (Std), and three quartiles (Q1 for 25\% quantile, Q2 for median, and Q3 for 75\% quantile).

\begin{table}[!htb]
    \centering 
    
    \begin{tabular}{c|c|*{5}{c}}
        \toprule
        \multicolumn{2}{c}{} & B30 & E100 & E30 & E5 & T5 \\
        \midrule
        \multirow[c]{5}{*}{\br}
         & Q1 & $0.00 \%$ & $0.00 \%$ & $0.00 \%$ & $0.00 \%$ & $4.36 \%$ \\
         & Q2(Med) & $1.66 \%$ & $1.83 \%$ & $1.28 \%$ & $1.60 \%$ & $25.01 \%$ \\
         & Q3 & $4.19 \%$ & $3.64 \%$ & $2.72 \%$ & $4.24 \%$ & $50.01 \%$ \\
         & Mean & $3.16 \%$ & $3.64 \%$ & $2.56 \%$ & $3.63 \%$ & $30.51 \%$ \\
         & Std & $5.30 \%$ & $6.98 \%$ & $4.58 \%$ & $5.88 \%$ & $25.23 \%$ \\
        \hline
        \multirow[c]{5}{*}{\ne}
         & Q1 & $0.00 \%$ & $0.00 \%$ & $0.00 \%$ & $0.00 \%$ & $5.01 \%$ \\
         & Q2(Med) & $2.47 \%$ & $2.16 \%$ & $2.74 \%$ & $3.74 \%$ & $40.37 \%$ \\
         & Q3 & $6.19 \%$ & $4.96 \%$ & $5.37 \%$ & $8.74 \%$ & $50.01 \%$ \\
         & Mean & $4.40 \%$ & $4.26 \%$ & $4.43 \%$ & $6.51 \%$ & $34.47 \%$ \\
         & Std & $6.23 \%$ & $7.24 \%$ & $6.19 \%$ & $8.57 \%$ & $25.81 \%$ \\
        \hline
        \multirow[c]{5}{*}{\ine}
         & Q1 & $13.87 \%$ & $19.21 \%$ & $26.32 \%$ & $11.73 \%$ & $5.29 \%$ \\
         & Q2(Med) & $26.50 \%$ & $50.37 \%$ & $50.72 \%$ & $20.47 \%$ & $20.03 \%$ \\
         & Q3 & $47.04 \%$ & $68.16 \%$ & $71.58 \%$ & $58.09 \%$ & $50.01 \%$ \\
         & Mean & $31.24 \%$ & $44.57 \%$ & $48.66 \%$ & $32.22 \%$ & $31.92 \%$ \\
         & Std & $20.72 \%$ & $25.61 \%$ & $26.53 \%$ & $24.90 \%$ & $31.64 \%$ \\
        \hline
        \multirow[c]{5}{*}{\rne}
         & Q1 & $0.00 \%$ & $0.00 \%$ & $0.04 \%$ & $0.00 \%$ & $0.00 \%$ \\
         & Q2(Med) & $1.18 \%$ & $0.00 \%$ & $2.47 \%$ & $2.35 \%$ & $0.00 \%$ \\
         & Q3 & $4.17 \%$ & $2.40 \%$ & $5.12 \%$ & $5.79 \%$ & $0.95 \%$ \\
         & Mean & $3.19 \%$ & $2.31 \%$ & $4.13 \%$ & $4.52 \%$ & $2.41 \%$ \\
         & Std & $5.14 \%$ & $5.34 \%$ & $6.09 \%$ & $6.69 \%$ & $7.80 \%$ \\
        \hline
        \multirow[c]{5}{*}{\yday}
         & Q1 & $99.68 \%$ & $99.53 \%$ & $99.54 \%$ & $99.60 \%$ & $100.00 \%$ \\
         & Q2(Med) & $99.92 \%$ & $99.88 \%$ & $99.85 \%$ & $99.89 \%$ & $100.00 \%$ \\
         & Q3 & $100.00 \%$ & $100.00 \%$ & $100.00 \%$ & $100.00 \%$ & $100.00 \%$ \\
         & Mean & $99.66 \%$ & $99.69 \%$ & $99.65 \%$ & $99.53 \%$ & $99.99 \%$ \\
         & Std & $2.49 \%$ & $0.45 \%$ & $1.59 \%$ & $2.69 \%$ & $0.57 \%$ \\
        \bottomrule
    \end{tabular}
    \caption{Overlap between real-world action (\gt) and game-strategic actions.}
    \label{tab:olap-gt}
\end{table}

\begin{table}[!htb]
    \centering 
    \begin{tabular}{c|c|*{5}{c}}
        \toprule
        \multicolumn{2}{c}{} & B30 & E100 & E30 & E5 & T5 \\
        \midrule
        \multirow[c]{5}{*}{\gt}
         & Q1 & $0$ & $0$ & $0$ & $0$ & $0$ \\
         & Q2(Med) & $0$ & $0$ & $10$ & $34$ & $0$ \\
         & Q3 & $45$ & $7$ & $194$ & $722$ & $6$ \\
         & Mean & $-18$ & $-5$ & $-185$ & $-392$ & $19$ \\
         & Std & $2247$ & $288$ & $47350$ & $19739$ & $93$ \\
        \hline
        \multirow[c]{5}{*}{\br}
         & Q1 & $420$ & $16$ & $2619$ & $9442$ & $0$ \\
         & Q2(Med) & $852$ & $66$ & $4786$ & $15316$ & $4$ \\
         & Q3 & $1612$ & $293$ & $8854$ & $26390$ & $24$ \\
         & Mean & $1445$ & $323$ & $8944$ & $22370$ & $45$ \\
         & Std & $2208$ & $724$ & $26329$ & $22750$ & $137$ \\
        \hline
        \multirow[c]{5}{*}{\ne}
         & Q1 & $325$ & $14$ & $1922$ & $8043$ & $0$ \\
         & Q2(Med) & $716$ & $49$ & $3775$ & $13186$ & $3$ \\
         & Q3 & $1466$ & $271$ & $7509$ & $23664$ & $20$ \\
         & Mean & $1300$ & $306$ & $7707$ & $20321$ & $40$ \\
         & Std & $2134$ & $713$ & $25807$ & $22167$ & $131$ \\
        \hline
        \multirow[c]{5}{*}{\ine}
         & Q1 & $-5$ & $0$ & $-32$ & $-26$ & $0$ \\
         & Q2(Med) & $8$ & $0$ & $65$ & $150$ & $0$ \\
         & Q3 & $32$ & $8$ & $224$ & $460$ & $5$ \\
         & Mean & $-3$ & $-3$ & $628$ & $-170$ & $16$ \\
         & Std & $351$ & $225$ & $16826$ & $7738$ & $68$ \\
        \hline
        \multirow[c]{5}{*}{\rne}
         & Q1 & $-11$ & $0$ & $-194$ & $-209$ & $0$ \\
         & Q2(Med) & $4$ & $0$ & $82$ & $155$ & $0$ \\
         & Q3 & $36$ & $2$ & $649$ & $1095$ & $0$ \\
         & Mean & $-45$ & $-25$ & $67$ & $-585$ & $0$ \\
         & Std & $719$ & $496$ & $5872$ & $17347$ & $1$ \\
        \hline
        \multirow[c]{5}{*}{\nea}
         & Q1 & $6$ & $1$ & $81$ & $429$ & $0$ \\
         & Q2(Med) & $22$ & $3$ & $229$ & $915$ & $1$ \\
         & Q3 & $75$ & $22$ & $638$ & $2234$ & $7$ \\
         & Mean & $127$ & $65$ & $1557$ & $2840$ & $19$ \\
         & Std & $576$ & $236$ & $18312$ & $7260$ & $82$ \\
        \bottomrule
    \end{tabular}
    
    \caption{Utility (\$) of each game strategic action (except \nea) when applied to the real-world game; other players are assumed not to deviate from their real-world actions.
    In \nea, every player follow the Nash equilibrium. }
    \label{tab:util}
\end{table}

\begin{table}[!htb]
    \centering 
    \begin{tabular}{c|c|*{5}{c}}
        \toprule
        \multicolumn{2}{c}{} & B30 & E100 & E30 & E5 & T5 \\
        \midrule
        \multirow[c]{5}{*}{\gt}
         & Q1 & $0.000 \%$ & $0.000 \%$ & $0.000 \%$ & $0.000 \%$ & $0.000 \%$ \\
         & Q2(Med) & $0.000 \%$ & $0.000 \%$ & $0.002 \%$ & $0.003 \%$ & $0.002 \%$ \\
         & Q3 & $0.023 \%$ & $0.016 \%$ & $0.029 \%$ & $0.050 \%$ & $0.007 \%$ \\
         & Mean & $-0.002 \%$ & $-0.006 \%$ & $-0.005 \%$ & $-0.008 \%$ & $0.010 \%$ \\
         & Std & $0.154 \%$ & $0.200 \%$ & $0.266 \%$ & $0.372 \%$ & $0.023 \%$ \\
        \hline
        \multirow[c]{5}{*}{\br}
         & Q1 & $0.176 \%$ & $0.045 \%$ & $0.433 \%$ & $0.582 \%$ & $0.003 \%$ \\
         & Q2(Med) & $0.429 \%$ & $0.198 \%$ & $0.821 \%$ & $0.958 \%$ & $0.010 \%$ \\
         & Q3 & $0.758 \%$ & $0.713 \%$ & $1.345 \%$ & $1.517 \%$ & $0.028 \%$ \\
         & Mean & $0.573 \%$ & $0.519 \%$ & $0.995 \%$ & $1.121 \%$ & $0.036 \%$ \\
         & Std & $0.581 \%$ & $0.754 \%$ & $0.818 \%$ & $0.745 \%$ & $0.080 \%$ \\
        \hline
        \multirow[c]{5}{*}{\ne}
         & Q1 & $0.138 \%$ & $0.035 \%$ & $0.326 \%$ & $0.519 \%$ & $0.003 \%$ \\
         & Q2(Med) & $0.373 \%$ & $0.182 \%$ & $0.643 \%$ & $0.858 \%$ & $0.009 \%$ \\
         & Q3 & $0.686 \%$ & $0.637 \%$ & $1.090 \%$ & $1.316 \%$ & $0.024 \%$ \\
         & Mean & $0.500 \%$ & $0.459 \%$ & $0.769 \%$ & $0.977 \%$ & $0.026 \%$ \\
         & Std & $0.515 \%$ & $0.675 \%$ & $0.642 \%$ & $0.634 \%$ & $0.060 \%$ \\
        \hline
        \multirow[c]{5}{*}{\ine}
         & Q1 & $-0.003 \%$ & $-0.004 \%$ & $-0.007 \%$ & $-0.002 \%$ & $0.000 \%$ \\
         & Q2(Med) & $0.005 \%$ & $0.003 \%$ & $0.013 \%$ & $0.012 \%$ & $0.002 \%$ \\
         & Q3 & $0.012 \%$ & $0.019 \%$ & $0.036 \%$ & $0.025 \%$ & $0.008 \%$ \\
         & Mean & $-0.001 \%$ & $-0.003 \%$ & $-0.002 \%$ & $-0.003 \%$ & $0.011 \%$ \\
         & Std & $0.034 \%$ & $0.095 \%$ & $0.138 \%$ & $0.107 \%$ & $0.031 \%$ \\
        \hline
        \multirow[c]{5}{*}{\rne}
         & Q1 & $-0.004 \%$ & $-0.000 \%$ & $-0.025 \%$ & $-0.010 \%$ & $0.000 \%$ \\
         & Q2(Med) & $0.001 \%$ & $0.000 \%$ & $0.011 \%$ & $0.007 \%$ & $0.000 \%$ \\
         & Q3 & $0.015 \%$ & $0.005 \%$ & $0.095 \%$ & $0.059 \%$ & $0.000 \%$ \\
         & Mean & $-0.033 \%$ & $-0.003 \%$ & $-0.023 \%$ & $-0.030 \%$ & $0.001 \%$ \\
         & Std & $0.337 \%$ & $0.438 \%$ & $0.613 \%$ & $0.489 \%$ & $0.006 \%$ \\
        \hline
        \multirow[c]{5}{*}{\nea}
         & Q1 & $0.004 \%$ & $0.004 \%$ & $0.016 \%$ & $0.030 \%$ & $0.001 \%$ \\
         & Q2(Med) & $0.012 \%$ & $0.017 \%$ & $0.036 \%$ & $0.062 \%$ & $0.003 \%$ \\
         & Q3 & $0.027 \%$ & $0.064 \%$ & $0.087 \%$ & $0.113 \%$ & $0.009 \%$ \\
         & Mean & $0.027 \%$ & $0.052 \%$ & $0.072 \%$ & $0.083 \%$ & $0.009 \%$ \\
         & Std & $0.054 \%$ & $0.092 \%$ & $0.118 \%$ & $0.076 \%$ & $0.015 \%$ \\
        \bottomrule
    \end{tabular}
    \caption{Return on investment (ROI, \%) of each game strategic action (except \nea) when applied to the real-world game; other players are assumed not to deviate from their real-world actions.
    In \nea, every player follow the Nash equilibrium. }
    \label{tab:roi}
\end{table}

\begin{table}[!htb]
    \centering 
    \begin{tabular}{c|c|*{5}{c}}
        \toprule
        \multicolumn{2}{c}{} & B30 & E100 & E30 & E5 & T5 \\
        \midrule
        \multirow[c]{5}{*}{\gt}
         & Q1 & $0.181 \%$ & $0.053 \%$ & $0.442 \%$ & $0.588 \%$ & $0.002 \%$ \\
         & Q2(Med) & $0.434 \%$ & $0.203 \%$ & $0.818 \%$ & $0.973 \%$ & $0.005 \%$ \\
         & Q3 & $0.770 \%$ & $0.714 \%$ & $1.344 \%$ & $1.488 \%$ & $0.017 \%$ \\
         & Mean & $0.575 \%$ & $0.525 \%$ & $1.000 \%$ & $1.129 \%$ & $0.026 \%$ \\
         & Std & $0.568 \%$ & $0.744 \%$ & $0.826 \%$ & $0.781 \%$ & $0.066 \%$ \\
        \hline
        \multirow[c]{5}{*}{\ne}
         & Q1 & $0.011 \%$ & $0.000 \%$ & $0.037 \%$ & $0.043 \%$ & $0.000 \%$ \\
         & Q2(Med) & $0.033 \%$ & $0.004 \%$ & $0.105 \%$ & $0.105 \%$ & $0.000 \%$ \\
         & Q3 & $0.081 \%$ & $0.029 \%$ & $0.263 \%$ & $0.196 \%$ & $0.001 \%$ \\
         & Mean & $0.073 \%$ & $0.060 \%$ & $0.226 \%$ & $0.144 \%$ & $0.010 \%$ \\
         & Std & $0.122 \%$ & $0.185 \%$ & $0.367 \%$ & $0.139 \%$ & $0.032 \%$ \\
        \hline
        \multirow[c]{5}{*}{\ine}
         & Q1 & $0.179 \%$ & $0.053 \%$ & $0.444 \%$ & $0.610 \%$ & $0.001 \%$ \\
         & Q2(Med) & $0.432 \%$ & $0.203 \%$ & $0.824 \%$ & $0.976 \%$ & $0.006 \%$ \\
         & Q3 & $0.761 \%$ & $0.714 \%$ & $1.358 \%$ & $1.507 \%$ & $0.020 \%$ \\
         & Mean & $0.574 \%$ & $0.522 \%$ & $0.998 \%$ & $1.125 \%$ & $0.025 \%$ \\
         & Std & $0.570 \%$ & $0.733 \%$ & $0.799 \%$ & $0.729 \%$ & $0.057 \%$ \\
        \hline
        \multirow[c]{5}{*}{\rne}
         & Q1 & $0.184 \%$ & $0.048 \%$ & $0.427 \%$ & $0.606 \%$ & $0.003 \%$ \\
         & Q2(Med) & $0.438 \%$ & $0.197 \%$ & $0.765 \%$ & $0.972 \%$ & $0.010 \%$ \\
         & Q3 & $0.787 \%$ & $0.664 \%$ & $1.296 \%$ & $1.437 \%$ & $0.026 \%$ \\
         & Mean & $0.606 \%$ & $0.522 \%$ & $1.019 \%$ & $1.151 \%$ & $0.035 \%$ \\
         & Std & $0.652 \%$ & $0.819 \%$ & $0.989 \%$ & $0.874 \%$ & $0.078 \%$ \\
        \bottomrule
    \end{tabular}
    
    \caption{Normalized optimality gap of each game strategic action when applied to the real-world game; other players are assumed not to deviate from their real-world actions.}
    \label{tab:gap}
\end{table}

\end{document}